\documentclass[10pt,journal,compsoc]{IEEEtran}

\ifCLASSOPTIONcompsoc
  \usepackage[nocompress]{cite}
\else
  \usepackage{cite}
\fi
\ifCLASSINFOpdf
\else
\fi
\usepackage{lipsum} 
\usepackage{tabularx}
\usepackage{lipsum,amsmath,multicol}
\usepackage{multirow}
\usepackage{array,booktabs}
\usepackage{lipsum} 
\usepackage{tabularx}
\usepackage{adjustbox}
\usepackage{lipsum,amsmath,multicol}
\usepackage{multirow}
\usepackage{array,booktabs}
\usepackage{threeparttable} 
\usepackage{array}
\usepackage{makecell}

\usepackage{pifont}
\usepackage[usestackEOL]{stackengine}
\strutlongstacks{T}
\usepackage{hyperref}
\usepackage{subcaption,siunitx,booktabs}
\newcommand{\comment}[1]{}
\usepackage[table]{xcolor}
\definecolor{lightgray}{gray}{0.9}

\usepackage{float}
\usepackage{ragged2e}
\usepackage{epstopdf}
\AppendGraphicsExtensions{.tif}

\begin{document}
%
\title{Resource Utilization of  Distributed Databases in Edge-Cloud Environment}
%
%
%
%

\author{Yaser Mansouri,
        Victor Prokhorenko,
        Faheem Ullah,
        and  Muhammad Ali Babar
\IEEEcompsocitemizethanks{\IEEEcompsocthanksitem Authors are with Centre for Research on Engineering Software Technology (CREST) Lab. School of Computer Science, The University of Adelaide, Adelaide, Australia.\protect\\
E-mail: yaser.mansouri@adelaide.edu.au
\IEEEcompsocthanksitem J. Doe and J. Doe are with Anonymous University.}
}

%
%

\markboth{Journal of \LaTeX\ Class Files,~Vol.~14, No.~8, December~2020}%
{Shell \MakeLowercase{\textit{et al.}}: Bare Advanced Demo of IEEEtran.cls for IEEE Computer Society Journals}
%



\IEEEtitleabstractindextext{%
\justify
\begin{abstract}
A benchmark study of modern distributed databases (e.g., Cassandra, MongoDB, Redis, and MySQL) is an important source of information for selecting the right technology for managing data in edge-cloud deployments. While most of  the existing studies have investigated  the performance and scalability of distributed databases in cloud computing, there is a lack of focus on  resource utilization (e.g., energy, bandwidth, and storage consumption) of workload offloading for distributed databases deployed in edge-cloud environments. For this purpose, we conducted experiments on various  physical and virtualized computing nodes including variously powered servers, Raspberry Pi, and hybrid cloud (OpenStack and Azure). Our extensive experimental results reveal insights into which database under which offloading scenario is more efficient in terms of energy, bandwidth, and storage consumption.
\end{abstract}

}

\maketitle

\IEEEdisplaynontitleabstractindextext

%
\IEEEpeerreviewmaketitle

\ifCLASSOPTIONcompsoc
\IEEEraisesectionheading{\section{Introduction}\label{sec:introduction}}
\else
\section{Introduction}\label{sec:introduction}
\fi
\IEEEPARstart{H}arnessing the power of cloud computing can improve the usage of computing, storage, networking, and multi-tenant applications and databases over the Internet \cite{rimal2009}.
The centralization of cloud computing introduces delays for time-critical processing over Wide Area Networks (WANs). This downside has led to the deployment of an edge computing paradigm that enables storing and processing data close to data sources rather than sending data to the cloud for processing. Such an approach aims to improve response times and reduce bandwidth consumption which may be critical for IoT applications. Relying on solely edge computing to deploy data-intensive applications, however, might not be always achievable due to limited resources in terms of computing, networking, data storage, and energy \cite{wang2017enorm}. Therefore, the usage of a combined edge-cloud framework may be a viable solution in certain scenarios.

Running databases on the edge-cloud framework is challenging because it should be highly efficient in both performance and utilization of resource-constrained devices. This rises the main question relating to the applicability boundary of distributed databases deployment on the edge-cloud framework in terms of \textit{energy, bandwidth}, and \textit{storage consumption per operations}. We intend to fill this research gap and shed light on the efficiency hierarchy in terms of energy, storage, and bandwidth consumption for Cassandra, MongoDB, Redis, and MySQL. In addition, we share the experience related to the challenges we faced and the lessons we learned throughout resource measurement experiments.

Deployment of distributed databases on edge-cloud framework enables \textit{task offloading} from resource-constrained devices to powerful servers. The task offloading concept involves the questions of \textit{when}, \textit{where} and \textit{what} tasks should be offloaded (i.e. executed remotely).\footnote{The answer to ``when'' a task should be transferred is beyond the scope of this paper. Interested readers are referred to \cite{Jiang2019}.}

As an answer to \textit{where} a task should be offloaded, we considered three options: \textit{edge device}, \textit{adjacent server}, and  \textit{remote server}. In our work, a laptop and a cluster of Raspberry Pis (RPi) are considered as edge devices. A high-performance server with a distance of several meters from the edge devices is an adjacent server. VMs in the hybrid cloud are considered as \textit{remote servers}. All edge devices, adjacent server, and remote servers have been connected through an overlay WireGuard\footnote{WireGuard: \url{https://www.wireguard.com}} network. These three offloading destinations allow us to investigate offloading workloads under different scenarios in which resource richness and the distance between \textit{database client} running workloads and \textit{databases servers} hosting data are varied.  

Our main criteria to select NoSQL and relational databases in this study are popularity, usage, and commercialization by well-known cloud providers. Thus, we selected Cassandra\footnote{Cassandra: \url{https://cassandra.apache.org/}}, MongoDB\footnote{Mongo/MongoDB: \url{https://www.mongodb.com/}},  Redis\footnote{Redis: \url{https://redis.io/}}, and MySQL\footnote{MySQL: \url{https://www.mysql.com/}}. 
These databases are often evaluated only in terms of throughput, response time, and scalability in both private and public clouds \cite{Rabl2012}\cite{Kuhlenkamp2014}\cite{Li2013}\cite{Mansouri20}\cite{Li2014}. These metrics are not enough for database selection because resource consumption is crucial for low-powered devices in edge-cloud scenarios. Therefore, we measured the resource utilization of these databases during workload offloading from edge nodes to powerful computing nodes. For the purpose of our study, we primarily focus on database client node resource utilization.

The consumption of \textit{resources}  we focus on is \textit{energy}, \textit{network bandwidth}, and \textit{storage} \footnote{We also presented resource utilization in terms of per-operation efficiency. This simplifies comparing efficiencies of different databases in addition to raw performance.}. Energy consumption is a key cost function in offloading because edge devices commonly have limited battery life, which depletes quicker under high load \cite{chen2018multi}. We measure the energy consumption of CPU, RAM, and the rest of the system (i.e., SSD, ports, screen, and so on). Bandwidth consumption of the database client node refers to the amount of data transferred during the task offloading \cite{yu2017survey}. The amount of bandwidth consumed impacts both response time and potential traffic costs. Storage cost is another essential metric in the edge-cloud framework due to increasing volumes of data generated by IoT devices. This metric refers to the data storage consumption of edge node or remote servers where the offloaded task is performed \cite{shi2016edge}.  
Therefore, we investigate \textit{how efficient is a database in terms of resource consumption (\textit{energy}, \textit{bandwidth}, and \textit{storage}) for offloading various workloads under different scenarios that are different in resource richness, connection types, and distance between database client and servers.} 

To conduct the above investigation, we leveraged multiple RPis, a laptop (termed \textit{edge node} hereafter), a high-performance adjacent server (termed \textit{edge server node} henceforth), and a cluster of VMs in a hybrid cloud. We also considered both WiFi and cable connections between database client and servers. Our experimental scenarios are defined in two categories: (a) \textit{offloaded scenarios} in which the client node is deployed in resource-constrained nodes and database servers are hosted at richer computing nodes, and (ii) Non-offloaded (local) scenarios in which database client and servers are residing on the same computing node. We evaluated these scenarios from a resource consumption perspective using different tools.  
To measure energy consumption, we relied on Intel's Running Average Power Limit (RAPL) technology \cite{Khan2018}. We also used  \textit{iperf3}\footnote{Iperf3: \url{https://iperf.fr}}, and \textit{iftop}\footnote{iftop: https://linux.die.net/man/8/iftop} network tools to measure the traffic between database client that runs the YCSB workloads \cite{Cooper2010} and database servers that host data. We used the standard  \textit{df} utility to measure storage consumption on database servers.

Our contributions are threefold: (1)  We present a modular edge-cloud framework in which the whole process of cloud infrastructure deployment/destruction, database installation, and database cluster configuration are performed in a fully automated manner; (2) We evaluate resource usage in terms of energy, network bandwidth, and storage to explore the feasibility of workloads offloading for distributed databases in the edge-cloud framework; (3) we finally discuss our experimental findings.

\begin{table*}[t]
\begin{threeparttable}
  \centering
  \caption{Comparison of empirical studies on the evaluation of  Distributed DataBases (DDB) and big data frameworks in edge-cloud paradigms. In this table, \textbf{E} stands for Energy, \textbf{R} for Run-time, \textbf{B} for Bandwidth, and \textbf{S} for Storage.}    
   \label{tab:relatedwork}%
     \begin{tabular}{p{1.3cm}p{1.5cm}p{2.5cm}p{5.2cm}p{0.5cm}p{3.6cm}p{0.5cm}p{0.5cm}}
     
    \toprule
          &    & &    & \multicolumn{2}{c}{\textbf{Evaluation metrics}} \\\hline
    \midrule
    \multicolumn{1}{p{1cm}}{\textbf{Paper}}
    &\multicolumn{1}{p{0.8cm}}{\textbf{Application}}
    &\multicolumn{1}{p{1cm}}{\textbf{Infrastructure}}
    &\multicolumn{1}{p{2cm}}{\textbf{Databases}}
    &E  &R &B &S   \\\hline
    \cite{Rabl2012}   &DDB$^{\dagger}$ &Private cloud   & \Longunderstack{Cassandra, HBase, Redis,\\ Voldemort, VoltDB, MySQL}      &\ding{56} &\ding{51}(throughput,latency)  &\ding{56} &\ding{51}\\\hline
    \cite{Kuhlenkamp2014}   &NDB$^{\ddagger}$ &Public cloud   &Cassandra and HBase 
    &\ding{56} &\ding{51}(throughput,scalability)  &\ding{56} &\ding{56}\\\hline
	\cite{Klein2015}   &NDB &Public cloud  &MongoDB, Cassandra, Riak 
	&\ding{56} &\ding{51}(throughput vs. consistency)  &\ding{56} &\ding{56}\\\hline
	\cite{Li2013} &DDB &Private cloud  & MongoDB, RavenDB, CouchDB, MySQL Cassandra,Hypertable, Couchbase
	&\ding{56} &\ding{51}(throughput,latency)  &\ding{56} &\ding{56}\\\hline
	\cite{Abramova2013} &NSDB   &Private cloud   &Cassandra, MongoDB 
	&\ding{56} &\ding{51}(latency)  &\ding{56} &\ding{56}\\\hline
	\cite{Mansouri20} &DDB      &Hybrid cloud &\Longunderstack{Cassnadra, MongoDB, Riak,\\ CouchDB, Redis, MySQL} 
	&\ding{56} &\ding{51}(throughput, latency)  &\ding{56} &\ding{56}\\\hline
    \cite{Mansouri20dis} &DDB      &Hybrid cloud &\Longunderstack{Cassnadra, MongoDB, Riak,\\ CouchDB, Redis, MySQL} 
    &\ding{56} &\ding{51}(throughput vs. distance)  &\ding{56} &\ding{56}\\\hline
    \cite{Mahajan2017} &NDB     &Server(s)  &\Longunderstack{Cassnadra, MongoDB} 
    &\ding{51} &\ding{51}(latency)  &\ding{56} &\ding{56}\\\hline
    \cite{Bani2016} &DDB     &NA  & MongoDB, MySQL, PostgresSQL 
    &\ding{51} &\ding{51}(response time)  &\ding{56} &\ding{56}\\\hline
    \cite{Li2014} &DDB,BD$^{\ast}$ &Single node  & Cassandra, HBase, Hive, Hadoop 
    &\ding{51} &\ding{51}(response time)  &\ding{56} &\ding{56}\\\hline
    \cite{Morabito2017} &General &Edge(RPis)  & NA 
    &\ding{51} &\ding{51}(response time)  &\ding{56} &\ding{56}\\\hline
    \cite{Lin2007} &RDB &Fog  & PostgresSQL
    &\ding{56} &\ding{51}(CPU usage)  &\ding{51} &\ding{56}\\\hline
    \cite{Hajji2016}  &General &RPis  &Hadhoop, Spark
    &\ding{51} &\ding{51}(CPU usage)  &\ding{51} &\ding{56}\\\hline
    \cite{Scolati2019}  &General &RPis  &Hadhoop, Spark &\ding{56} &\ding{51}(CPU and RAM usage)  &\ding{51} &\ding{56}\\\hline
    \cite{Liu2021}  &DDB &RPis  &MongoDB, SQLite, LevelDB  &\ding{51} &\ding{51}(latency)$^{+}$  &\ding{51} &\ding{56}\\\hline
    \textbf{Our work}  &\textbf{DDB} &\textbf{Edge-cloud}  &\textbf{MongoDB, Cassandra, Redis, MySQL} &\ding{51} &\ding{51}(\textbf{Run-time})  &\ding{51} &\ding{51}\\\hline
    
    \bottomrule
    \end{tabular}%
    \begin{tablenotes}
      \tiny
       \item $\dagger$ DDB stands for distributed databases and includes both relational and NoSQL databases. $\ddagger$ NDB stands for NoSQL databases and includes only NoSQL databases.$\ast$ BD stands for Big Data. $+$Storage I/O bandwidth has been measured.
    \end{tablenotes}
\end{threeparttable}
\vspace{-5mm}
\end{table*}%

\section{Related Work}
To position the novelty of our work with respect to the state-of-the-art, we divided the related studies into the following categories. Table \ref{tab:relatedwork} compares these notable studies.

\textbf{Performance Evaluation of Distributed Databases on Clouds:} 
With the advent of NoSQL databases, researchers conducted a variety of experimental evaluations and achieved notable results from a performance perspective. Rabl et al. \cite{Rabl2012} presented a comprehensive performance evaluation in terms of throughput, latency, and disk usage for six modern databases on two different private clusters using the YCSB workloads.  Kuhlenkamp et al. \cite{Kuhlenkamp2014} evaluated the correlation between scaling speed and throughput for Cassandra and HBase\footnote{HBase: \url{https://hbase.apache.org/}} on different Amazon EC2 infrastructure configurations. Klein et al. \cite{Klein2015} analyzed the impact of consistency models (eventual, quorum-based, and strong) of MongoDB, Cassandra, and Riak running on a single node and a cluster of nodes at Amazon EC2. In \cite{Li2013}, the authors investigated the read and write performance, and concluded that not all No-SQL databases have outperformed SQL databases. In \cite{Abramova2013}, the study compared  MongoDB and Cassandra in read and write performance on VMware Player. We recently evaluated the performance of six distributed databases on a hybrid cloud \cite{Mansouri20}. Also, we measured the impact of distance on the performance of distributed databases as the vertical and horizontal scalability of a hybrid cloud are changed \cite{Mansouri20dis}. Differently, our work expands previous evaluations to consider constrained resources, which are crucial for workload offloading in edge-cloud scenarios.

\textbf{Energy Consumption of Distributed Databases on edge-Cloud computing:} Several studies evaluated the energy efficiency of NoSQL and relational databases. Mahajan et al. \cite{Mahajan2017}\cite{MAHAJAN2019}  evaluated the impact of the optimized queries on performance, power, and energy efficiency for  MongoDB, Cassandra, and MySQL using a single server. Authors in \cite{balaji2015} measured the power consumption and performance of Cassandra and the impact of different power management techniques on the power consumption of the Cassandra cluster. 
Bani \cite{Bani2016} presented an empirical study on the impact of cloud applications (i.e., Local Database Proxy, Local Sharding-Based Router, and Priority Message Queue) on the performance and energy consumption of MongoDB, MySQL, and PostgreSQL.  Li et al. \cite{Li2014} studied a benchmark of energy consumption of Selection, Grep, Aggregation, and Join operations for Cassandra, HBase, Hive\footnote{Hive: \url{https://hive.com/}}, and Hadoop\footnote{Hadoop: \url{https://hadoop.apache.org}} on a single node. 
Liu et al.  \cite{Liu2021} studied energy consumption benchmarks for MongoDB, SQLite, and LevelDB on RPi3, RPi4, and ODROIDC2\footnote{ODROIDC2: \url{https://www.hardkernel.com/}}. This work is the closest to ours, however, it only considered resource consumption of non-offloading scenarios with standalone databases devices.

\textbf{Performance Evaluation of Distributed Databases in Edge Computing:} 
Some researchers studied the deployment of data-intensive applications in edge computing.  In \cite{Morabito2017}, authors deployed different models of RPis in the form of native (bare metal) and Docker virtualization to evaluate energy, network, disk, and RAM consumption under compute-intensive and network-intensive scenarios. Several studies made effort to select or adapt cloud-based distributed databases for the edge computing paradigm. Alelaiwi et al. \cite{ALELAIWI2018} explored an analytic hierarchy process to evaluate the usability, portability, and support ability of database development tools for IoT databases in edge computing. Mayer et al. \cite{Mayer2017} tailored distributed data store 
for fog computing and deployed the MaxiNet network emulator \cite{szymaniak2005} on a server with 8 cores to simulate 6 fog nodes to measure latency for conducting operations in Cassandra based on the proposed policy. Lin et al. \cite{Lin2019} presented a protocol to measure CPU and bandwidth usage of read-only and update transactions in PostgresSQL.  \cite{Hajji2016} presented the performance of HDFS (Hadoop Distributed File System) on a single RPi and a 12-node RPi cluster and  \cite{Scolati2019} demonstrated CPU and RAM usage for the same frameworks, however, on a containerized RPis cluster. We measured the resource consumption of distributed databases to provide insight into the suitability of workload offloading in an edge-cloud environment.

\textbf{Computation Offloading Towards Edge Computing:}
Offloading compute-intensive tasks has attracted researchers' attention to optimize response time and reduce energy consumption through different optimization techniques. 
Authors in \cite{vu2020}  provided a joint offloading and resource allocation framework for hierarchical cooperative fog computing nodes to optimize energy consumption using an improved branch-and-bound algorithm. Pei et al. \cite{Pei2020} studied energy-efficient resource allocation through latency-sensitive tasks offloading in Mobile Edge Computing (MEC). Ghmary et al. \cite{Khan2020} used Integer-Linear programming to offload tasks from a mobile node to MEC to optimize energy consumption and latency. \cite{Li2019} and \cite{Mehrabi2021} studied numerical optimization offloading approaches and \cite{Huynh2021} combined such approaches with data caching to reduce battery energy and latency in MEC.   In contrast to these studies, Canete et al. \cite{canete2021} proposed the implementation of offloading decisions based on tasks and infrastructure for mobile IoT applications to reduce energy usage. 

Recently, researchers proposed AI-based offloading approaches to optimize energy consumption and response time. In \cite{Duy2019}, authors developed human- and device-driven intelligent algorithms for offloading tasks to reduce energy consumption and latency in edge computing.  Zhou et al . \cite{Zhou2021} proposed ML-based dynamic offloading and resource scheduling to save energy in mobile edge nodes. \cite{Breitbach2021} presented a ML-based code offloading to reduce energy for edge devices. \cite{Lan2020}\cite{Chen2022} achieved a reduction in energy consumption and latency through a combination of ML-based tasks offloading and caching data. Reinforcement Learning (DRL) \cite{wang2021}\cite{QU2021}\cite{Wu2020}, Markov-based \cite{CHEN2021}\cite{LIU2022}, partial code offloading \cite{Lin2019}, intelligent collaboration for computation offloading \cite{Kang2020}, data synchronization and management via offloading techniques \cite{Wang2019} \cite{Qiu2022} have proposed to improve response time, energy, and bandwidth consumption. 

All the above studies investigated optimization techniques to make decisions on either partial or full offloading for computational tasks including video rendering, gaming, etc. Also, these solutions have been evaluated through simulation in which tasks are defined based on the required CPU cycles, amount of memory, and bandwidth. In contrast, we investigated resource consumption for emerging NoSQL databases. The closest work to ours is \cite{Liu2021}, which differs in infrastructure scale and databases selected (\cite{Liu2021} in Table \ref{tab:relatedwork}). The rest of the studies listed in Table \ref{tab:relatedwork}, primarily focus on performance and scalability rather than measuring resource consumption in terms of energy, bandwidth, and storage as the key offloading factors. Our work is complementary to these works, as we evaluated the resource consumption of distributed databases for offloaded workloads from resource-constrained to resource-rich devices. 

\section{Design and Implementation of Edge-Cloud Framework}
This section discusses the design and implementation of our edge-cloud framework.

\begin{figure*}[t]
\centering
\includegraphics[width=1\textwidth]{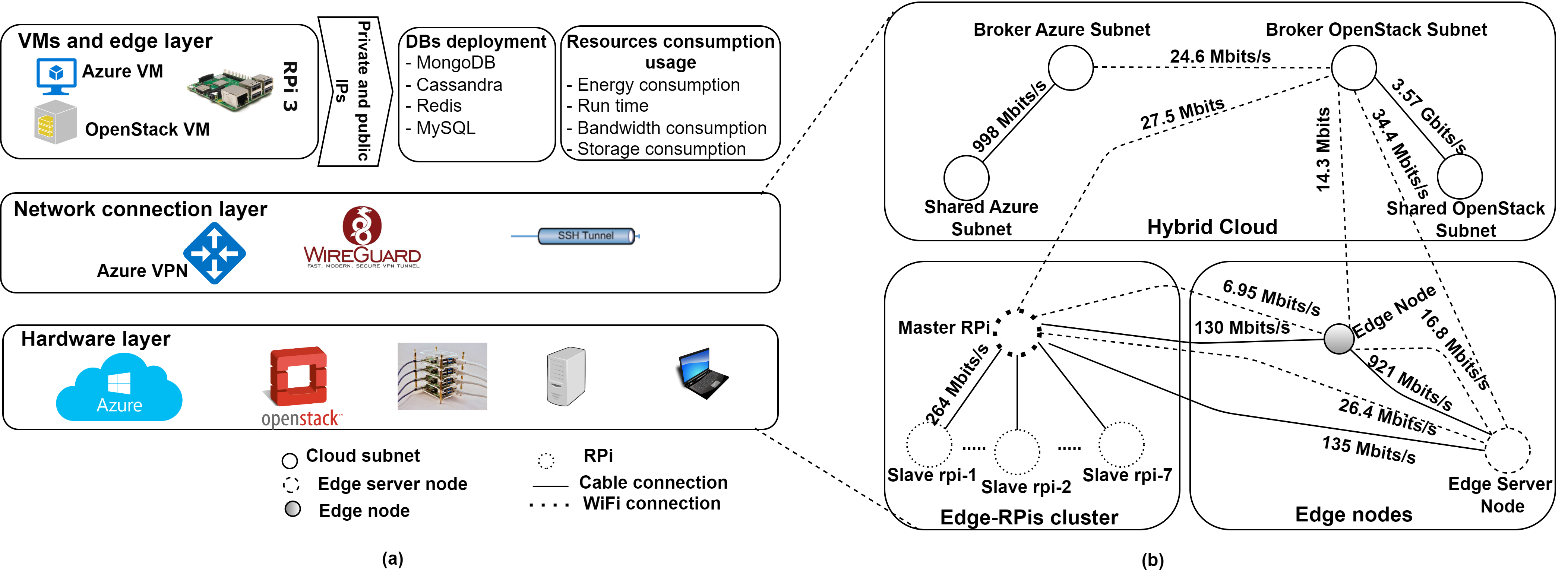}
\vspace{-3mm}
\caption{(a) A hierarchical architecture of our edge-cloud framework, (b) Overview of the implemented edge-cloud framework. Links label shows bandwidth.}
\label{fig:cloud-edge-arch}
\vspace{-3mm}
\end{figure*}

\vspace{-3mm}
\subsection{Implementation of Edge-cloud Framework}
We designed a layered edge-cloud framework (Fig. \ref{fig:cloud-edge-arch}(a)). The bottom layer is \textit{hardware infrastructure} that consists of edge nodes, RPis, and VMs in a hybrid cloud. 
The middle layer is \textit{network connection} that includes WireGuard to build an overlay network across different nodes. The top layer of the framework is \textit{VMs and edge deployment} in which we used  Terraform to deploy VMs in a hybrid cloud. The output of this layer is a set of VM IPs, which enables the deployment of distributed databases across computing nodes to measure resource utilization. We discuss two bottom layers in this section and the topmost layer in the next section. 

The network connectivity topology and individual link throughput used within our experiment sets are detailed in Fig. \ref{fig:cloud-edge-arch}(b).  To have reproducible resource deployment in a hybrid cloud, we used Terraform\footnote{Terraform: \url{https://www.terraform.io/}}. Despite the illusion of unlimited resources available in clouds, the increased network latency may negatively impact the real-time analysis of large amounts of data. In addition, potential costs increase as well as privacy challenges inherent to cloud environments need to be taken into account. Thus, we deployed RPis and edge nodes, where their computing and storage resources form a hierarchy with regard to resource richness (Fig. \ref{fig:cloud-edge-arch} (b)). To have richer resources in edge computing, we built a cluster of  8  RPis connected through a Gigabit switch.

To make a network connection across all computing nodes in the edge-cloud framework, we leveraged WireGuard which is faster and more cost-efficient compared to the VPNs provided by commercial cloud providers \cite{Mansouri20}. 
A key metric of network connection strength is the network throughput (measured in terms of data transferred and received per second). 
We leveraged Iperf3 \footnote{Iperf3: \url{https://iperf.fr}}  to measure the throughput between the two end nodes in both directions.  
We installed this tool on all nodes and ran it for 10 minutes to record the throughput between each pair of computing nodes, as labeled on links in Fig. \ref{fig:cloud-edge-arch}(b). As can be seen, the network connection between VMs in the private cloud achieves the highest throughput of 3.57 Gbits/sec,  whereas this value for VMs in the public cloud holds the second rank and is of 998 Mbits/sec. The reason behind such values is the VMs in the private cloud may reside on the same server, while in the public cloud VMs might be provisioned in different servers or even different racks. In contrast, the lowest network throughput is observed across private and public clouds (24.6 Mbits/sec), and the master RPi and the broker VM in the private cloud (34.4 Mbits/sec). 

\begin{table*}[t]
	\caption{A summary of resource consumption probes.}\label{tab:probs}
	\centering
	\vspace{-3mm}
	\begin{tabular}{p{4.7cm} p{7.1cm} p{3.2cm}}
		\hline
		Probe          &Functionality     &Device/command utility  \\\hline\hline
		Edge-node-energy-consumption     &The energy consumption of edge node/edge server node   & RAPL \\
		Battery     &The energy consumption of battery for edge node         & Upower \\
		USB-energy-consumption     & The energy consumption of master RPi    & USB Power Meter    \\
		Power-socket-energy-consumption     &The energy consumption of  RPis cluster   &Energy Cost Meter \\\hline
		Bandwidth consumption     &The transferred data between all computing nodes  & \textit{iftop} utility command\\
		Storage consumption     &The required storage to run the YCSB Workload &\textit{df} utility command \\\hline
	\end{tabular}
	\vspace{-5mm}
\end{table*}

\subsection{Implementation of Resources Consumption Probes}\label{sec:resource-consumption}
We discuss the following resource consumption probes as summarized in Table \ref{tab:probs}.

\textbf{Energy consumption probes:} These probes are implemented through both software and hardware tools, which depend on the facilities provided by the computing nodes. For the edge node and edge server node, we provided an \textit{edge-node-energy-consumption-probe} that leverages energy Running Average Power Limit (RAPL) to measure the energy consumption of CPU and RAM \cite{Khan2018}. For the edge node, we implemented a \textit{battery-probe}, which exploits Upower\footnote{Upower: \url{https://www.commandlinux.com/man-page/man1/upower.1.html}}  command to measure the battery depletion of the edge node. Based on these two probes, we measured the energy consumed by the rest of the system (i.e., storage, ports, screen, etc.) in the edge node.  For the master RPi, we implemented a \textit{USB-energy-consumption-probe} in which the energy consumption of the master RPi is recorded with the help of the USB Power Meter (UPM) -- WEB U2 model. UPM can provide voltage readings down to 0.01V and current to 0.001A, which can be either displayed on the built-in LCD or recorded in a file. We exploited Energy Cost Meter (ECM) to implement \textit{power-socket-energy-consumption-probe} to measure the energy of the whole cluster of RPis.  ECM measures voltage and the current range of 200-276V AC and 0.01-10A, respectively. For the virtualized resources in the hybrid cloud,  we did not provide energy measurement probes for two main reasons. (i) The depletion of energy resources of the edge computing nodes is crucial in the context of edge computing. (ii) It is almost impossible to measure the energy consumption of a server for individual tasks in a cloud since each server provides multi-tenant services.

\textbf{Bandwidth consumption probe:} This probe captures the amount of data transferred and received between nodes. This measurement is implemented through  \textit{iftop}, which monitors the ingress and egress bandwidth of a network interface. This service is termed \textit{bandwidth-consumption-probe} and we activated it on the network interfaces of computing nodes issuing and receiving operations. 

\textbf{Storage consumption probe:}
This probe measures the consumed storage during workloads execution against a particular database. We used the standard \textit{df} to implement the \textit{storage-consumption-probe}. We activated this service during experiments on the disk hosting the database. 

Once the workload runs on the client node, the probes start to measure the consumed resources. Upon finishing the execution of a workload, the probes are stopped and results are collected for analysis.

\section{Distributed Databases and Workloads}
We discuss databases under evaluation and workloads used.
\vspace{-3mm}
\subsection{NoSQL and Relational Databases}
We evaluate Mongo and Cassandra as document-based NoSQL databases \cite{Mansouri2017a}\cite{Han2011}  and MySQL as the most-used relational database in the industry sector. In addition, we chose Redis as an in-memory database for evaluation.

\begin{table}[t]
	\begin{threeparttable}
	\caption{Core workloads in YCSB.}\label{tab:ycsb}
	\centering
	\begin{tabular}{p{1.7cm} p{3.4cm} p{2cm}}
		\hline
		Type &         Operations &Label    \\\hline\hline
		Workload A     & 50\% Read + 50\% write & Write-intensive  \\
		Workload B     & 95\% Read + 5\% Update  & Read-intensive  \\
		Workload C     & 100\% Read              & Read-only    \\\hline
		Workload D     &95\% Read + 5\% Insert   & Read-latest\\
		Workload E     &95\% Scan + 5\% Insert   & Scan\\
		Workload F     &50\% Read + 50\% RMW        &RMW$^{\dagger}$ \\\hline
	\end{tabular}
	\begin{tablenotes}
      \tiny
      \item $\dagger$ RMW stands for read-modify-write
      \end{tablenotes}
	\end{threeparttable}
	\vspace{-3mm}
\end{table}

\vspace{-3mm}
\subsection{Workloads}
We used  YCSB workload (v0.15.0) \footnote{YCSB Workload: \url{https://github.com/brianfrankcooper/YCSB}} to evaluate both NoSQL and relational databases. 
The YCSB workload facilitates a set of tunable parameters and acts on a loose schema including a string key assigned to a collection of fields, which themselves are the string to binary blob key-value pairs. The YCSB workload consists of elementary operations such as read, write, and insert for a record based on a single key. YCSB also supports a  complicated “scan” operation, which refers to a paging operation starting from a particular key. Due to these advantages, our experiments targeted 6 core workloads as summarised in Table \ref{tab:ycsb}.

We used the default YCSB workload configuration values except for two parameters: the number of records and operations. We adjusted them based on our hardware infrastructure support. For RPis and the edge node, we set up 10K records, while for the edge server node, which is more powerful, we set this parameter to a value of 10M records. Nevertheless, we used a variable value for the number of operations in each workload for RPi, edge node, and edge server node. The reason behind such a setting is that the information about battery depletion of the edge node is updated every two minutes. If we set the number of operations with a small value, then the implemented battery-depletion-probe might record zero for energy consumption. This implies that the workload runs out before updating data regarding to battery depletion. To avoid such an issue, we initially ran the workload for 10K operations and then the number of operations was calculated as throughput achieved for 10K operations multiplied by 1200 seconds (20 minutes). This duration time of 20 minutes for running the YCSB workload gives a good enough precision with respect to the battery depletion information. 
\section{Performance Evaluation}
In this section, we describe the setup of our edge-cloud framework and delineate our experimental results.
\begin{table}[t]
	\caption{A summary of infrastructure setup.}\label{tab:infras}
	\centering
	\vspace{-3mm}
	\begin{tabular}{p{2.3cm} p{0.8cm} p{1.2cm} p{1cm} p{1cm}}
		\hline
		Computing node     &Number     &CPU(cores)         &RAM     &Disk     \\\hline\hline
		Private VM         &1-8       &2        &4 GiB  &40 GiB                \\
		Public VM          &0-7       &1         &2 GiB  &30 GiB                \\
		Server edge node   &1         &8        &16 GiB  &1 TB                \\
		Edge node          &1         &4        &8 GiB   &250 GiB    \\
		RPi                &7        &4        &1 GiB   &16 GiB          \\\hline
	\end{tabular}
	\vspace{-3mm}
\end{table}

\vspace{-3mm}
\subsection{Testbed Setup}
The edge-cloud framework consists of the following computing components as summarized in Table \ref{tab:infras}.  
\textbf{Hybrid cloud:} We built the hybrid cloud on the on-premises infrastructure virtualized through OpenStack at Adelaide University and Azure datacenter in the Sydney region \cite{Mansouri20}.  We exploited clusters of VMs in the hybrid cloud with a size of ($n\_m$), where $n$ $(1\leq n \leq8)$ and $m$ $(0\leq m \le7)$ are the number of nodes on the private and public clouds, respectively. Based on the cluster size, we considered 3 combinations of the hybrid cloud configuration settings: $(8\_0)$, $(4\_4)$, $(1\_7)$. This allowed us to evaluate the hybrid cloud when (a) most nodes sit on either the private or public cloud, or (b) nodes are equally distributed on each cloud side. Each VM in the private cloud has  2 vCPUs,  4 GiB RAM, and 40 GiB HDD,  and the size of each VM in the public cloud is Standard B1m (1 vCPU, 2 GiB, 30 GiB HDD). 

\textbf{RPis cluster:} We built a homogeneous cluster of 8 RPis 3 Model B+, where each RPi is equipped with a Quad-core CPU, 1 GiB RAM, and 16 GiB microSD storage.\footnote{Please note that a single VM on both private and public clouds possesses a CPU with fewer cores compared to the RPi, but a cluster of VMs provides more CPU cores. The impact of horizontal and vertical scalability of VMs on energy consumption remains as future work.}

\textbf{Edge node:} We deployed two different edge nodes. (i) a laptop, referred to as \textit{edge node}, has a Quad-core CPU, 16 GiB RAM, and 256 GiB SDD. (ii) The high-performance edge server, refereed as \textit{edge server node}, provides an 8-core CPU, 32 GiB RAM, and 1 TB SSD. 

\begin{table}[t]
	\begin{threeparttable}
	\caption{Experimental scenarios.}\label{tab:scenarios}
	\centering
	\tiny
	\begin{tabular}{p{1.7cm} p{1.7cm} p{1.8cm} p{1.7cm}}
		\hline
		Scenario\#      &Database worker             &Database servers &Concept    \\\hline   
		Scenario 1      &RPi                              &RPi                    &Non-offloading         \\   	
		Scenario 2      &RPi                              &Edge node (C$^{\dagger}$) &Offloading       \\
		Scenario 3      &RPi                              &Edge node (W$^{\ddagger}$)   &Offloading    \\
		Scenario 4      &RPi               &Edge server node (C) &Offloading \\
		Scenario 5      &RPi                         &Edge server node (W) &Offloading\\
		Scenario 6      &RPi                         & Hybrid cloud &Offloading\\\hline
		Scenario 7      &Edge node                   &Edge node &Non-offloading \\
		Scenario 8      &Edge node                   &Edge server node (C)  &Offloading\\
		Scenario 9      &Edge node                   &Edge server node (W)  &Offloading\\
		Scenario 10     &Edge node                   &Hybrid cloud  &Offloading\\\hline
		Scenario 11     &Edge server node            &Edge server node &Non-offloading \\
		Scenario 12     &Edge server node            &Hybrid cloud  &Offloading\\\hline
		Scenario 13     & RPi                        & Cluster of RPis &Offloading \\\hline
	\end{tabular}
	\begin{tablenotes}
      \tiny
      \item $\dagger$ C stands for a cable connection. 
      \item $\ddagger$ W stands for a WiFi connection.
      \end{tablenotes}
	\end{threeparttable}
	\vspace{-3mm}
\end{table}

\textbf{Experimental scenarios:} 
We considered three types of workers and four types of database servers (Table \ref{tab:scenarios}). We generally offloaded data from the resource-constrained to the more powerful computing resources (the database servers).  This concept of offloading includes all scenarios except scenarios 1, 7, and 11 in which the worker and the server are the same computing node.  Such scenarios termed non-offloading (local) scenarios, provide more insight into the databases in terms of energy consumption when databases are utilized locally. 
Furthermore, we considered different connection types for the RPi and edge node that give us insight into the effectiveness of databases from an energy consumption perspective as a faster connection (cable vs. WiFi) is used (Scenarios 2, 3, 4, 5, 8, and 9). For simplicity of presentation, a scenario of (A $\rightarrow$  B (C/W)) indicates that A is a worker and B is a database server, and the connection between them is either Cable or WiFi. The extensive nature and high flexibility of our framework enable us to investigate, evaluate, and provide recommendations on resource utilization of distributed storage systems occurring in real-world scenarios. This includes examples such as distributed monitoring, data logging for delayed processing, and predictive analytics in the context of smart farming, mobile field operations, and electric power grids. 

\begin{figure}[t]
\centering
\includegraphics[width=0.7\columnwidth]{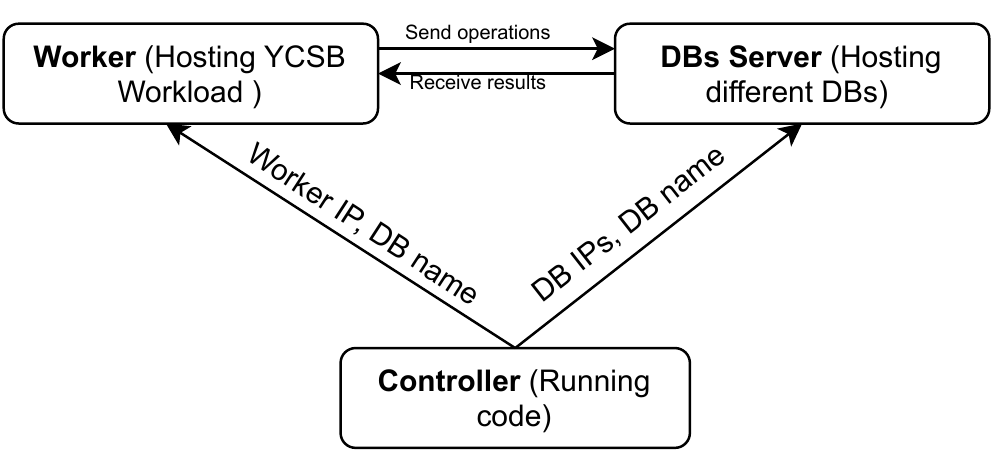}
\vspace{-3mm}
\caption{A schematic of modular components for running experimental scenarios in our edge-cloud framework.}
\label{fig:mcomponents}
\vspace{-5mm}
\end{figure}

To evaluate the experimental scenarios, we implemented a modular approach including three components: \textit{controller node},  \textit{worker/database client node}, and \textit{database server nodes} (Fig. \ref{fig:mcomponents}).  The controller node initially receives  IPs of computing nodes as input and then runs installation and cluster configuration of databases across those database servers if needed. At the same time, the controller node communicates with the worker node to set up the probes and runs the YCSB workload.  Once the database workloads are sent to the DB server nodes, all resource consumption probes are activated to record the consumed energy, bandwidth, and storage of the worker and server(s). It should be noted that uploading the probes consumes energy, and we thus exclude it from the experimental results. 
We also ran all scenarios without running YCSB for 20 minutes and measured only the idle energy consumption. Then, this idle energy consumption is subtracted from the one for the corresponding scenario in which the YCSB workload was run.

\vspace{-3mm}
\subsection{Experimental Results}
This section explains energy, bandwidth, and storage consumption for the scenarios listed in Table \ref{tab:scenarios}.

\vspace{-3mm}
\subsubsection{Energy Consumption}
We investigate the energy usage (in Joules per Million Operations (J/MOPs)) of different databases \footnote{Note that there is a direct correlation between energy consumption and database throughput in all experiments.  Though, we did not plot database throughput here due to space constraints.}.  

\begin{figure*}[ht!]
  \centering
  \subfloat[Cassandra]{\label{fig:rpi-cass-af}\includegraphics[height=5cm,width=0.5\textwidth]{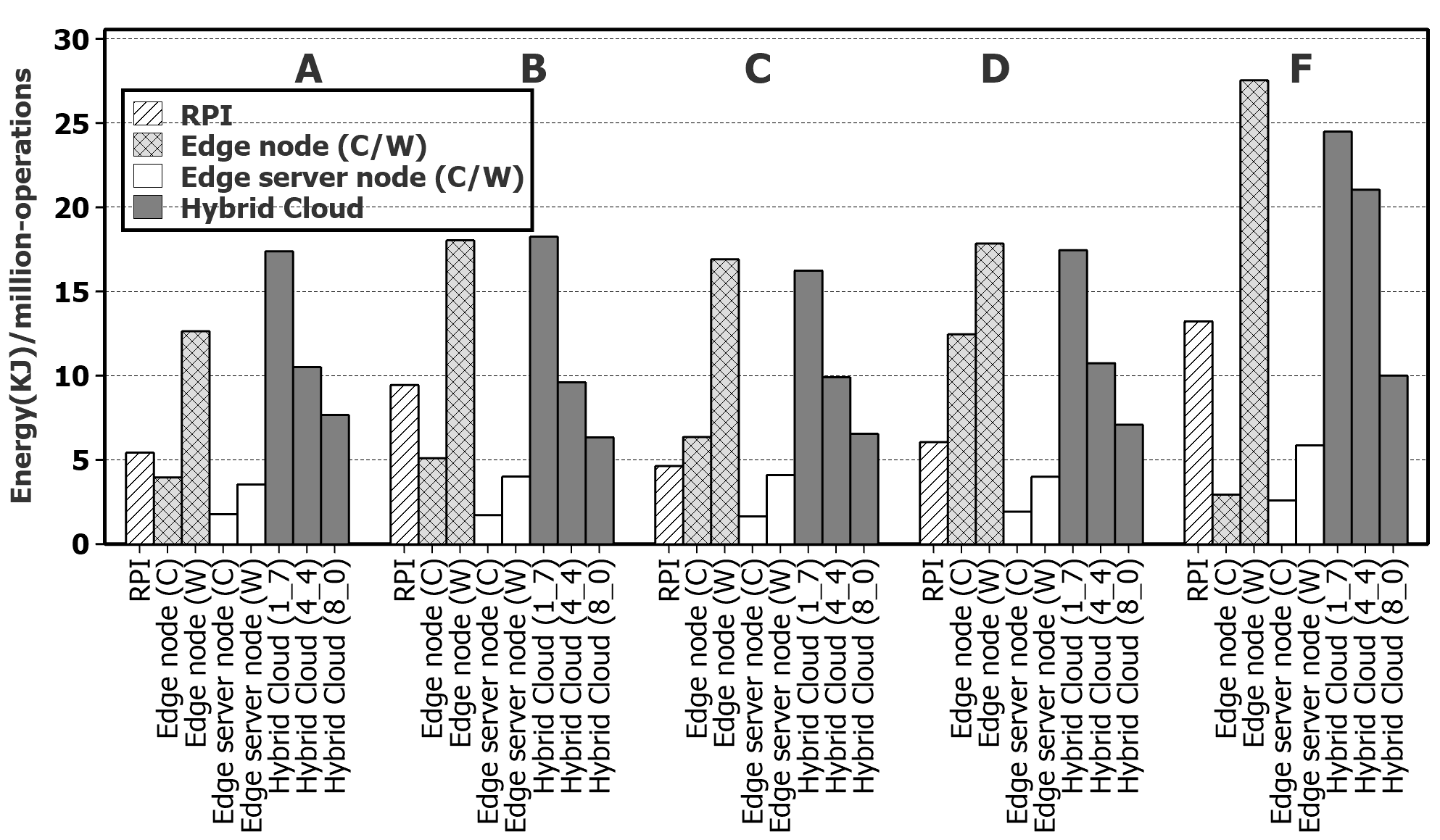}}
  \subfloat[Mongo]{\label{fig:rpi-mongo-af}\includegraphics[height=5cm,width=0.5\textwidth]{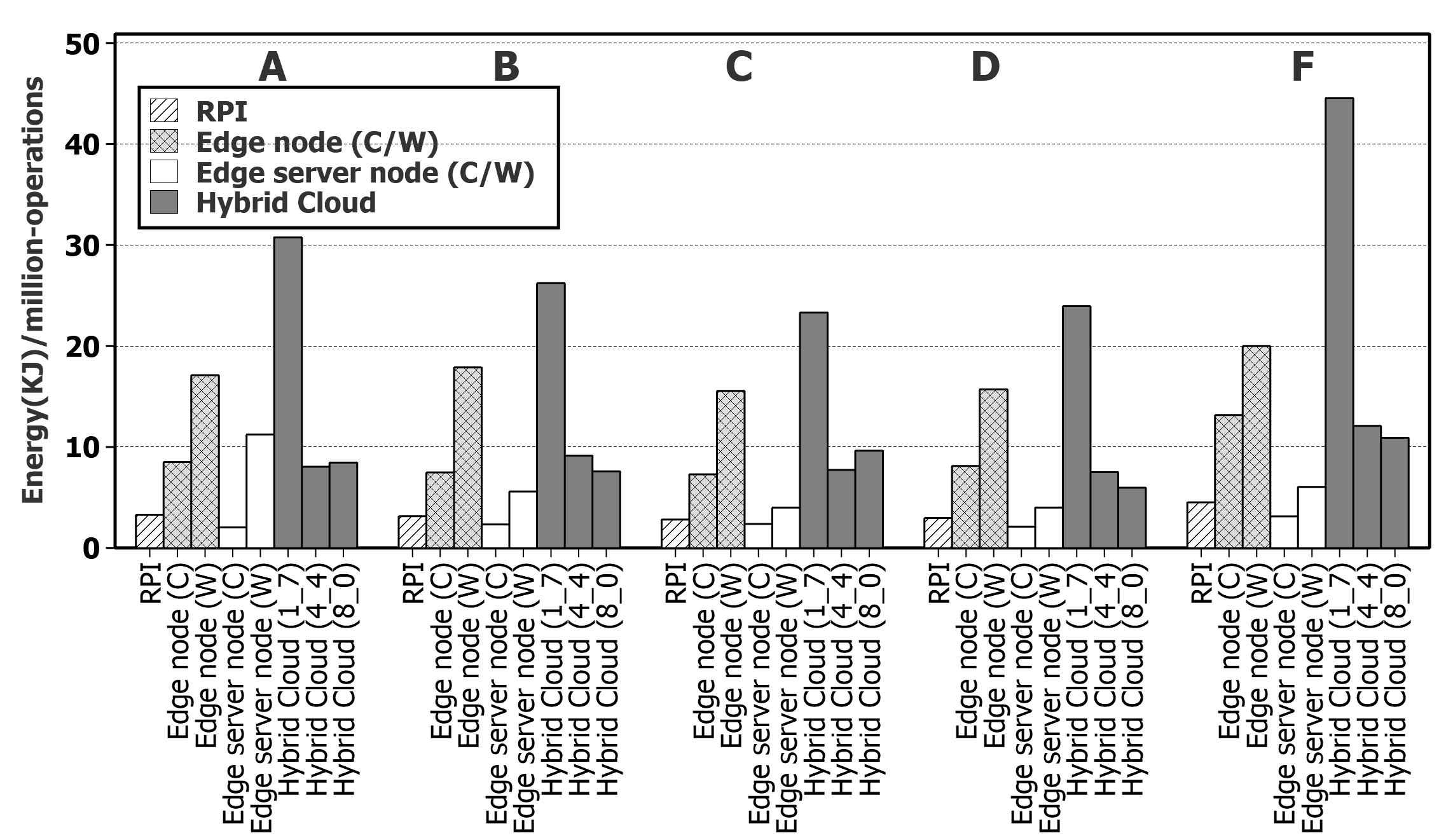}}\\
  \subfloat[Redis]{\label{fig:rpi-redis-af}\includegraphics[height=5cm,width=0.5\textwidth]{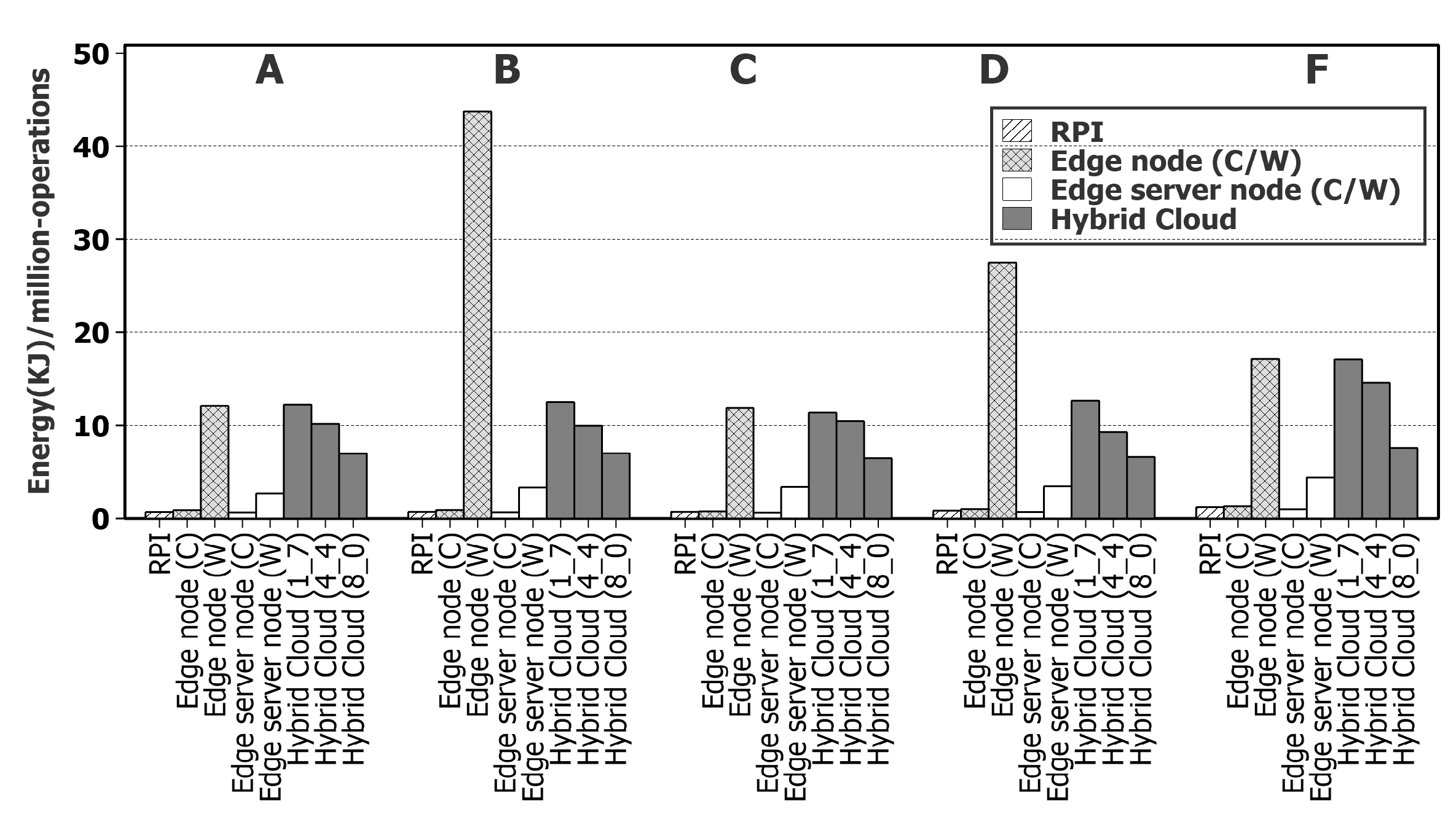}}
  \subfloat[MySQL]{\label{fig:rpi-mysql-af}\includegraphics[height=5cm,width=0.5\textwidth]{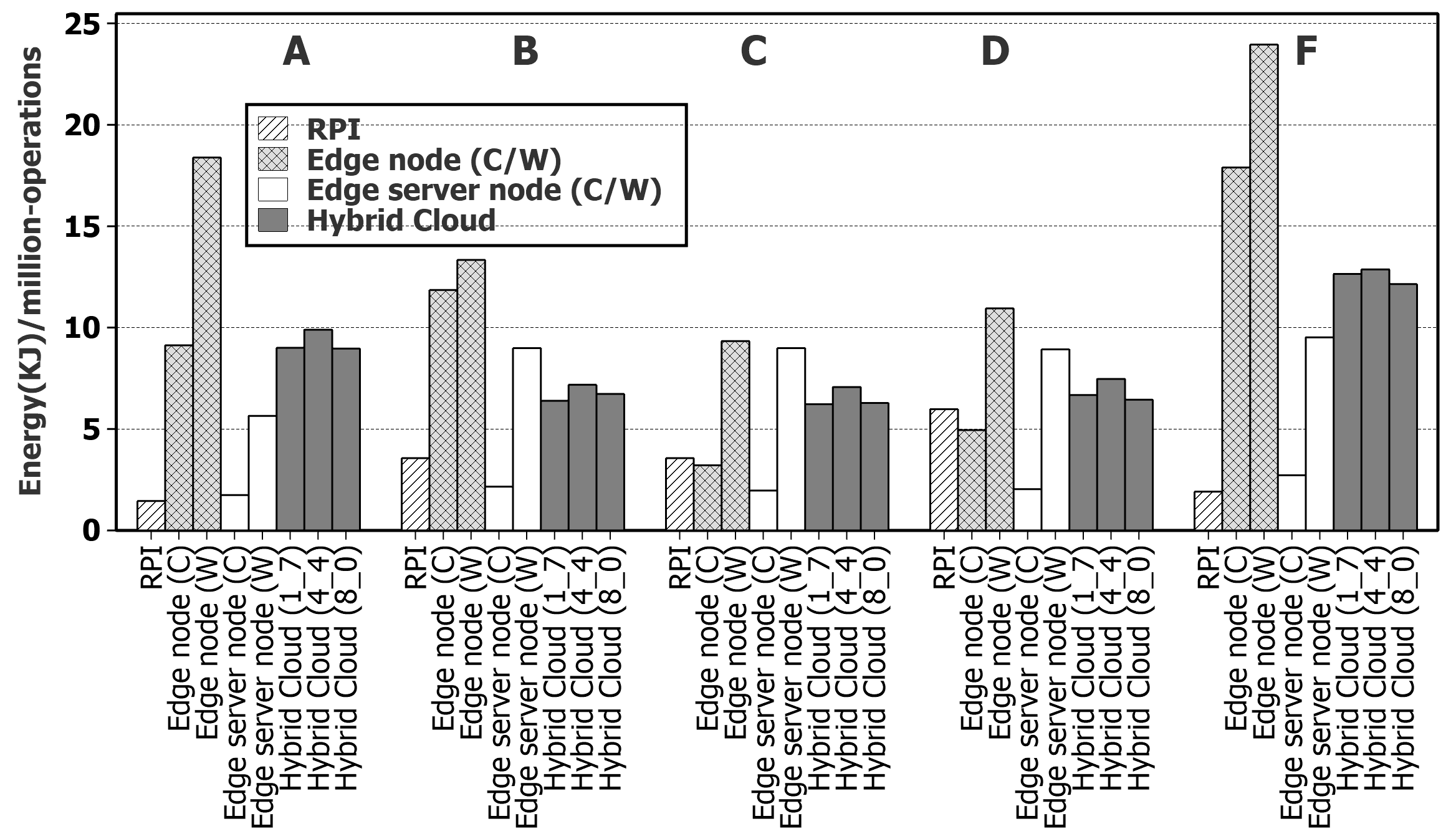}}
  \caption{Energy consumption of data offloading from \textbf{RPi} to the computing nodes for Workloads \textbf{A}, \textbf{B}, \textbf{C}, \textbf{D}, and \textbf{F}. (m\_n) indicates $m$ and $n$ nodes in the private and public clouds respectively. C/W denotes a Cable/WiFi connection. }
\label{fig:rpi-af}
\end{figure*}

\textbf{(A) The energy consumption of a single RPi (Scenarios 1-6).}
Fig. \ref{fig:rpi-cass-af} shows the energy consumption of \textbf{Cassandra}. For (RPi $\rightarrow$ RPi),  the energy consumption is about 5000 J/MOPs for workloads (A, C, and D) and about 2 and 2.5 times this value for workloads B and F, respectively. As we move to (RPi $\rightarrow$ edge node (C)), the energy consumption for workloads A, B, and F respectively reduces by 28\%, 46\%, and 78\%  compared to the ones for (RPi $\rightarrow$ RPi). In contrast, in the same scenario with the WiFi (W) connection, the energy consumption increases by 190-363\%  for all workloads compared to the ones for (RPi $\rightarrow$ RPi). This implies that faster connections cause less energy consumption. For (RPi $\rightarrow$ edge server node (C/W)), Cassandra requires less energy to serve workloads as compared to both discussed scenarios.  The value for this scenario decreases between  50 (Workload A - Cable)-82\% (Workload B - WiFi) in contrast to the values for (RPi $\rightarrow$ RPi).  This indicates that more powerful computing resources at a close distance from the worker allow for saving energy. For (RPi $\rightarrow$  hybrid cloud),  as the number of VMs in the public cloud increases, the energy consumption of all workloads raises from 10 KJ/MOPs for (8\_0)  to 25 KJ/MOPs for (1\_7). This means the worker requires more time to receive responses from database servers due to a longer distance. All workloads (except F) have the same energy consumption (more than 15 KJ/MOPs) for (1\_7), which implies that the energy consumption is dominated by the distance between nodes regardless of the workload. 
\textit{In summary, two out of six scenarios are energy-efficient in offloading for Cassandra (Table \ref{tab:scenariocomparison-rpi}). }  

Fig. \ref{fig:rpi-mongo-af} illustrates the energy consumption of \textbf{Mongo}.  For (RPi $\rightarrow$ RPi), RPi consumed energy between 2800 (workload C)- 4500 J/MOPs (workload F), which is 39-65\% less than the consumed energy for Cassandra. This fact can be explained by memory swapping occurring on RPi to operate Cassandra due to RAM constraints.  
In contrast, the relaxation of this constraint through hosting Mongo on the edge node (i.e., RPi $\rightarrow$ edge node (C)) with more memory capacity, the energy consumption grows by a factor of  (1.05 - 2.41) against Cassandra. This is because Cassandra utilizes the CPU effectively compared to Mongo, which results from the internal design and implementation of these databases \cite{lakshman2010cassandra}.  For WiFi connection, there is no obvious supremacy of Mongo and Cassandra over each other because the network fluctuations have an impact on the execution time of databases, which leads to the increment/decrement of energy consumption. Similarly, we observe the same trend for (RPi $\rightarrow$ edge server node (C)) in which Mongo requires more energy by a factor of (at most) 1.44  for workload C in comparison with Cassandra.  For (RPi $\rightarrow$ hybrid cloud), Mongo, compared to Cassandra, increases energy usage by  (30-80\%) for (8\_0)  and by (9-47\%) for (1\_7). This is because Cassandra is balancing data placement, while Mongo is not\footnote{Due to space constraint, we did not present the bandwidth usage across participant nodes for (RPi $\rightarrow$ hybrid cloud). }. \textit{ In summary, Mongo consumes more energy than Cassandra on average except for the scenario with a memory shortage. Also, the hierarchy of scenarios for Mongo has changed slightly in comparison to Cassandra (Table \ref{tab:scenariocomparison-rpi}).}

\begin{figure*}[t!]
  \centering
  \subfloat[Cassandra]{\label{fig:rpi-cass-e}\includegraphics[height=5cm,width=0.25\textwidth]{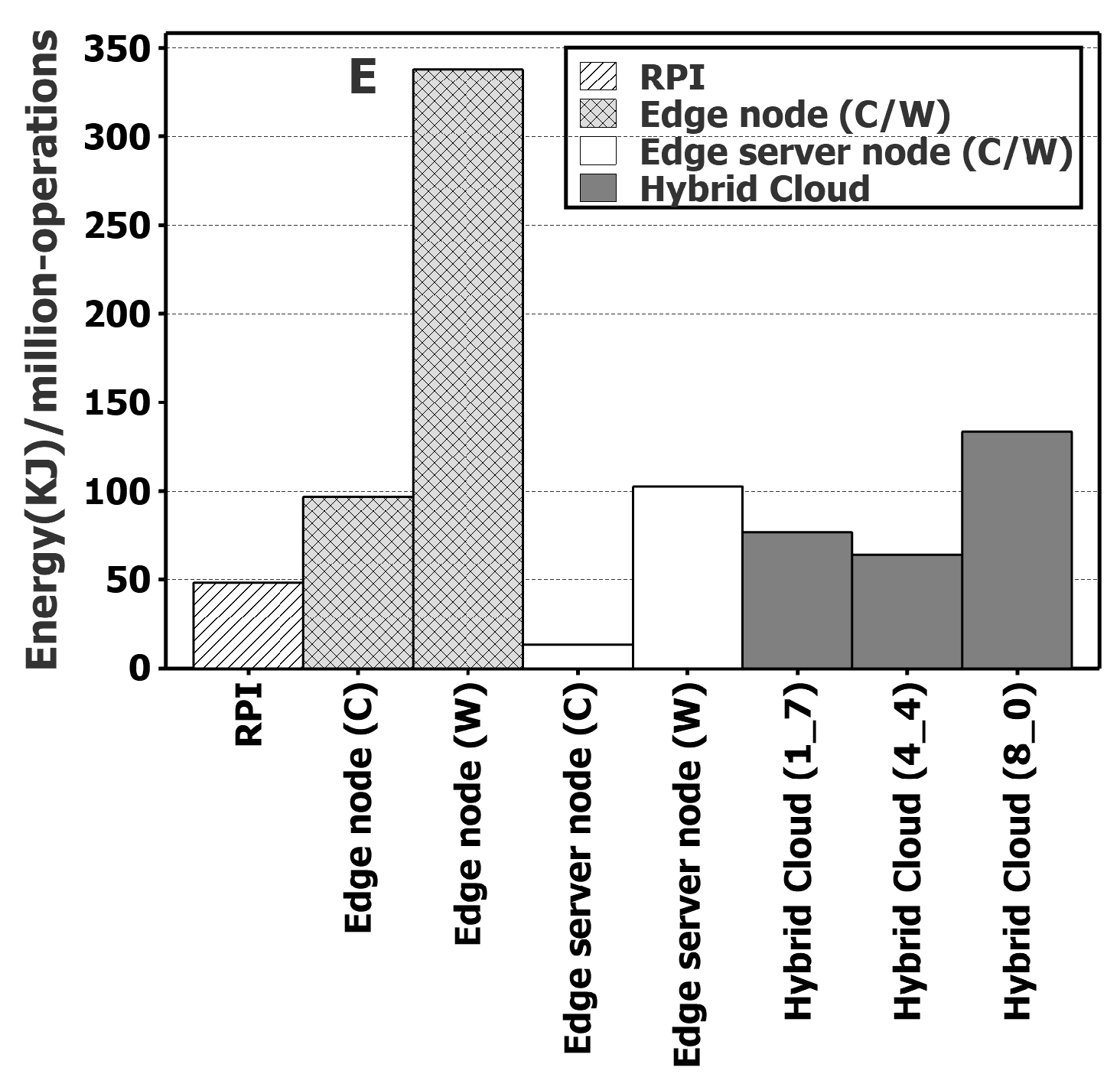}}
  \subfloat[Mongo]{\label{fig:rpi-mongo-e}\includegraphics[height=5cm,width=0.25\textwidth]{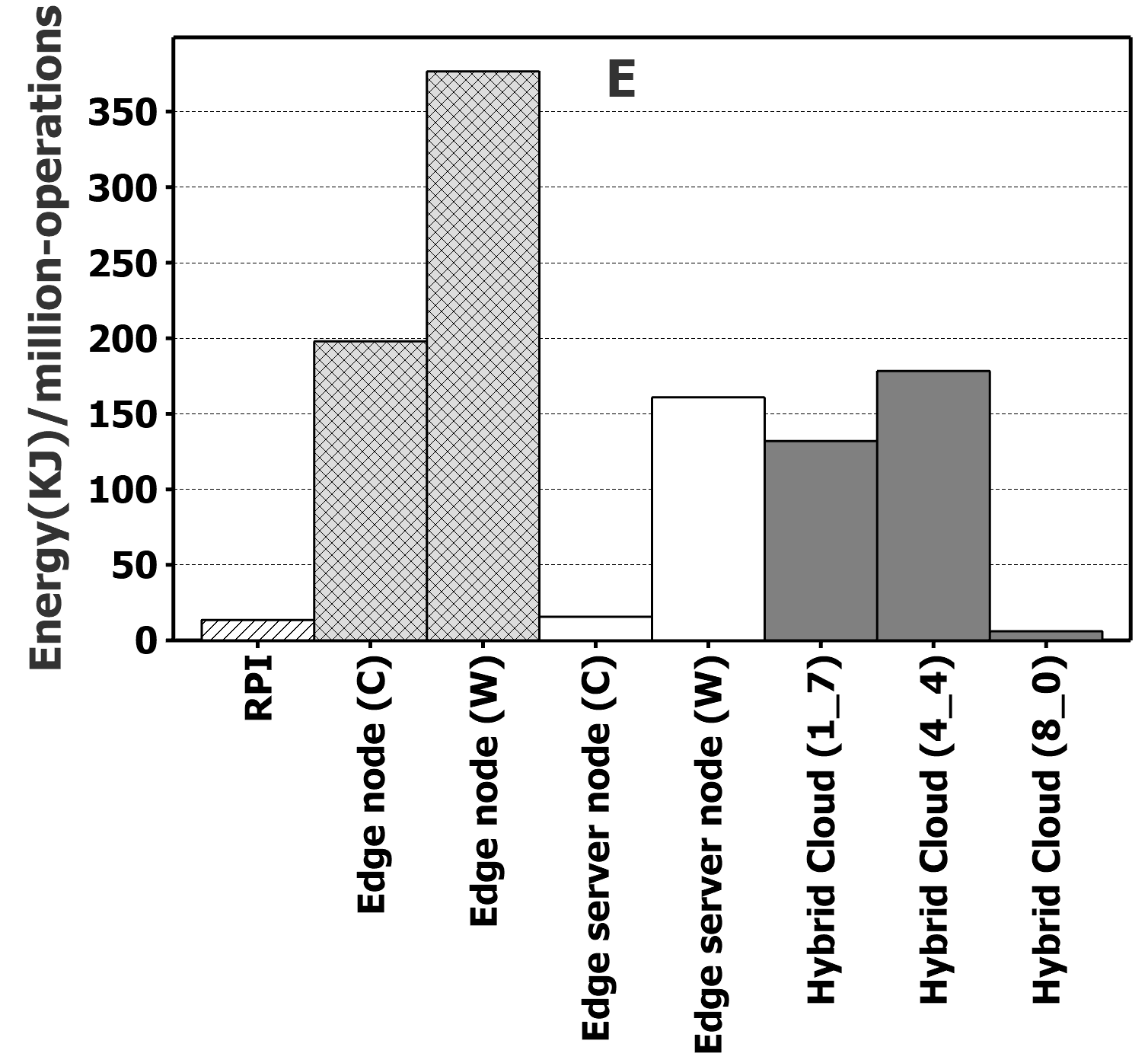}}
  \subfloat[Redis]{\label{fig:rpi-redis-e}\includegraphics[height=5cm,width=0.25\textwidth]{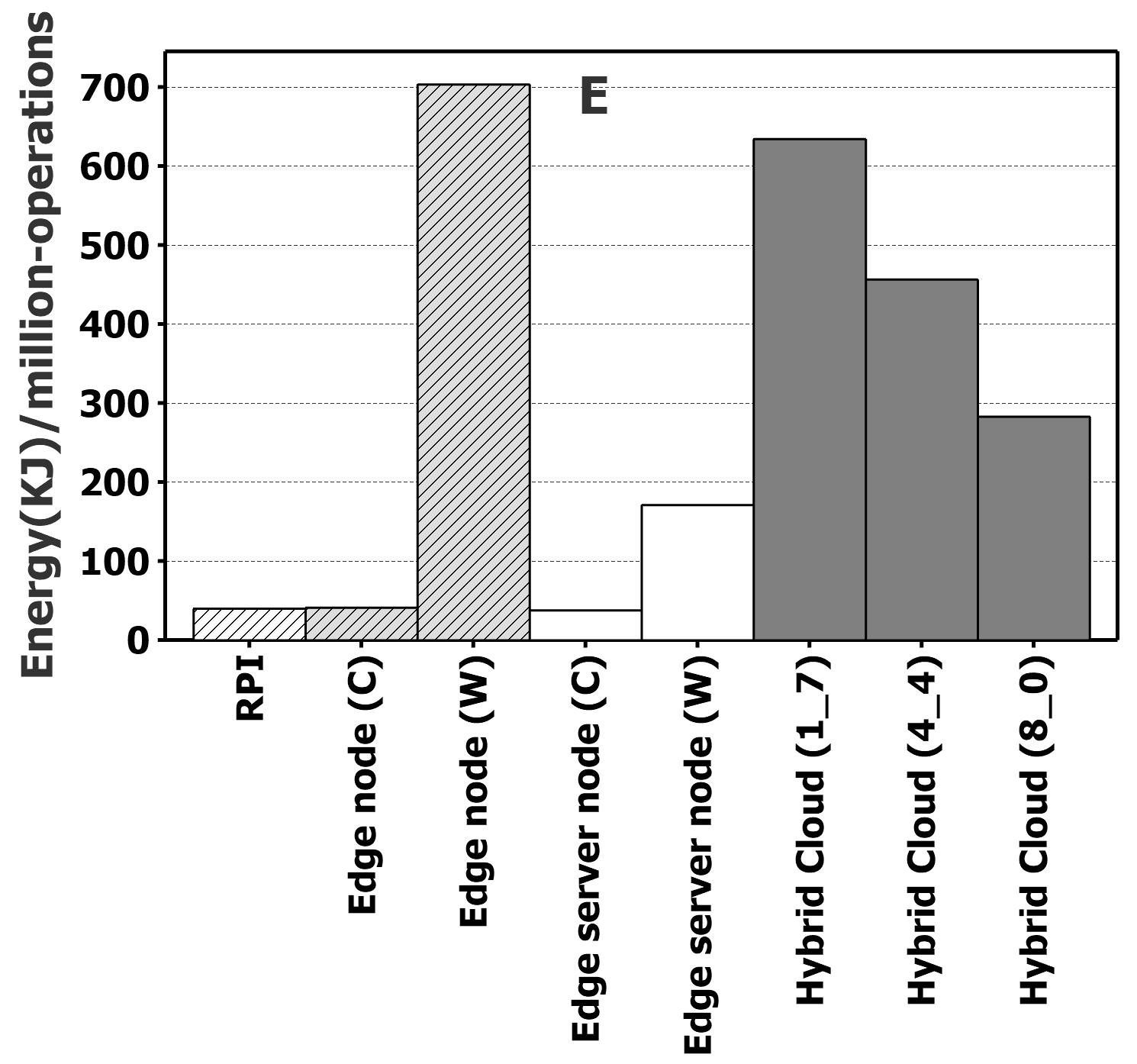}}
  \subfloat[MYSQL]{\label{fig:rpi-mysql-e}\includegraphics[height=5cm,width=0.25\textwidth]{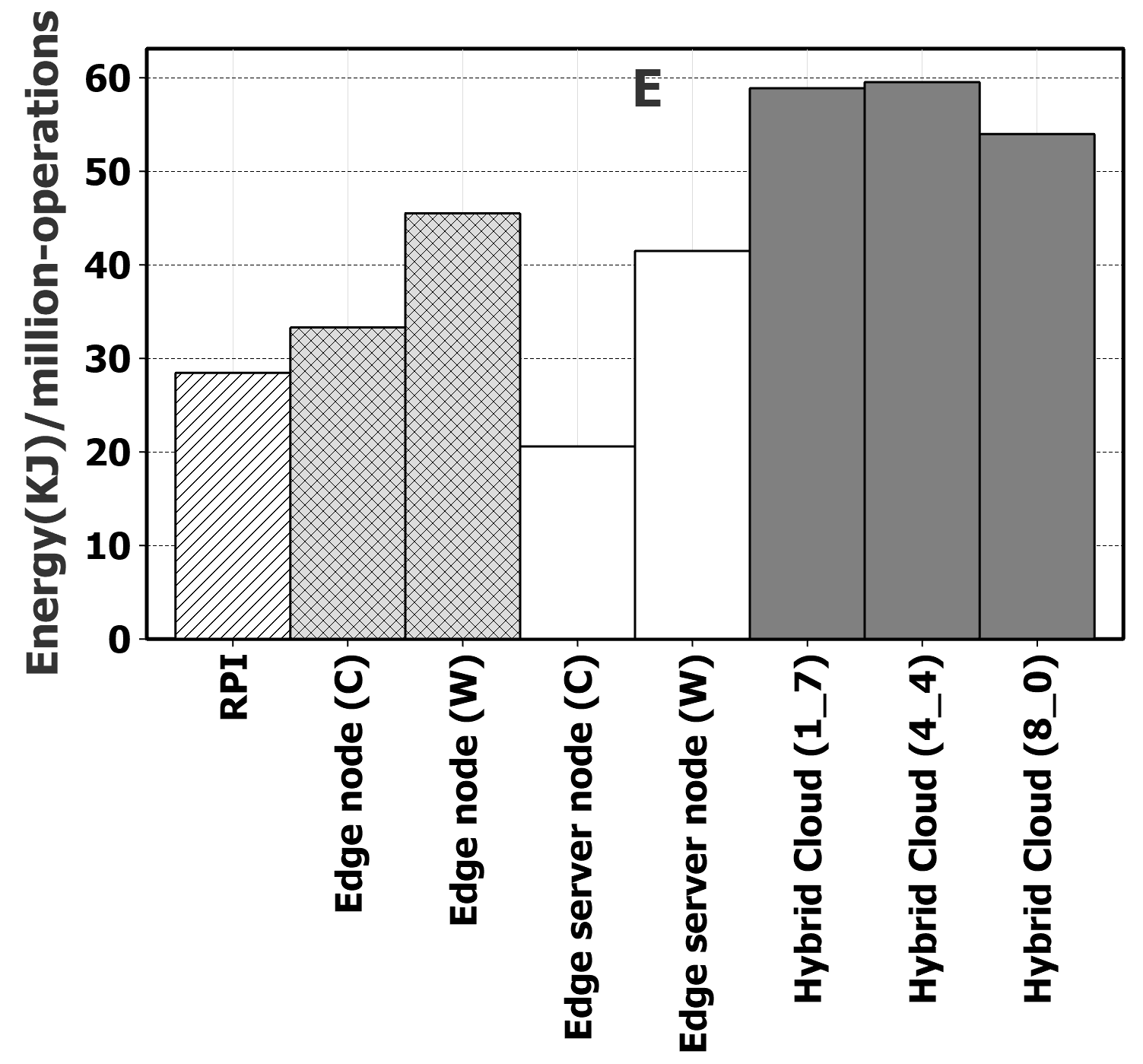}}
  \caption{Energy consumption of data offloading from \textbf{RPi} to the computing nodes for Workload \textbf{E}. (m\_n) indicates $m$ and $n$ nodes in the private and public clouds, respectively. C/W denotes a Cable/WiFi connection. }
\label{fig:rpi-e}
\vspace{-5mm}
\end{figure*}

Fig. \ref{fig:rpi-redis-af} depicts energy usage of \textbf{Redis}, which  
 is significantly less than Cassandra and Mongo for all scenarios (except for (RPi$\rightarrow$ edge node (W)). For  (RPi$\rightarrow$ RPi), Redis decreases energy consumption by (55-92\%) and (75-90\%) with respect to Cassandra and Mongo, respectively. We can also see the same trend for (RPi$\rightarrow$ edge node (C/W)). Redis consumes (88-90\%) and (55-92\%) less energy than Mongo and Cassandra for (RPi$\rightarrow$ edge node (C)); Likewise, (66-73\%) and (60-63\%) for (RPi$\rightarrow$ edge server node (C)). For the WiFi setting, Redis also outperforms Cassandra and Mongo except for workloads B and D (Fig. \ref{fig:rpi-redis-af}), where we observed instability in connection.  
In fact, Redis reduces energy consumption by (15-30\%) and (5-38\%) compared to Mongo and Cassandra respectively as it is hosted on the edge node (W). Likewise, (28-76\%) and (17-25\%) reduction in energy compared to Mongo and Cassandra when Redis is deployed on the edge server (W).
As data is offloaded to the hybrid cloud, Redis outperforms Cassandra and Mongo in energy consumption. For example, the maximum energy consumption by (1\_7) is slightly more than 10 KJ/MOPs for all workloads except F, while for Cassandra and Mongo, this value grows to 15 and 20 KJ/MOPs, respectively.  \textit{In summary, apart from (RPi$\rightarrow$ edge node (W)), Redis outperforms Mongo, which in turn, outweighs Cassandra in energy consumption. The hierarchy of scenarios for Redis is different from the one for Cassandra and Mongo (Table \ref{tab:scenariocomparison-rpi}}). 

Fig. \ref{fig:rpi-mysql-af} illustrates the energy usage of \textbf{MySQL}. Results show that for write-related workloads (A, F) under the (RPi $\rightarrow$ RPi) scenario,  MySQL consumes more energy than Redis by (1.5-1.9) times, while less energy than Cassandra and Mongo by (3.74-6.97) and (2.27-2.37) times respectively. This can be explained by the fact that Redis outperforms MySQL in response time due to its RAM-based nature.
By contrast, for the same scenarios, Mongo outperforms MySQL by (1.1-1.9) times in energy consumption. This shows the superiority of Mongo over MySQL in response time.  

The (RPi $\rightarrow$ edge node (C)) and  (RPi $\rightarrow$ edge server node (C)) scenarios respectively are more energy-efficient by 40-67\%   and  11-18\%  in offloading vs. non-offloading due to the fast CPU and network connection. In contrast,  using the same computing nodes with the Wifi connection makes offloading non-effective. Under the same scenarios, MySQL performs worse than Redis, and these scenarios consume (4.05-13.5) and (2.56-3.12) times more energy respectively. This is because more memory allows Redis to run faster. However, on average, MySQL saves 9\%  (resp. 18\%) energy compared to the deployment of  Mongo (resp. Cassandra) on the edge node (resp. the edge server node). 

For (RPi $\rightarrow$ hybrid cloud), unlike the other databases, the configuration of the hybrid cloud does not impact the energy consumption of MySQL. The results show that MySQL consumes 6-10 KJ/MOPs for workloads (A-D) and around 12.5 KJ/MOPs for workload F, which is less than the ones for Cassandra and Mongo and stays competitive with the energy consumption of Redis. The reason behind such results is that MySQL supports strong consistency in a data node group (i.e., two replicas on the private cloud) and then the updated data is asynchronously propagated to other data node groups. Thus, the only latency between RPi and the VMs in the private cloud is reflected in the energy consumption.  
\textit{In summary, Redis outperforms MySQL in almost all scenarios in terms of energy consumption, while MySQL is relatively effective in energy consumption compared to Mongo and Cassandra for (RPi $\rightarrow$ hybrid cloud).  Furthermore, Table \ref{tab:scenariocomparison-rpi} summarizes the energy consumption of different scenarios from lowest to highest, where the rank of (RPi$\rightarrow$ edge server node (W)) and (RPi$\rightarrow$ hybrid cloud) is exchangeable based on the workload.}

\begin{table}[t!]
	\caption{A sorted list of the lowest to the highest energy consumption for scenarios 1-6.}\label{tab:scenariocomparison-rpi}
	\centering
	\tiny
	\vspace{-3mm}
	\begin{tabular}{p{1.7cm} p{1.9cm} p{1.9cm} p{2cm}}
		\hline
		Cassandra                 &Mongo                           &Redis      &MySQL                \\\hline\hline
		Edge server node (C)     &Edge server node (C)             &  Energy server node (C)      &  Edge server node (C)  \\
		Edge server node (W)     &RPi (local)                      & RPi (local)                          &  RPi (local) \\
		RPi (local)                      &Hybrid cloud (4\_4, 8\_0)        & Edge node (C)                &  Edge server node(B,C,D)   \\
	    Edge node (C)	         &Edge server node (W)             & Edge server node (W)         &  Hybrid cloud (A,F)   \\
		Hybrid cloud (all)       &Edge node (C/W)                  & Hybrid clouds(all)           &  Edge node (C)     \\
	    Edge node (W)            &Hybrid cloud (1\_7)              & Edge node (W)                &  Edge node (W)  \\\hline
	\end{tabular}
	\begin{tablenotes}
      \small
      \item \tiny 
   \end{tablenotes}
	\vspace{-5mm}
\end{table}

Fig. \ref{fig:rpi-e} depicts energy consumption of \textbf{workload E} for scenarios 1-6.
Results show that the energy usage of workload E is higher than the one for the other workloads. This is because workload E is expensive in terms of operations. Cassandra and Mongo respectively consume the highest (48.15 KJ/MOPs) and the lowest (13.5 KJ/MOPs) energy under the non-offloading scenario. When other edge computing nodes host databases, only the edge server node (C) provides promising offloading, where its energy consumption decreases by 73\%  for Cassandra, 6\%  for Redis, and 28\% for MySQL.  Mongo suffers a 15\% energy usage increase under the same conditions. This shows that more capacity of RAM accelerates the response time of Cassandra and Redis, which results in energy consumption reduction. 
For (RPi $\rightarrow$ hybrid cloud), MySQL operates the best and Redis acts the worst in the case of energy usage with a value of (300-600) KJ/MOPs and (520-600) KJ/MOPs, respectively. This is because Redis transmits more data to the public, while MySQL requires the least.  In the same scenario, Cassandra (at most 140 KJ/MOPs)  and Mongo (at most 185 KJ/MOPs) achieves middle ranks in energy consumption.

\begin{figure*}[t!]
  \centering
  \subfloat[Cassandra]{\label{figur:laptop-cass-af}\includegraphics[height=5cm,width=0.5\textwidth]{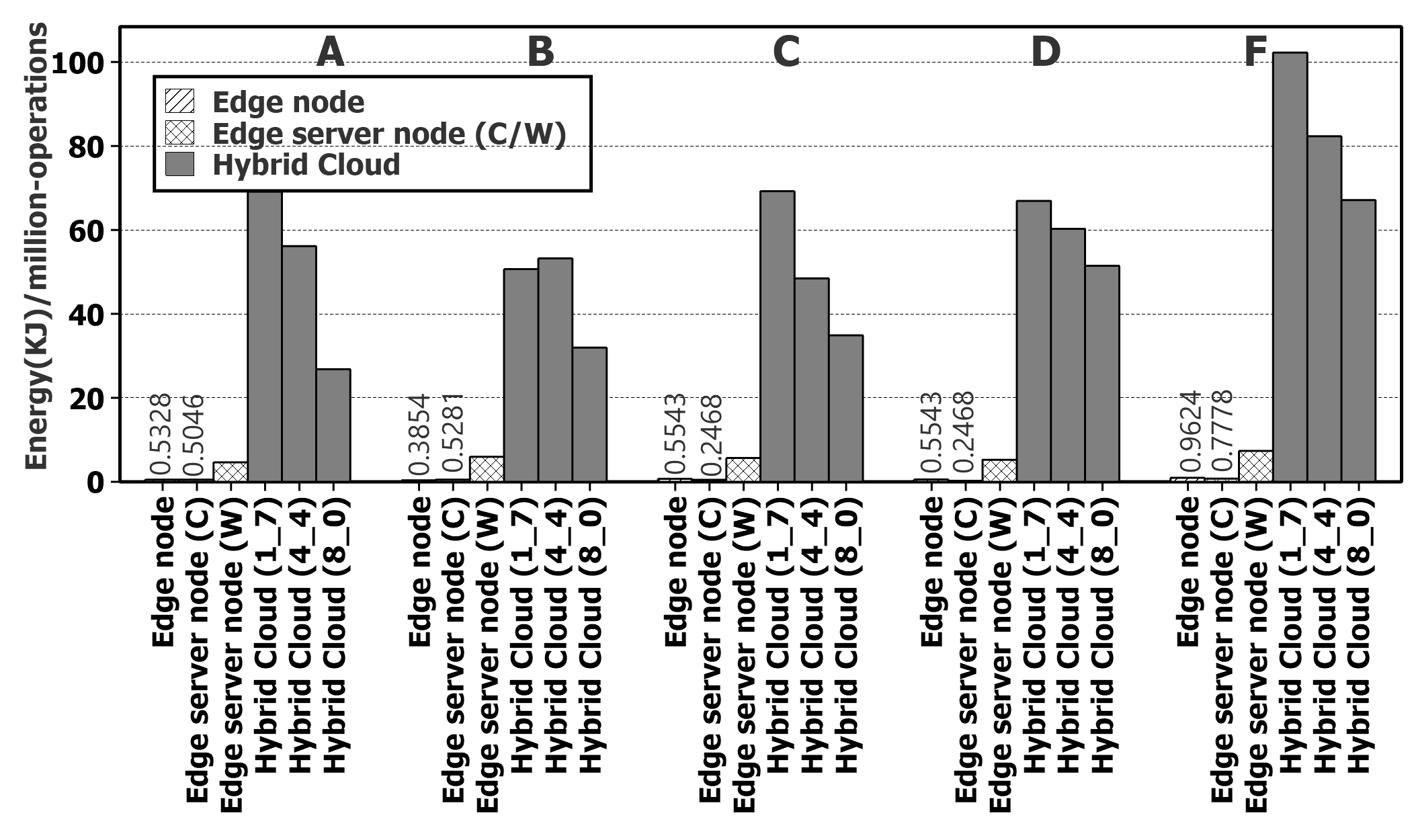}}
  \subfloat[Mongo]{\label{figur:laptop-mongo-af}\includegraphics[height=5cm,width=0.5\textwidth]{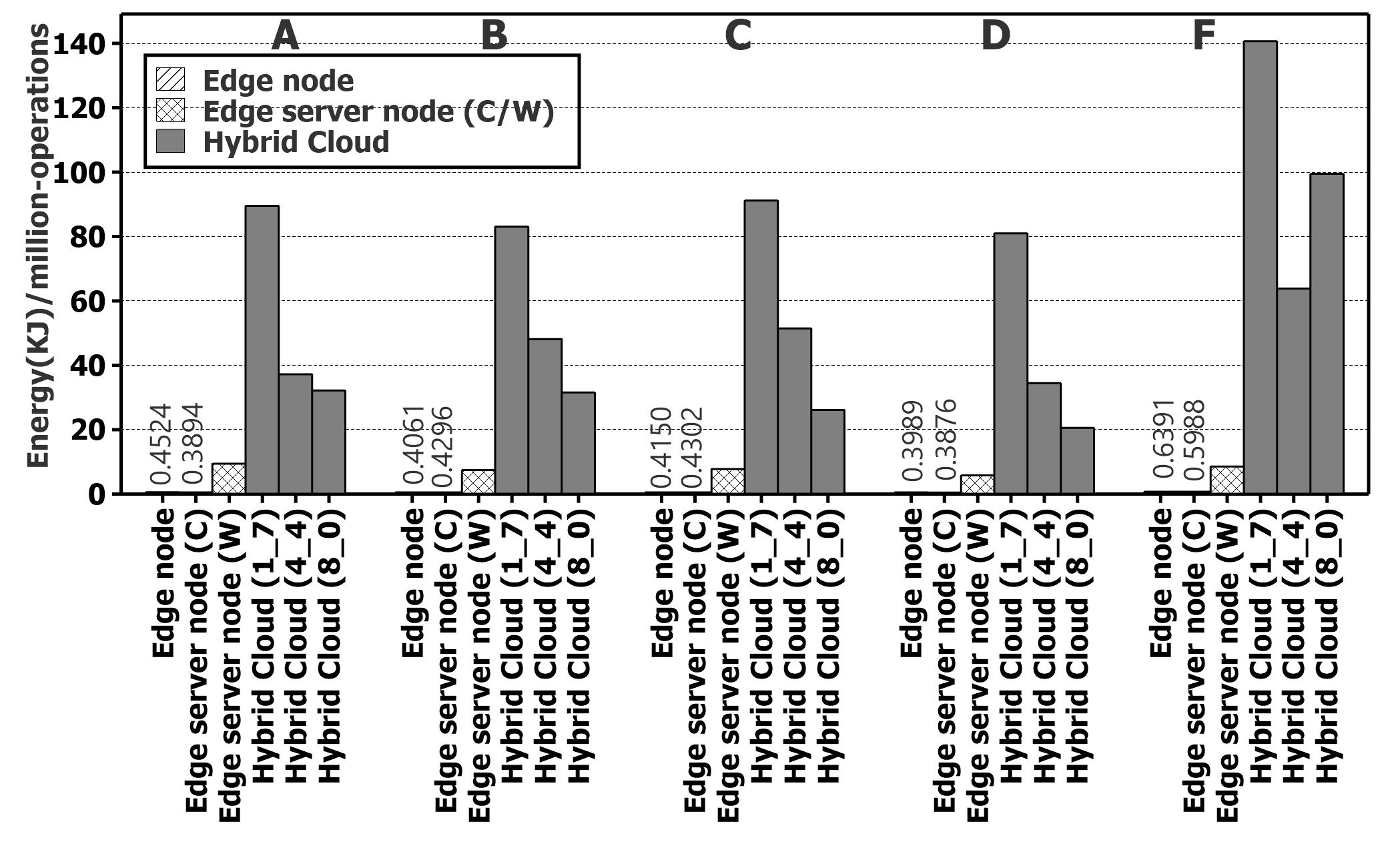}}\\
  \subfloat[Redis]{\label{figur:laptop-redis-af}\includegraphics[height=5cm,width=0.5\textwidth]{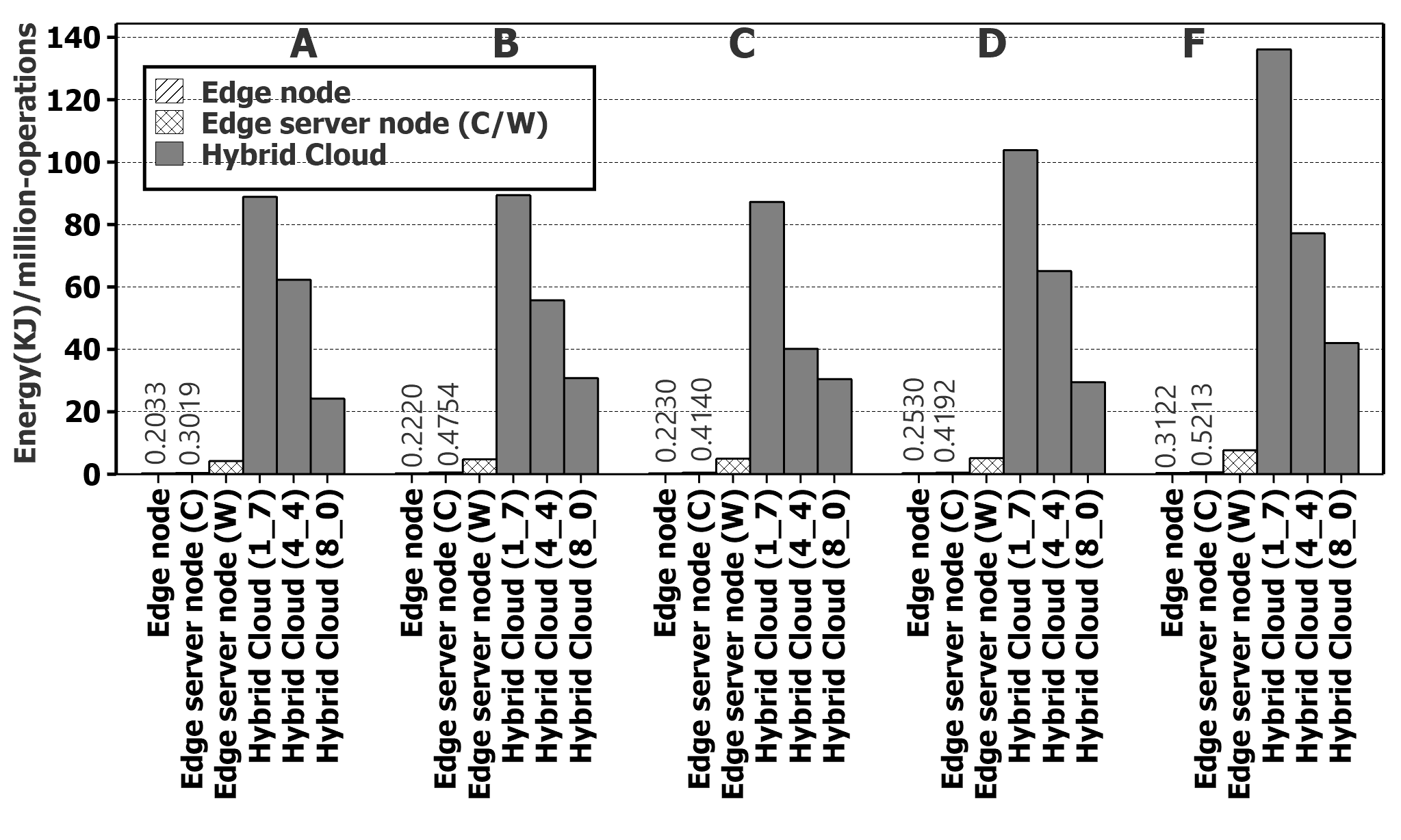}}
  \subfloat[MySQL]{\label{figur:laptop-mysql-af}\includegraphics[height=5cm,width=0.5\textwidth]{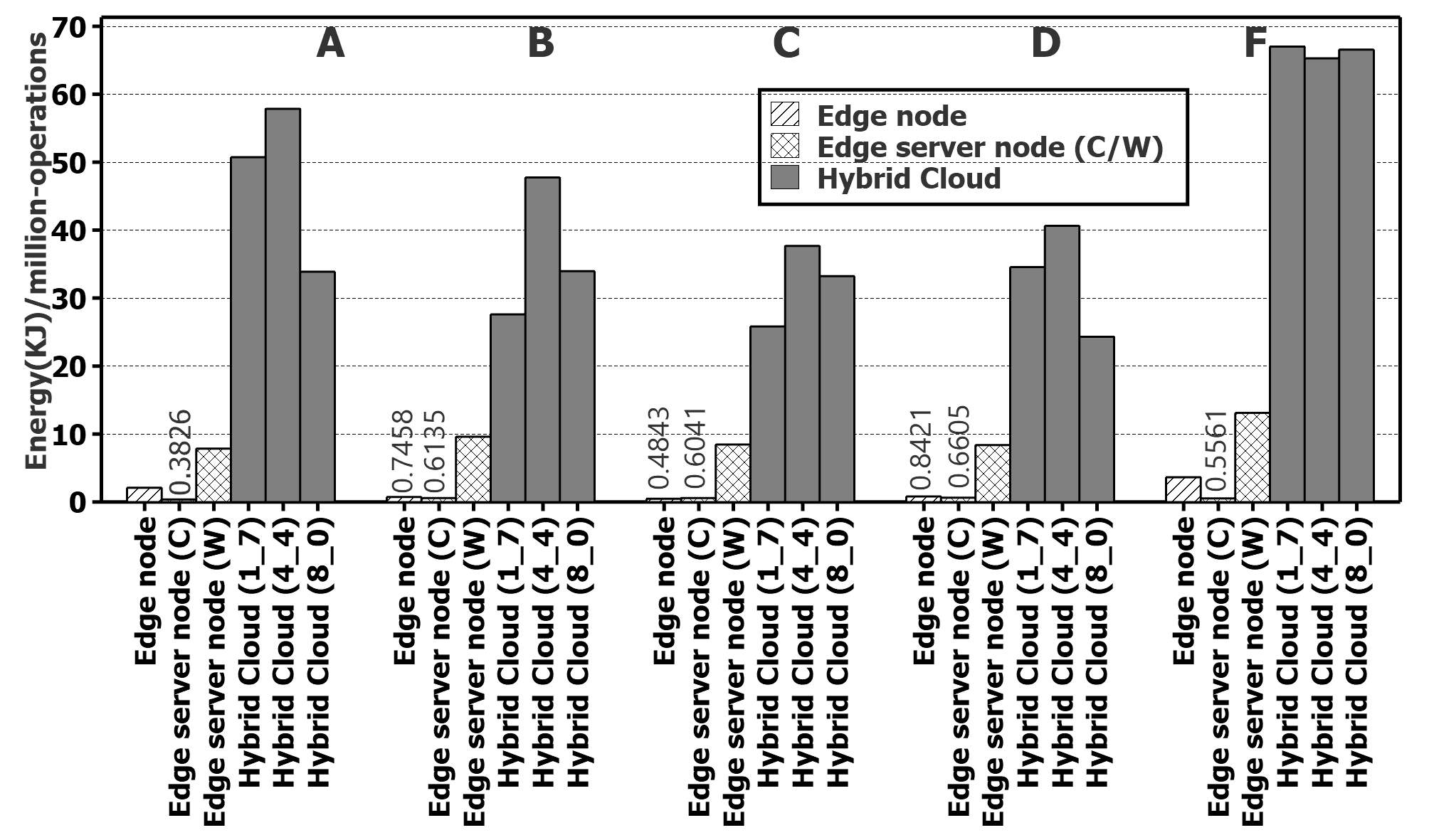}}
  \caption{Energy consumption of data offloading from \textbf{edge node} to the computing nodes for Workloads \textbf{A}, \textbf{B}, \textbf{C}, \textbf{D}, and \textbf{F}. (m\_n) indicates $m$ and $n$ nodes in the private and public clouds respectively. C/W denotes a Cable/WiFi connection.}
\label{fig:laptop-af}
\vspace{-3mm}
\end{figure*}

\textbf{(B) The energy consumption of the edge node (Scenarios 7-10).} We make the following observations from Fig. \ref{fig:laptop-af}. (i) Locally running YCSB and databases on the edge node exhibits the lowest energy consumption compared to the non-local running. As expected, Redis outperforms all databases in energy consumption (203-312 J/MOPs), while MySQL has the worst performance in energy consumption (480-3650 J/MOPs). This is because Redis is RAM-based, while MySQL has to update data and logs to disk regularly compared to NoSQL databases. (ii) As databases are deployed on the edge server node (C/W), Redis and MySQL still exhibit the lowest  (301-521 J/MOPs for cable and 4170-7610 J/MOPs for WiFi) and the highest (382-556 J/MOPs for cable and 7880-13110 J/MOPs for WiFi) energy consumption, respectively. The ratio of energy consumption for WiFi to  Cable is (10-25) times for Cassandra, (1.5-25) times for Mongo, (10-16.5) times for Redis and (1.5-10) times for MySQL. These values exhibit that the faster connection between the worker and the database servers is, the less energy consumption. (iii)
 The more nodes reside on the private cloud, the less energy is consumed for all databases except MySQL. 
The energy consumption of Cassandra, Mongo, and Redis respectively is at the level of 60, 80, and (80-100) KJ/MOPs for workloads (A-D) with the configuration of (7\_0). With the same condition and workloads, the energy consumption for  (8\_0) drops by (38-76\%) for Cassandra, (25-37\%) for Mongo, and (27-34\%) for Redis. These results show that the highest reduction happens for Cassandra since it requires reading and writing data on a quorum of replicas. The energy consumption of Workload F is more than workloads A-D so the ratio is (1.3-2.5) times for Cassandra, (3-4.8) times for Mongo, (1.36-1.73) times for Redis, and (1.96-2.7) times for MySQL. As more VMs are used on the public cloud, this factor drops significantly, which means running all workloads across WAN is expensive. 
 
 \textit{In summary, as databases are deployed on the edge and edge server nodes (C/W), Redis and MySQL consume the lowest and highest energy respectively, followed by Mongo and Cassandra. In contrast, for the edge server node (W), there is no preference between Mongo and Cassandra in energy consumption.  Furthermore, only under the (edge node $\rightarrow$ edge server node (C)) scenario, offloading is effective for MySQL (Table \ref{tab:scenariocomparison-edgenode})}.

\begin{table}[t]
	\caption{A sorted list of the lowest to the highest energy consumption for scenarios 7-10.}\label{tab:scenariocomparison-edgenode}
	\centering
	\tiny
	\vspace{-1mm}
	\begin{tabular}{p{1.7cm} p{1.7cm} p{1.7cm} p{1.7cm}}
		\hline
		Cassandra                 &Mongo                           &Redis              &MySQL                \\\hline\hline
		Edge node                 &Edge node                       &Edge node                  & Edge server node (C) \\
		Edge server node (C)      &Edge server node (C)            & Edge server node (C)      & Edge node\\
		Edge server node (W)      &Edge server node (W)            & Edge server node (W)      & Edge server node (W)  \\
	   Hybrid cloud (8\_0)	      &Hybrid cloud (8\_0)             &Hybrid cloud (8\_0)        &Hybrid cloud (all) \\
	   Hybrid cloud (4\_4)        &Hybrid cloud (4\_4)              &Hybrid cloud (4\_4)       & - \\
	   Hybrid Cloud (1\_7)        &Hybrid cloud (1\_7)             &Hybrid cloud (1\_7)        & -\\\hline
	\end{tabular}
	\begin{tablenotes}
      \small
      \item \tiny There is no particular hierarchy among hybrid cluster configurations for MySQL, and we denoted Hybrid cloud (all) in the table. 
   \end{tablenotes}
	\vspace{-3mm}
\end{table}

\begin{figure*}[ht!]
  \centering
  \subfloat[Cassandra]{\label{fig:laptop-cass-e}\includegraphics[height=5cm,width=0.25\textwidth]{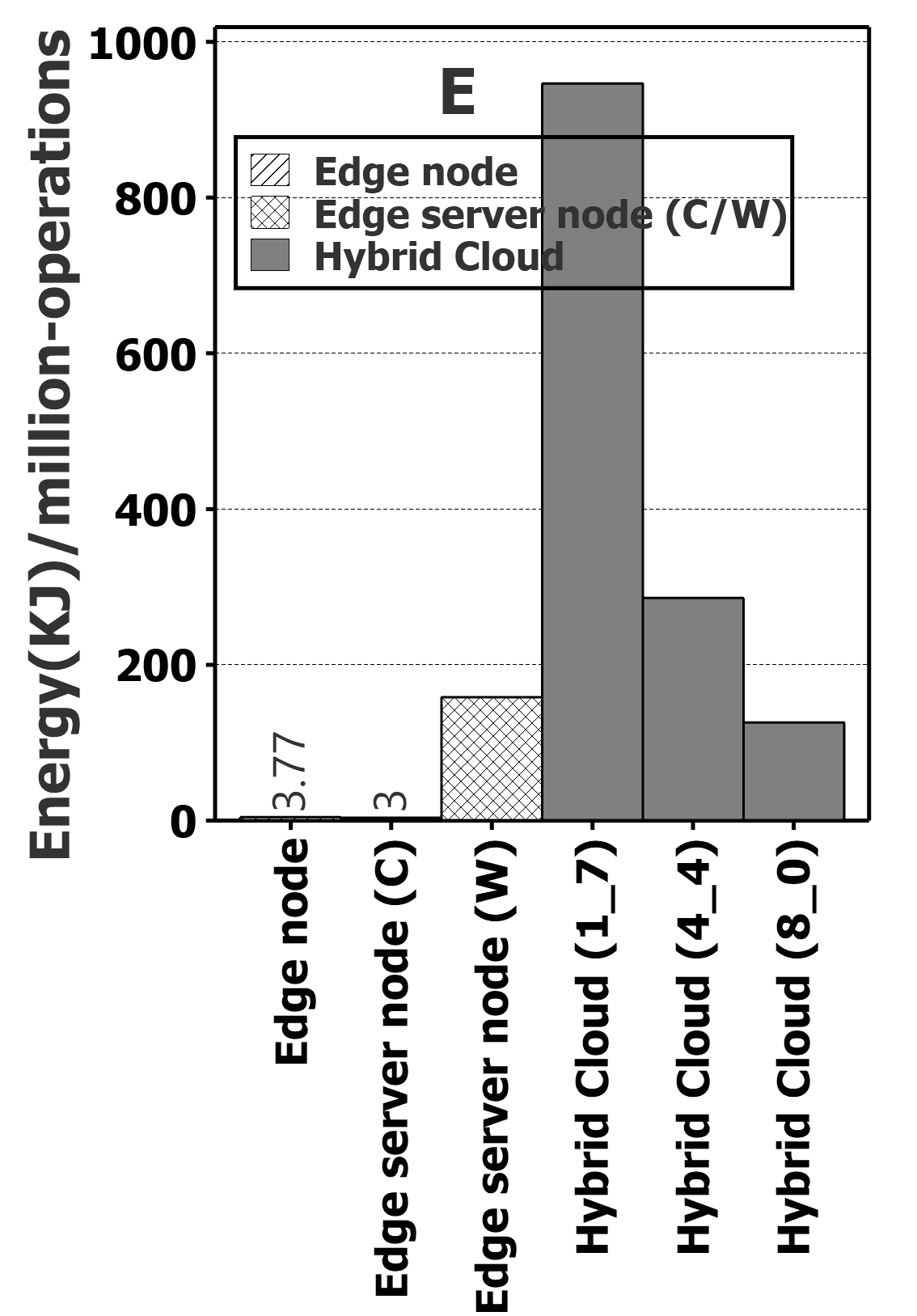}}
  \subfloat[Mongo]{\label{fig:laptop-mongo-e}\includegraphics[height=5cm,width=0.25\textwidth]{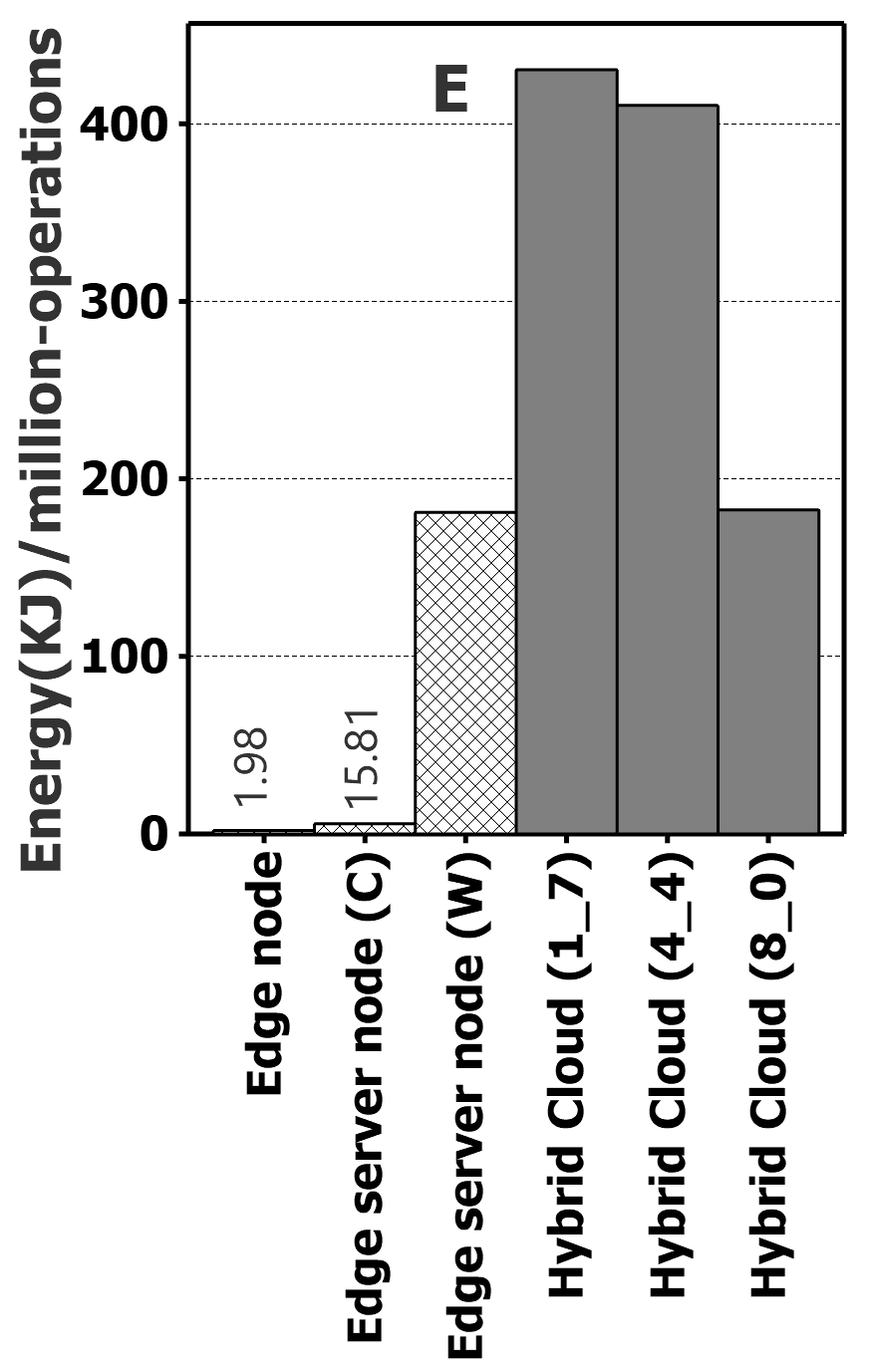}}
  \subfloat[Redis]{\label{fig:laptop-redis-e}\includegraphics[height=5cm,width=0.25\textwidth]{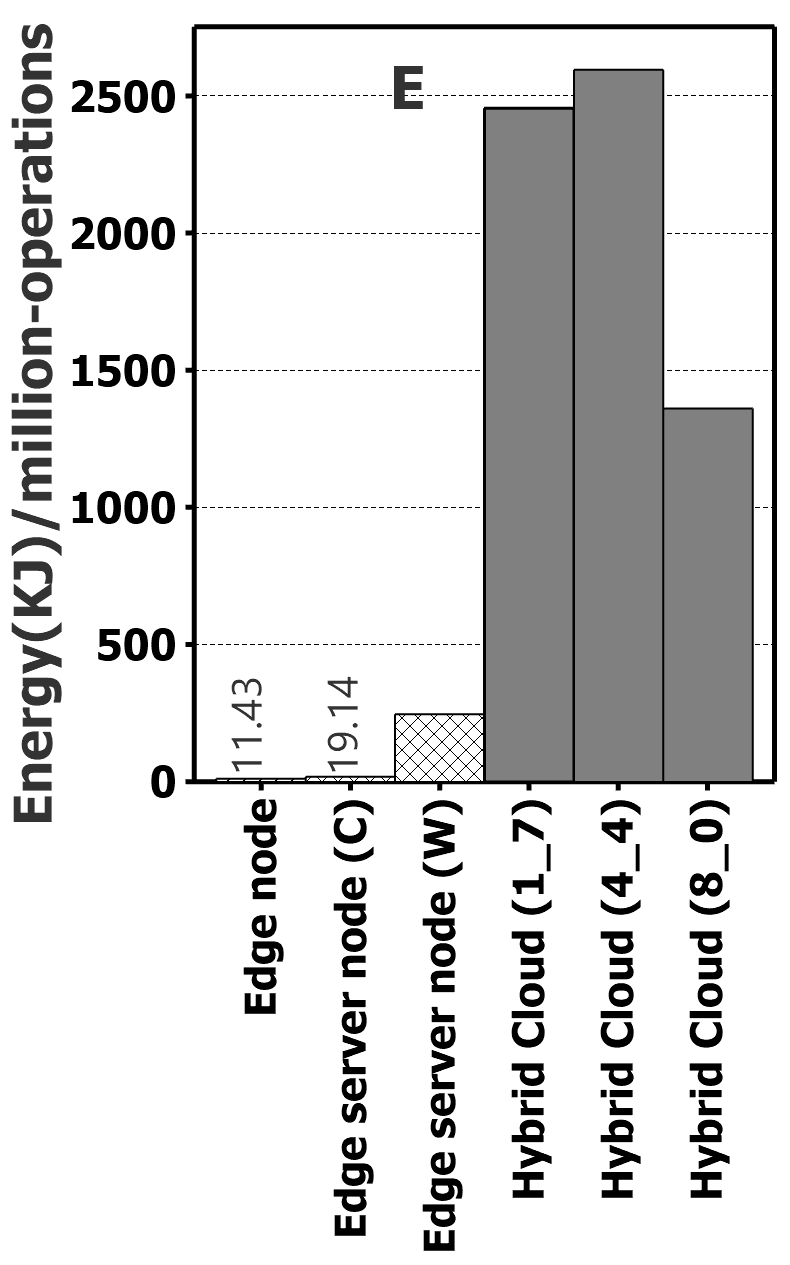}}
  \subfloat[MYSQL]{\label{fig:laptop-mysql-e}\includegraphics[height=5cm,width=0.25\textwidth]{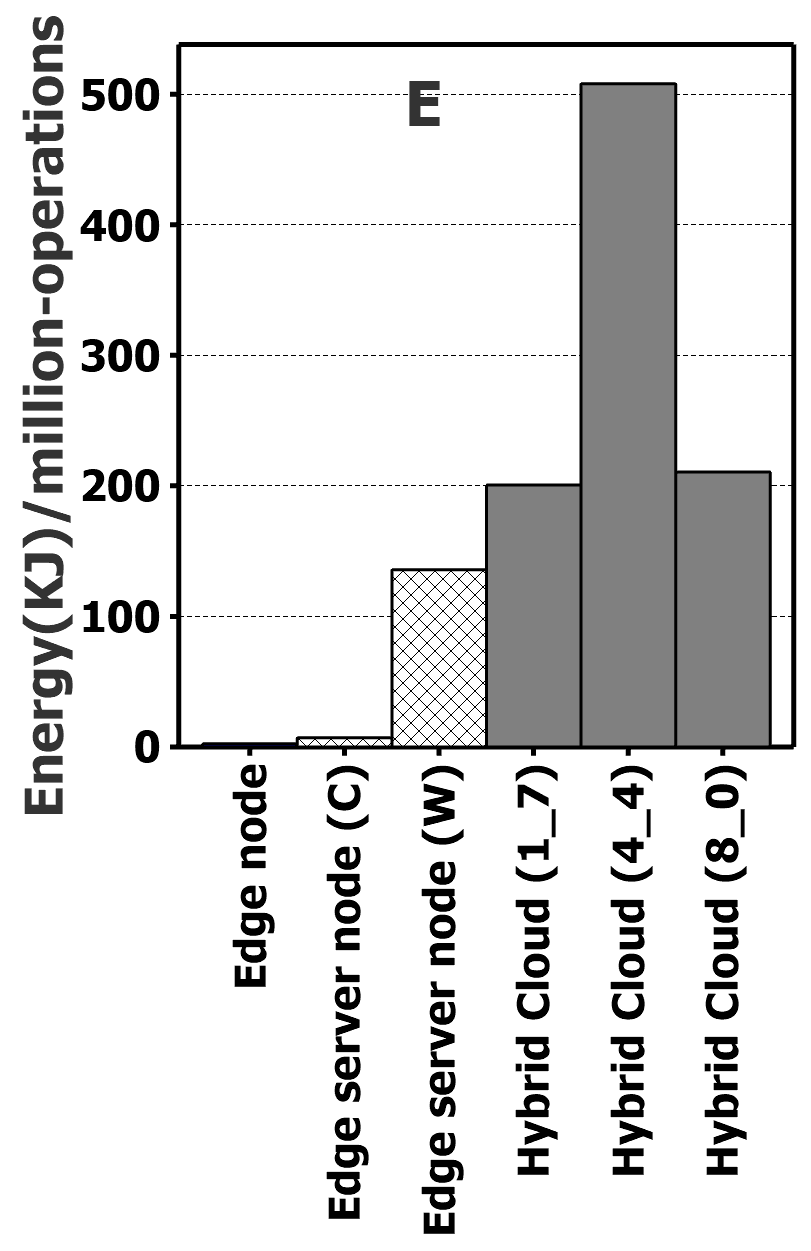}}
  \caption{Energy consumption of data offloading from \textbf{edge node} to the computing nodes for Workload \textbf{E}. (m\_n) indicates $m$ and $n$ nodes in the private and public  clouds respectively. C/W denotes a Cable/WiFi connection.}
\label{fig:laptop-e}
\vspace{-5mm}
\end{figure*}

Fig. \ref{fig:laptop-e} shows the energy consumption of workload E for scenarios 7-10.
Running workload E on the edge node consumes the lowest energy for Mongo (1980 J/MOPs) and the highest for Redis (11420 J/MOPs). For offloading data to other computing resources, Redis still needs the highest energy (19/246 KJ/MOPs) and even more (1361-2455 KJ/MOPs) as it is deployed on the edge server (C/W) and hybrid cloud, respectively.
Cassandra has the lowest energy consumption on the edge server (Fig. \ref{fig:laptop-cass-e}) because it transmits fewer data across WAN to satisfy quorum consistency (Appendix B, Table 1, workload E).

\textbf{(C) The energy consumption of the edge server node (Scenarios 11-12).} We evaluated the energy consumption of databases for scenarios 11 and 12, where the edge server node is the database worker. We observed the same trend of energy consumption for different databases so that the more computing nodes are close to the worker, the less energy is consumed. Similarly, workload E is the most expensive workload for all databases (see Appendix A).

\begin{figure}[ht!] 
\begin{subfigure}{.65\columnwidth}
  \includegraphics[width=1\textwidth, height=3.5cm]{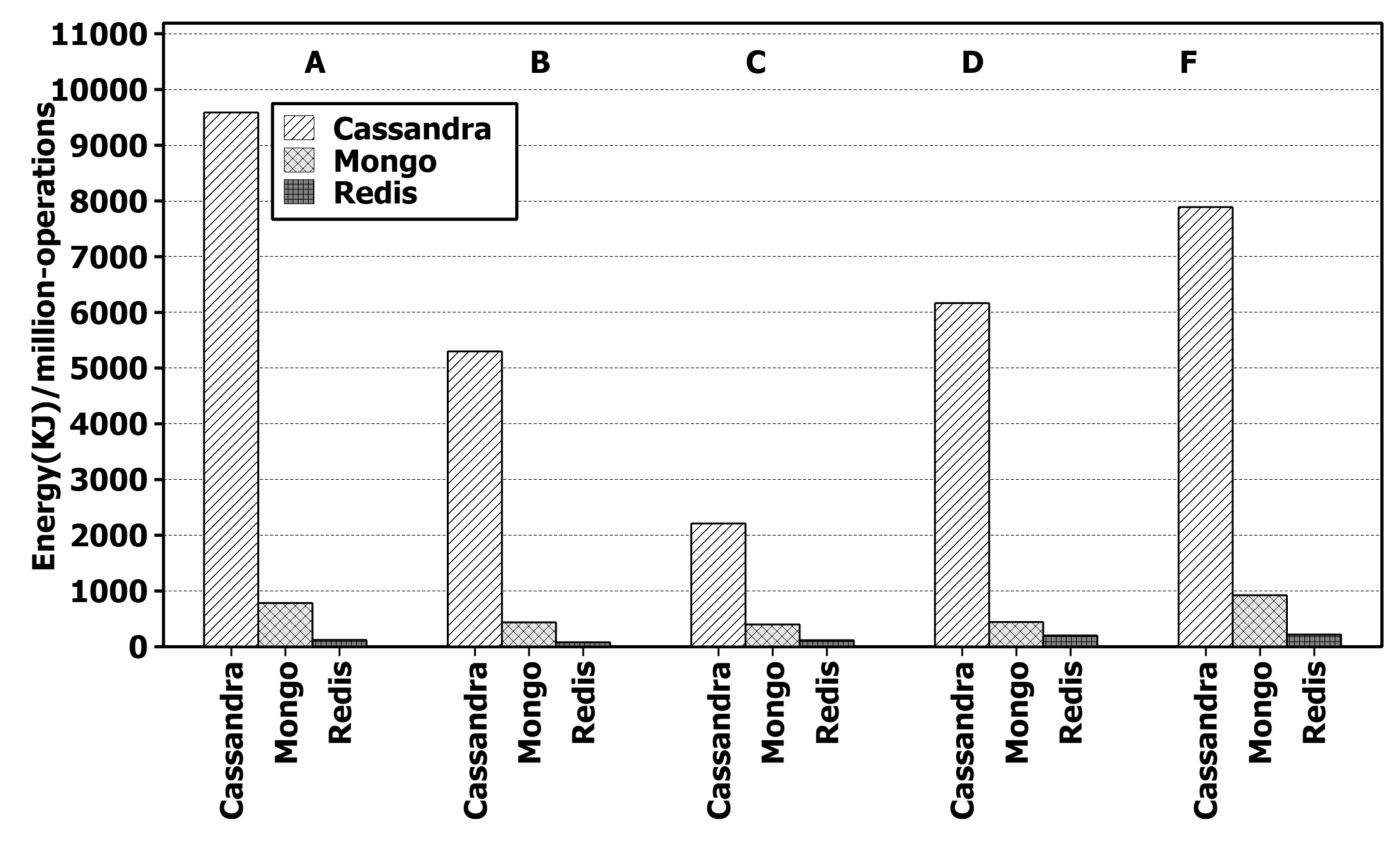}
  \caption{Workloads A, B, C, D, and F}
  \label{fig:rpi-cluster-af}
\end{subfigure}%
\begin{subfigure}{.35\columnwidth}
  \includegraphics[width=1\textwidth, height=3.5cm]{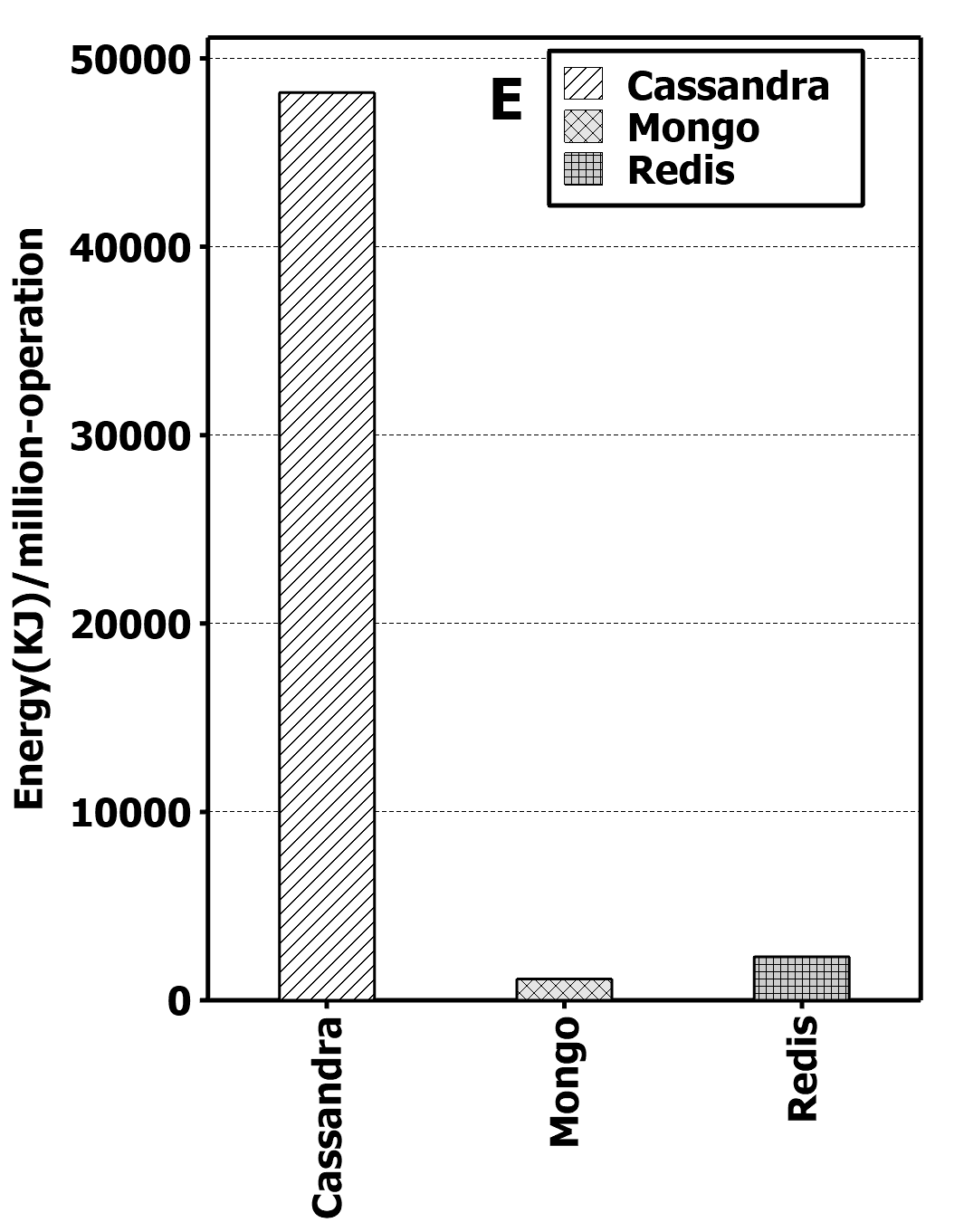}
  \caption{Workload E}
  \label{fig:rpi-cluster-e}
\end{subfigure}
\caption{Energy consumption of RPi $\rightarrow$ RPi cluster. }
\label{fig:rpi-cluster}
\vspace{-3mm}
\end{figure}

\begin{table}[t]
	\caption{A comparison of databases  listed from the lowest to the highest in energy consumption of the RPi-cluster }\label{tab:energy-comparison-rpi-cluster}
	\centering
	\vspace{-3mm}
	\begin{tabular}{p{3cm} p{3cm}}
		\hline
	    Workloads (A, F)  &Workload E\\\hline\hline
		Redis                   & Mongo  \\
		Mongo                   & Redis  \\
		Cassandra               & Cassandra    \\\hline
	\end{tabular}
	\vspace{-3mm}
\end{table}

\textbf{(D) The energy consumption of RPi Cluster (Scenario 13).} Fig. \ref{fig:rpi-cluster} plots the energy consumption of worker and database servers running on a cluster of 8-RPis\footnote{MySQL results are skipped (see \S \ref{sec:discussion} for details).}. 
Fig. \ref{fig:rpi-cluster-af} shows that Cassandra's energy usage is the highest compared to Mongo and Redis.  Cassandra requires 80-97 KJ/MOPs to run write-related workloads. This value drops by 2-6 KJ/MOPs for read-related workloads. This is likely due to the memory swapping required by Cassandra (2GiB), which reduces the speed of writing operations. In contrast, Redis is the most energy-efficient (74-214 J/MOPs), while Mongo is in the middle position (390-918 J/MOPs) for all workloads, except E.  For workload E, the position of Redis and Mongo changes, since Redis generally requires longer data transfer between computing nodes to serve workload E, which leads to longer execution time, which causes, higher energy consumption (see \ref{fig:rpi-cluster-e}). Table \ref{tab:energy-comparison-rpi-cluster} summarises the above discussion.

\begin{figure*}[ht!]
  \centering
  \subfloat[Edge node]{\label{fig:laptop-ele-e}\includegraphics[height=3cm,width=0.33\textwidth]{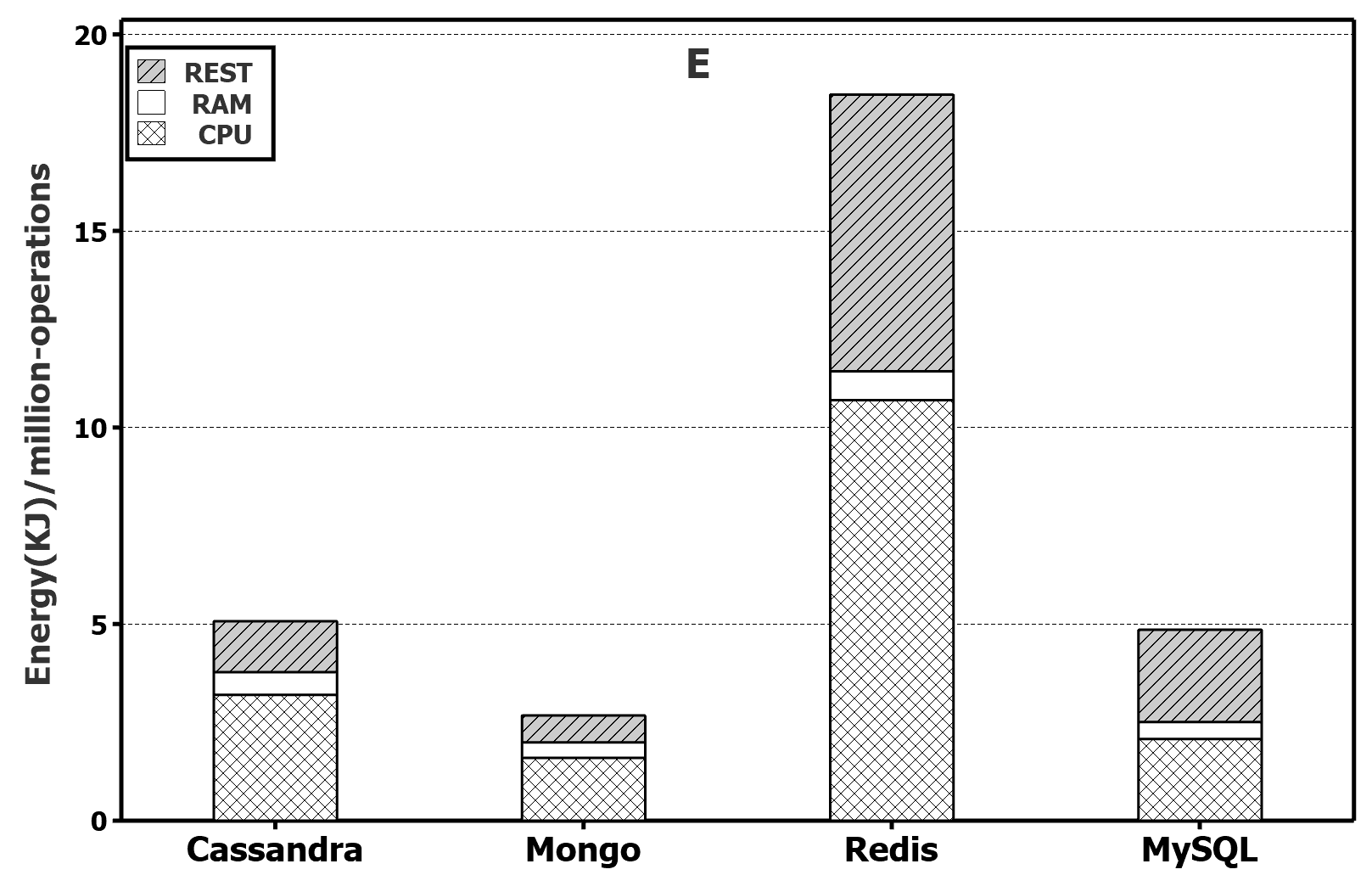}}
  \subfloat[Edge Server node (Cable)]{\label{fig:server(c)-ele-e}\includegraphics[height=3cm,width=0.33\textwidth]{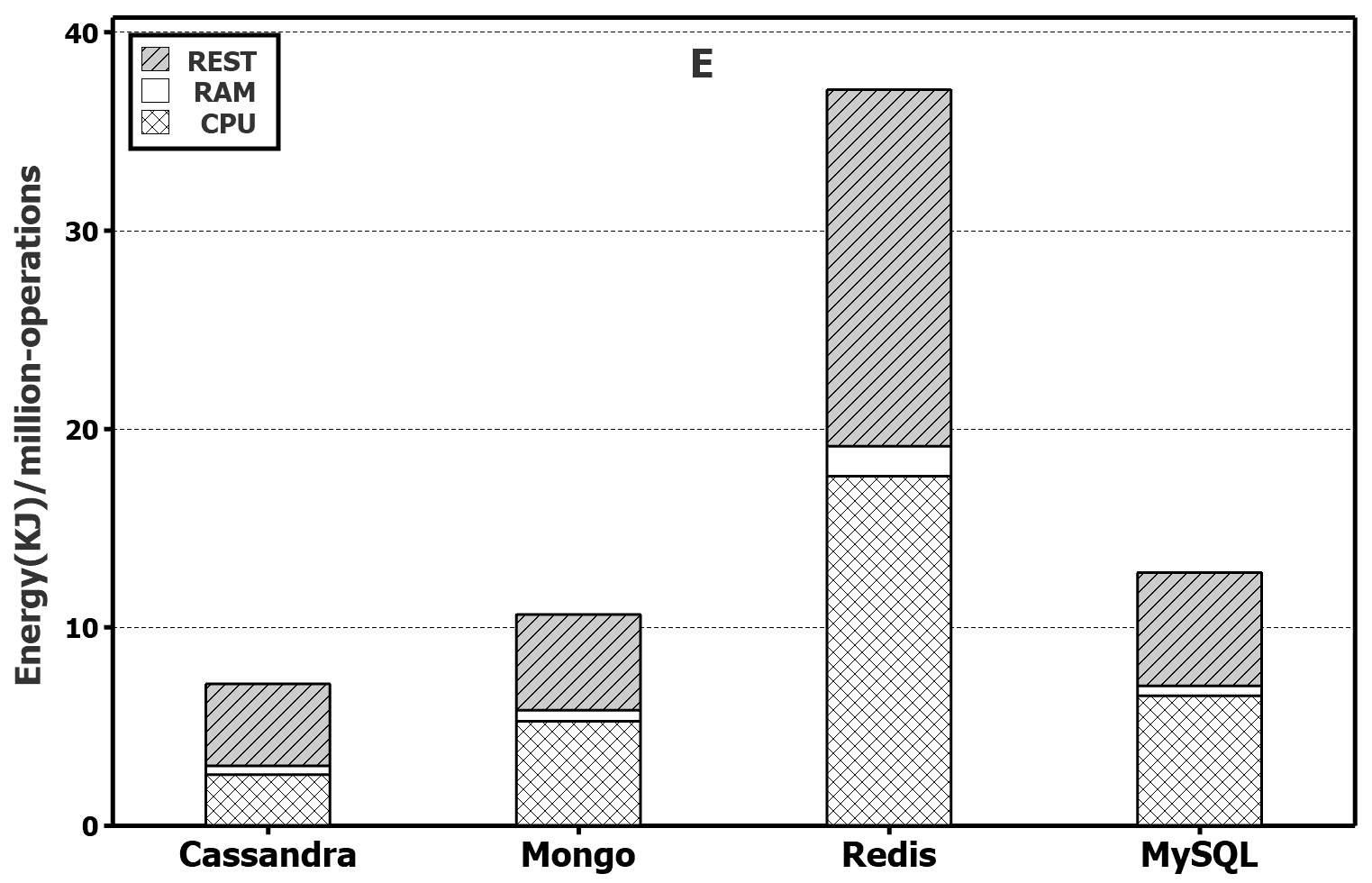}}
  \subfloat[Edge Server node (WiFi)]{\label{fig:server(w)-ele-e}\includegraphics[height=3cm,width=0.33\textwidth]{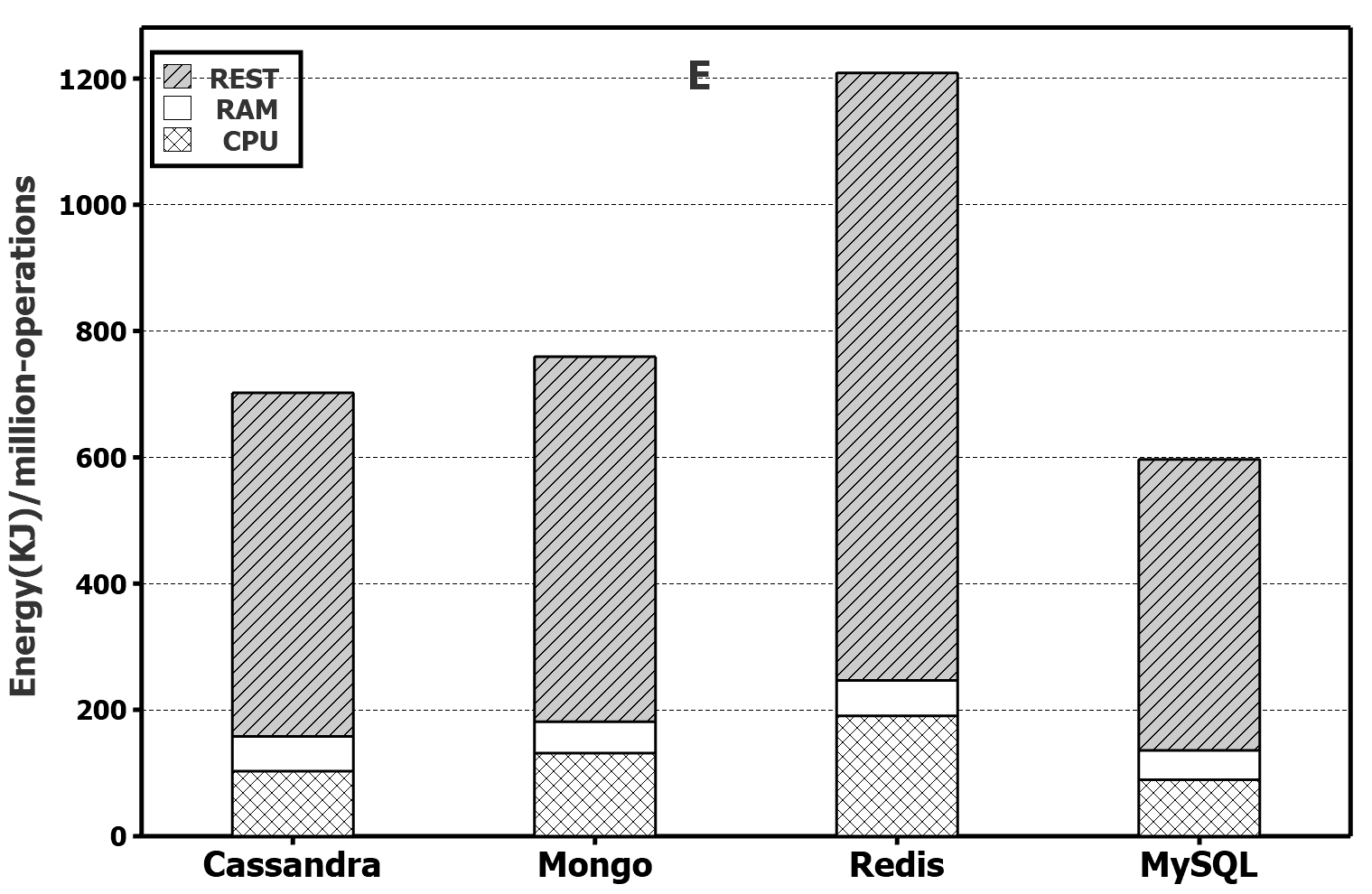}}\\
  \subfloat[Hybrid cloud (1\_7)]{\label{fig:hc-1_7-ele-e}\includegraphics[height=3cm,width=0.33\textwidth]{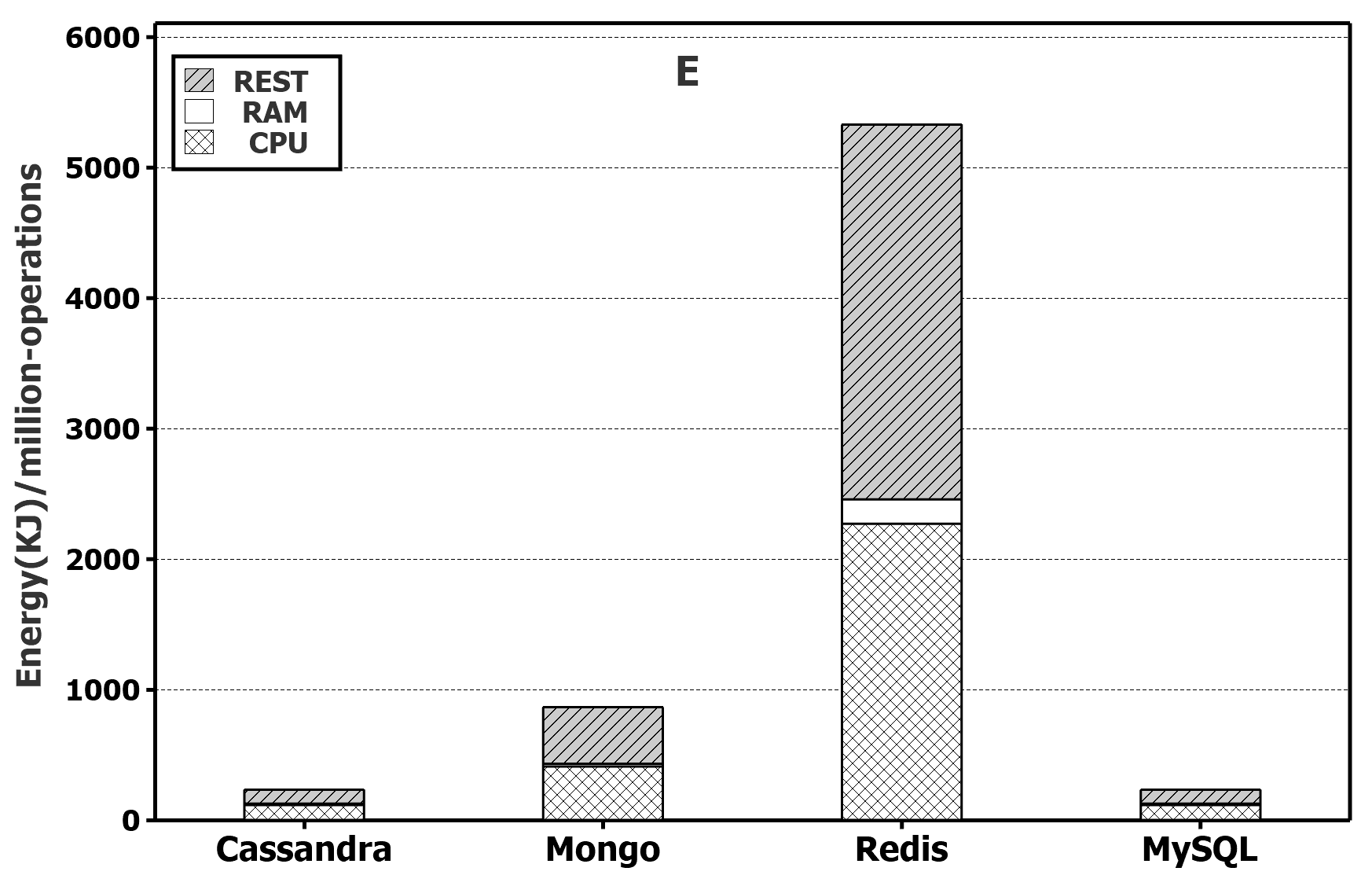}}
  \subfloat[Hybrid cloud (4\_4)]{\label{fig:hc-4_4-ele-e}\includegraphics[height=3cm,width=0.33\textwidth]{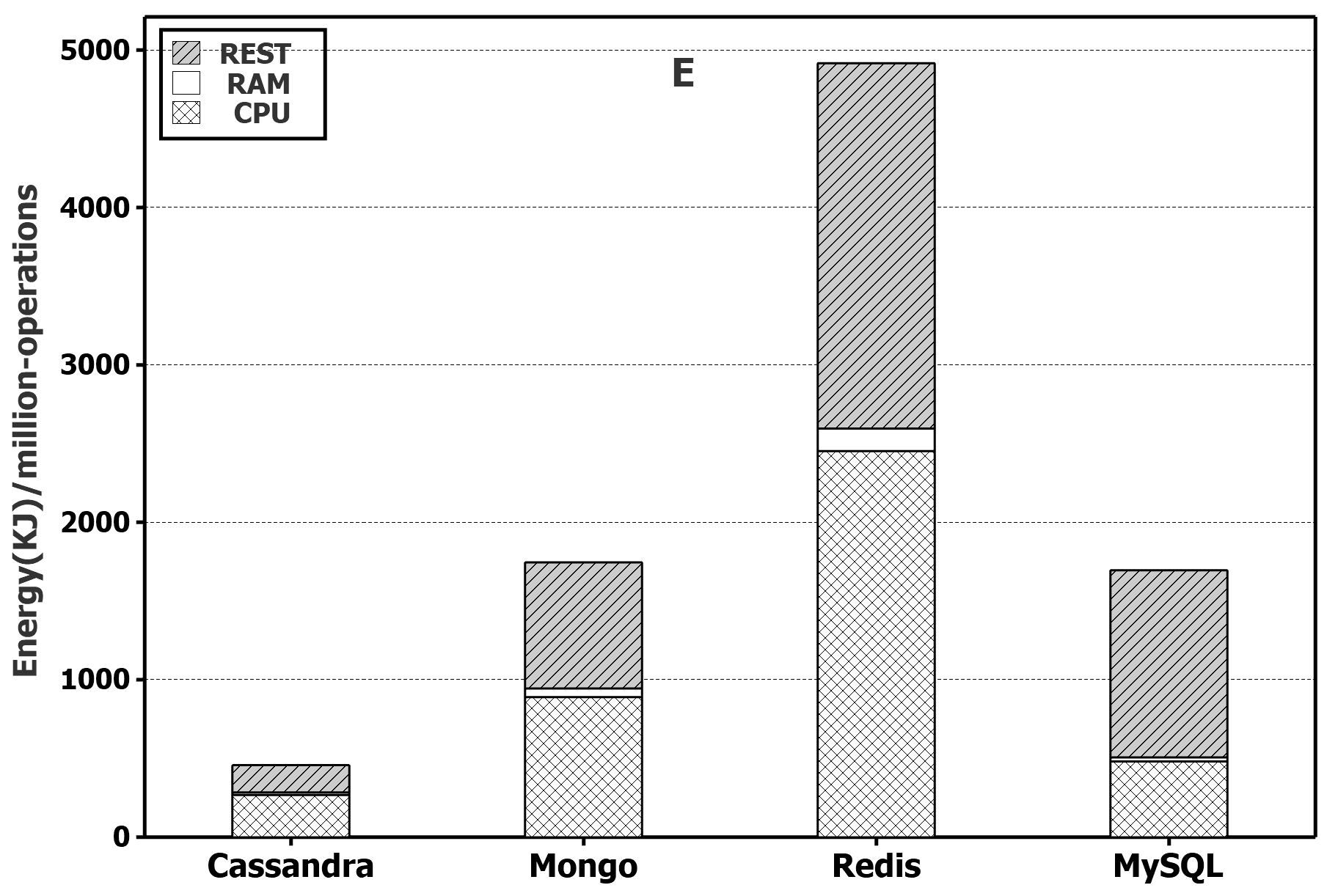}}
  \subfloat[Hybrid cloud (8\_0)]{\label{fig:hc-8_0-ele-e}\includegraphics[height=3cm,width=0.33\textwidth]{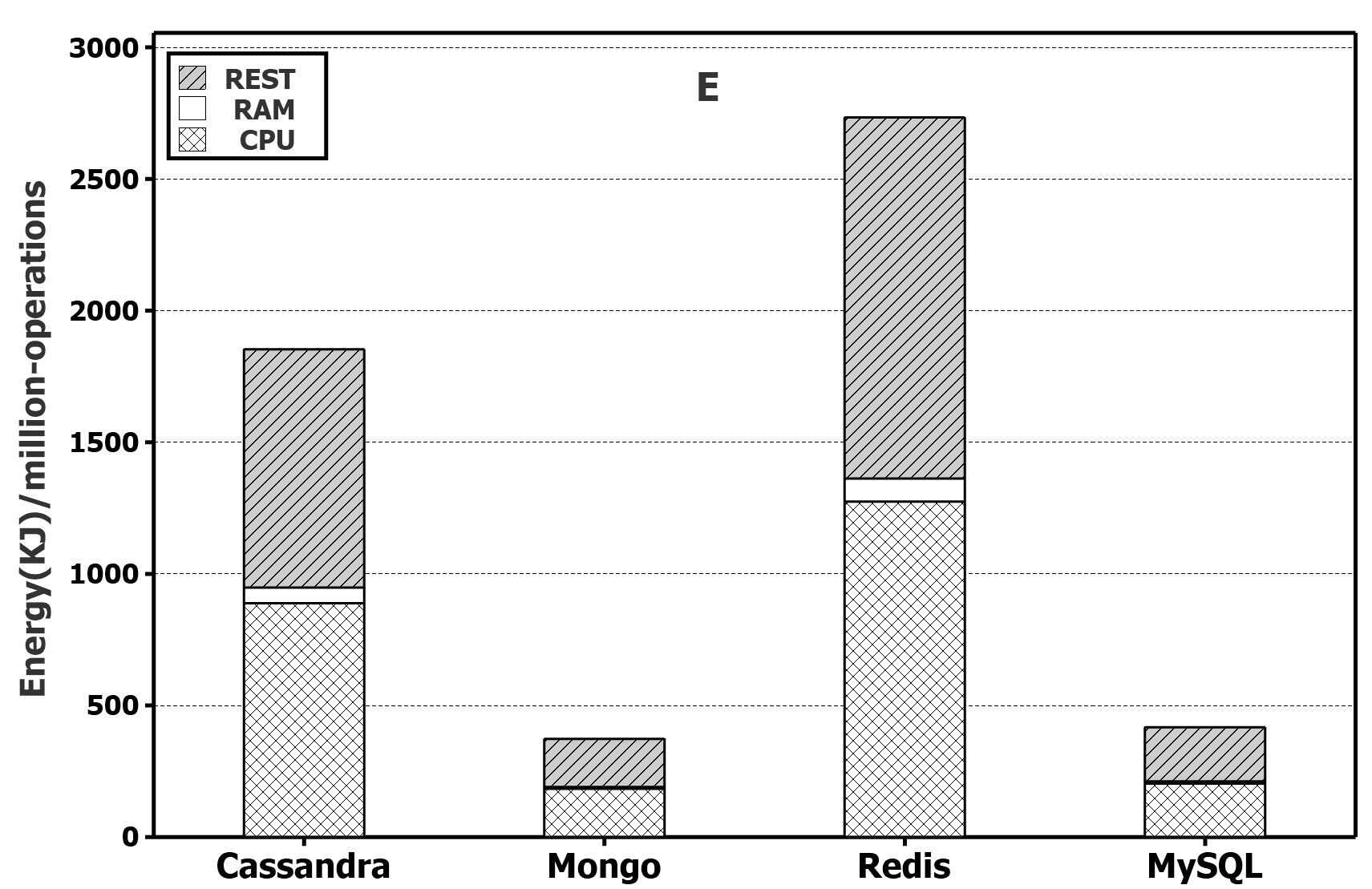}}
  \caption{The breakdown of energy consumption as the edge node runs workload E and sends requests to the hybrid cloud. (m\_n) indicates $m$ and $n$ nodes in the private and public clouds respectively.}
\label{fig:laptop-ele}
\vspace{-5mm}
\end{figure*}

\textbf{(E) Breakdown of the energy consumption of edge node.} Fig. \ref{fig:laptop-ele} breaks down the energy consumption of the edge node including CPU, RAM, and the rest of the system (monitor, peripheral devices, ports, etc.) - termed by REST - for workload E\footnote{Results for other workloads are skipped due to space constraints.}. 
Simply, the energy consumption of "REST" is the energy measured through Upower utility for battery depletion minus the one through RAPL for CPU and RAM. Results show that the energy consumption of RAM was the lowest (\textless7\%) for most of the scenarios and databases. Thus, CPU and REST have the most contribution to the energy consumption of the edge node. Interestingly, when databases are hosted locally, the energy consumption of the CPU made a significant contribution of $\approx$ 60\% to the whole consumed energy, while the energy consumption of REST is 25-39\%. As databases are moved into the edge server node and hybrid cloud, this percentage of energy consumption decreases for CPU and increases for REST. This is because the worker spends energy even during waiting to receive a response from database servers. For example, to run Cassandra on edge server node (C), CPU and REST  respectively are 35\% and 57\% of the whole energy consumption, while these values respectively changed to 14\% and 77\% as the edge server node (W) was deployed. This is because the worker waits longer to receive a response from the edge server node through WiFi compared to cable. This waiting time increases the energy consumption of REST, while CPU is idle without using significant energy. We also see the same trend with other cluster configurations, where more VMs in the public cloud wait for longer, thus, leading to increased REST energy consumption.

\begin{figure*}[h]
  \centering
  \subfloat[Workload A]{\label{fig:txrx-wa}\includegraphics[width=0.33\textwidth]{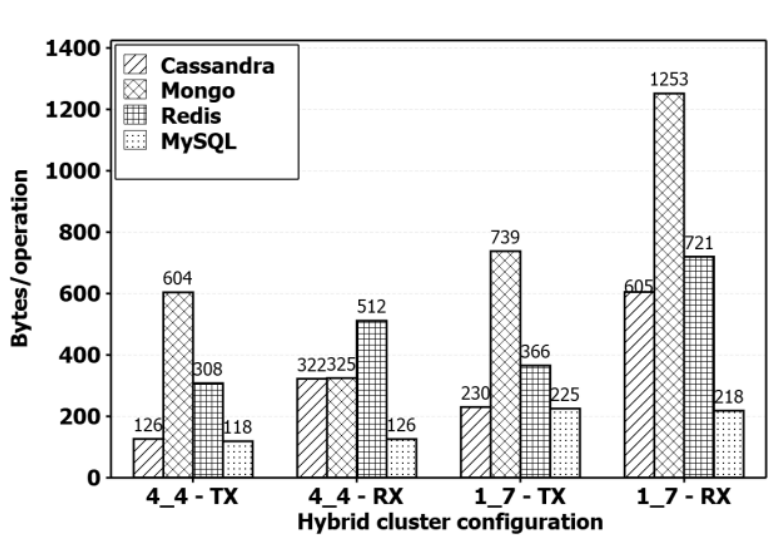}}
  \subfloat[Workload B]{\label{fig:txrx-wb}\includegraphics[width=0.33\textwidth]{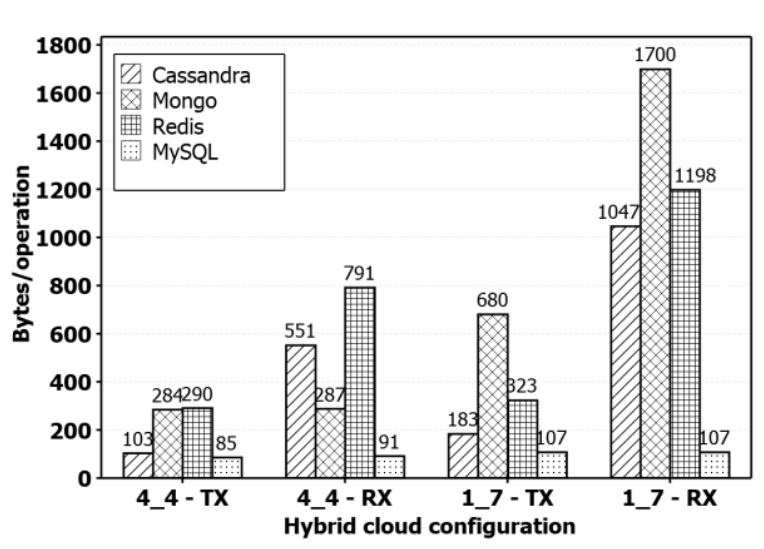}}
  \subfloat[Workload C]{\label{fig:txrx-wc}\includegraphics[width=0.33\textwidth]{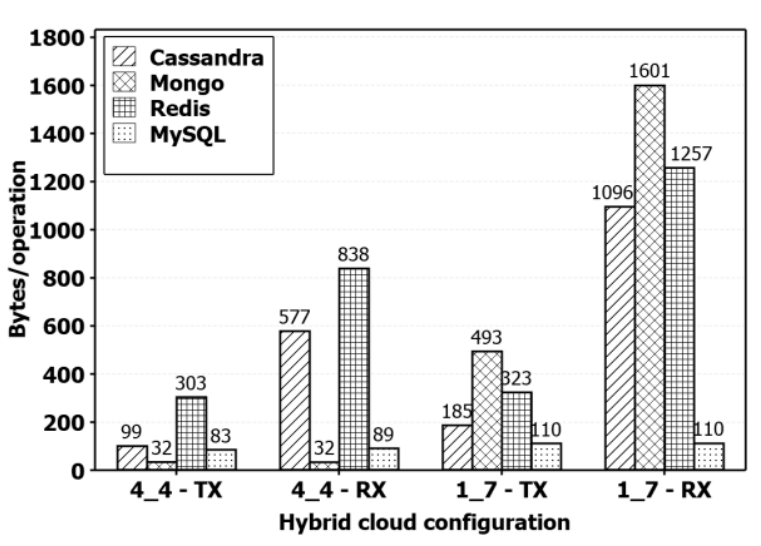}}\\
  \subfloat[Workload D]{\label{fig:txrx-wd}\includegraphics[width=0.33\textwidth]{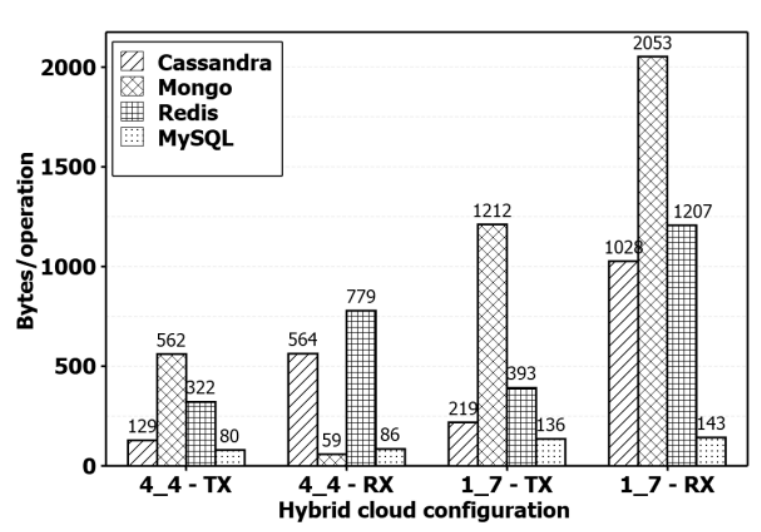}}
  \subfloat[Workload E]{\label{fig:txrx-we}\includegraphics[width=0.33\textwidth]{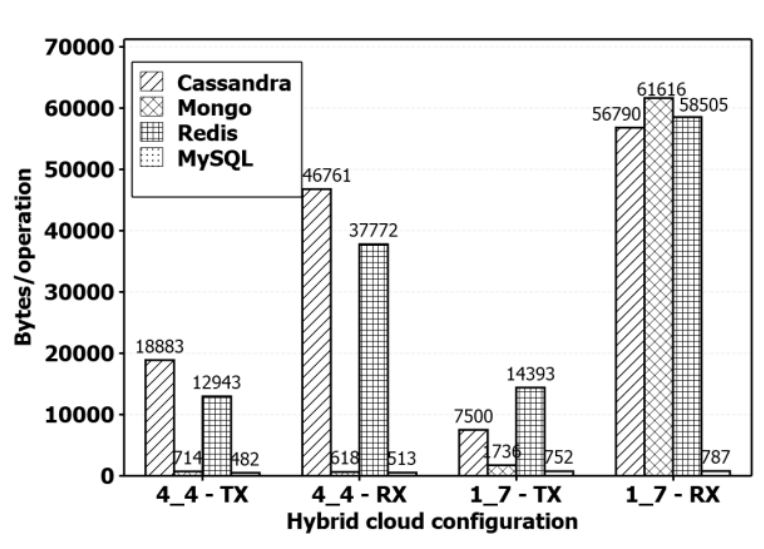}}
  \subfloat[Workload F]{\label{fig:txrx-wf}\includegraphics[width=0.33\textwidth]{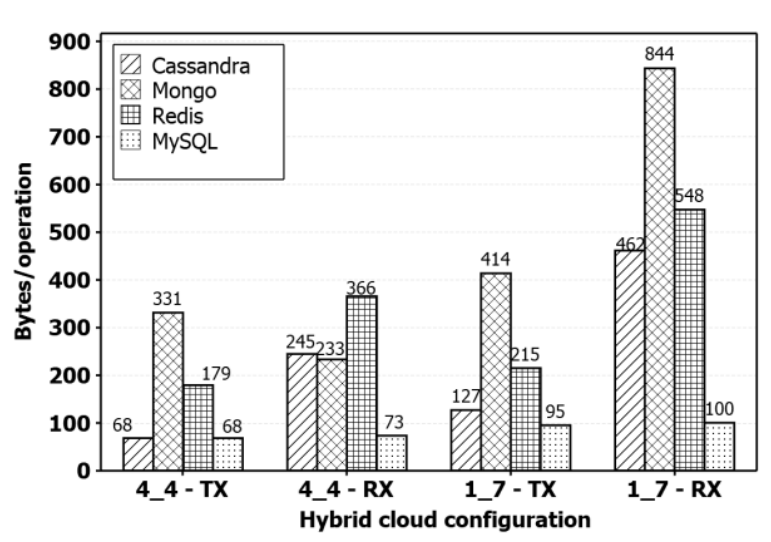}}
  \caption{Transmit (TX) and Receive (RX) data measured (bytes/operation) between private and public clouds as data is offloaded from the edge node to the hybrid cloud configurations of (4\_4) and (1\_7).}
\label{fig:TXRX}
\vspace{-5mm}
\end{figure*}
\vspace{-3mm}
\subsubsection{Bandwidth Consumption}
This section presents the amount of data transferred (TX) and received (RX) between the private and public clouds (Fig. \ref{fig:cloud-edge-arch}), where a worker is the edge node\footnote{Due to space constraints, we only consider edge node as a worker.}.  Fig. \ref{fig:TXRX} depicts  TX and RX in bytes per operation for the OpenStack broker sub-net only since these metrics are symmetric for both sub-nets. Clearly, the values of TX and RX are zero for (8\_0) due to all nodes being in the same cloud. 

Based on Fig. \ref{fig:TXRX}, we can make several observations. (i) As expected, TX and RX for workload E are at the level of several KB per operation, while for the other workloads, the values are at the level of several hundred bytes per operation. This confirms that workload E is the most expensive in execution time and energy usage. (ii) The values of TX and RX for (4\_4) are less than the ones for (1\_7), which implies that more nodes in the public cloud cause higher TX and RX values.  This is another confirmation of higher energy consumption for (1\_7) compared to (4\_4) and (8\_0).  (iii) The  TX values are less than TX values for most databases and cluster configurations since TX includes the request issued from the worker and RX is the response returning from the database server. Thus, we focus on  RX for (1\_7) and (4\_4).

MySQL has the lowest RX values compared to the other databases for all workloads for (1\_7). This is because, as summarized in Tables in Appendix B, the worker mostly exchanges data with the nodes in the private cloud. Hence, we have fewer data transferred across clouds for MySQL. 
For the same configuration, Mongo possesses the highest RX values, followed by Redis and Cassandra. This is because, as summarized in Appendix B - Table 1, Mongo mostly sends data to the public cloud (node 7) while Redis and Cassandra almost equally spread data across nodes in the hybrid cloud.

For (4\_4), Mongo and MySQL serve the read-related workloads through the private cloud since they should satisfy eventual and strong consistency, respectively. Strong consistency is provided by MySQL because the default replica number for MySQL is two (with both in the private cloud). For workloads A and F,  Mongo and Redis obtain the highest RX values. For workload E, Cassandra and Redis generate the most traffic on the WAN while MySQL and Mongo transmit less. This is because MySQL and Mongo serve workload E locally, while Cassandra and Redis spread data across nodes (Appendix B, Table 2, workload E).   

\subsubsection{Storage consumption}

\begin{figure*}[t!]
  \centering
  \subfloat[Workload A]{\label{fig:storage-wa}\includegraphics[width=0.33\textwidth]{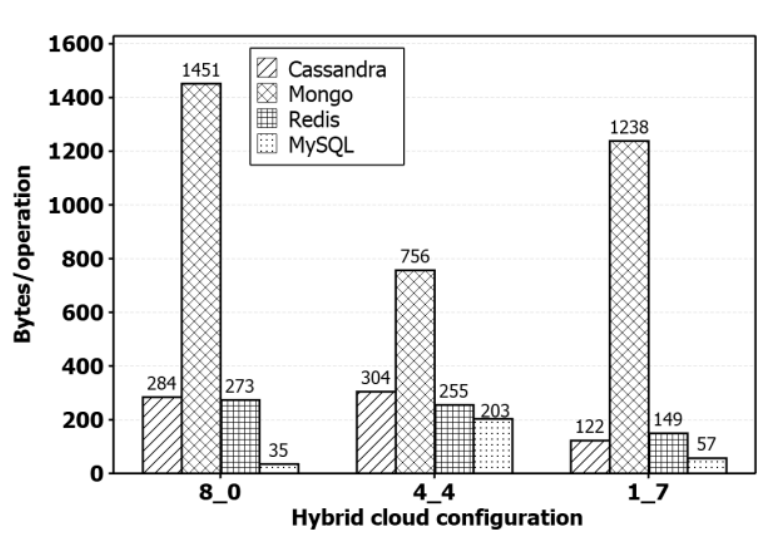}}
  \subfloat[Workload E]{\label{fig:storage-we}\includegraphics[width=0.33\textwidth]{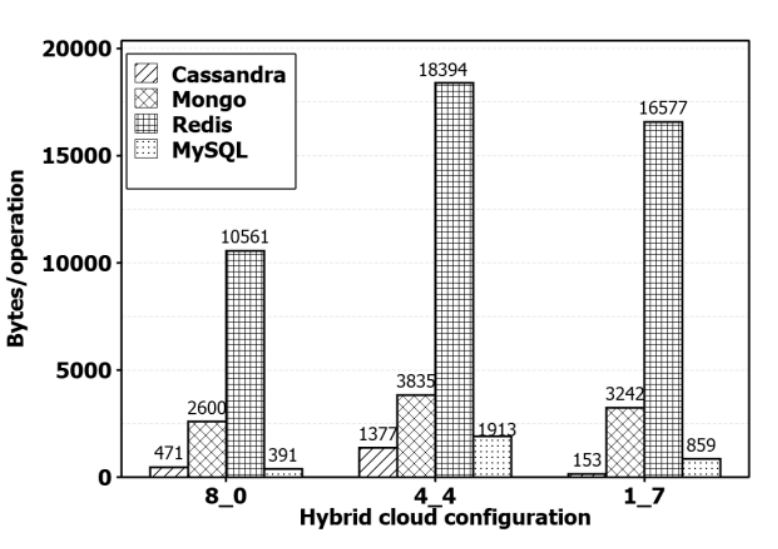}}
  \subfloat[Workload F]{\label{fig:storage-wf}\includegraphics[width=0.33\textwidth]{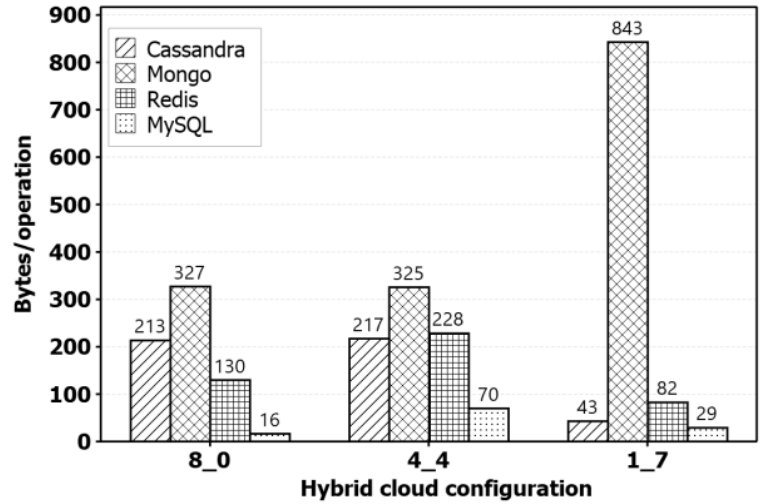}}
  \caption{Storage consumption (Bytes/operation) of data offloaded from the edge node to the hybrid cloud.}
\label{fig:storage}
\vspace{-5mm}
\end{figure*}

This set of experiments plots the storage consumption (measured in Bytes/operation), where the edge node is a worker and the hybrid cloud is a database server. Due to space constraints,  we only report results of write-related and scan workloads.  For  (1\_7) and (8\_0), Fig. \ref{fig:storage-wa} and \ref{fig:storage-wf} show that Mongo is the worst in terms of storage consumption compared to other databases for write-related workloads,  where workload A  uses storage space more than workload F. This is because Mongo uses a document-based data model with full replication as the default setting. By contrast, for the same configuration, MySQL is the most efficient database in storage consumption (35-57 Bytes/Ops for workload A vs. 29-43 Bytes/Ops for workload F) due to using two replicas rather than full replication for Mongo and three replicas for Cassandra.
Fig. \ref{fig:storage-we} exhibits the storage consumption of databases for workload E, which is more than the one for the write-related workloads. Redis uses the largest amount of storage, followed by Mongo with a 20\% reduction. This correlates with a high  RX value of workload E for Redis. Cassandra and MySQL stay close to each other with the lowest storage usage.  Table \ref{tab:storage-comparison} summarizes discussed results.

\begin{table}[t!]
	\caption{A comparison of different  databases  listed from the lowest to the highest in storage consumption }\label{tab:storage-comparison}
	\centering
	\vspace{-3mm}
	\begin{tabular}{p{3cm} p{3cm}}
		\hline
	    Workloads (A,F)  &Workload E\\\hline\hline
		MySQL                         & MySQL, Cassandra  \\
		Cassandra, Redis              & Mongo  \\
		Mongo                         & Redis    \\\hline
	\end{tabular}
	\vspace{-5mm}
\end{table}

\vspace{-3mm}
\section{Discussion}\label{sec:discussion}
We discuss findings, practical experiences, and technical challenges that we encountered during experimentation.

\textbf{Research findings:} From the discussed evaluated experiments, it is a challenging problem to select a specific database solution that incurs the lowest resource consumption (energy, bandwidth, and storage) in an edge-cloud framework for all workloads. However, from the results, we have extracted several insights as follows. (i) In terms of offloading, a few scenarios make data offloading profitable in terms of energy usage. Indeed, if database operations are offloaded from source-constrained edge nodes to powerful computing nodes with high bandwidth and low latency connection, then we expect to save energy for edge devices (e.g., RPi $\rightarrow$ edge server (C)). (ii) Connection bandwidth and latency have a direct impact on the energy usage of data offloading. Hence, all databases exhibit less energy consumption with a faster connection between workers and data servers. (iii) The limitation of memory can increase the energy consumption of disk-based databases such as  Cassandra because memory swapping further increases the response time, which directly impacts energy consumption. (iv) The distance between worker and database nodes, and the spread of data across computing nodes in a cluster of VMs in a hybrid cloud are two key factors that affect the response time, which results in energy consumption increment. In other words, the greater is the distance between worker and database servers, the more is energy consumption. The more data is distributed among nodes in the hybrid cloud, the less energy is consumed. This is because more operations can be served through the private cloud, as seen in the case of Cassandra and Redis. (v) The energy consumption of CPU and RAM has the highest and lowest contribution respectively in the total energy consumption. This is likely the reason why Redis is superior to disk-based databases in terms of energy consumption in most cases. 

With respect to the superiority of databases to each other, Redis consumes the least amount of energy followed by Cassandra if an edge computing node supports a high amount of memory capacity. This superiority is also valid when we run workloads A-F locally (i.e., on RPi, edge node, and edge server node),  and offload these workloads from the database worker to the edge node and edge server nodes. By contrast, for workload E, Redis performs the worst in energy consumption, while MySQL requires the least energy on average. For offloading data from the database worker to the hybrid cloud, MySQL consumes the lowest energy, followed by Redis particularly when more nodes are deployed on the public cloud. With regard to bandwidth usage across clouds, MySQL transmits the least amount of data irrespective of a cloud configuration. This correlates with MySQL using less energy in hybrid cloud scenarios compared to other databases. We can also see that MySQL and Cassandra require the lowest storage capacity.

\textbf{Practical experiences:} While we automated the installation and configuration of the databases across cloud and edge use cases, the ARM architecture of RPi caused some issues with MySQL. The default MySQL server package provided by Ubuntu 20.04.1 does not come with clustering components included. Thus, we had to compile our own version with the clustering explicitly enabled. Compiling on RPi node itself was failing as more than 12GB of RAM was required to complete the build process. Enabling a swap file allowed us to proceed, however, the resulting build performance was unacceptably slow, requiring several days to complete. Thus, we also attempted to cross-compile ARM binaries on a high-end x86\_64 server, which was significantly faster. Unfortunately, both produced packages crashed upon execution on RPi nodes due to the lack of L3 cache. Upon a brief MySQL source code inspection and assessing the time constraints, we skipped MySQL test for RPi nodes. Further investigation of this issue and related code changes might be useful in the future. This is a prime example of the reasons behind distributed databases being unsuitable in the context of resource-constrained devices. This can motivate further database development geared toward lightweight deployments.

We used RAPL which exploits a software power model to estimate energy usage of the edge node and edge server node through hardware performance.  The main issue with this utility is the maximum energy range of 65$^{+}$ Billion Micro-joules for its counter. This imposes constraints on the duration of the experiment for each workload because when the energy consumption reaches this value, the counter resets, and consequently the energy consumption probe records a wrong value. Hence, we had to take extra care to adjust the counter values to compensate for this limitation.

\vspace{-3mm}

\section{Conclusion and Future Work}
Selecting a suitable distributed database to deploy across the edge-cloud framework is not a trivial task as overall performance and energy efficiency highly depend on a multitude of factors. To disclose these factors, we conducted an extensive evaluation of distributed databases through a variety of scenarios in which operations are issued from resource-constrained computing nodes to more powerful ones via cable and WiFi connections. We implemented these scenarios through a modular framework to achieve flexibility and accuracy in experimental data. Our evaluation quantified the impact of connection speed, latency, and the computational power of database servers on various types of resource utilization. Notably, our results exhibit that the distance (and hence latency) between the database client issuing operations and the database servers hosting databases is a major factor that should be considered. Similarly, the bandwidth usage in the edge-cloud framework greatly impacts the client's energy consumption. We see that Redis generally consumes the least amount of energy for most workloads in local and edge-offloaded processing due to being RAM-based. For offloading data to the hybrid cloud (higher latency), MySQL is the most efficient in energy consumption for most workloads on average since it transmits fewer data across private and public clouds. Mongo and Cassandra hold a rank after MySQL and Redis in terms of energy usage, where Cassandra commonly outperforms Mongo when more nodes reside on the public cloud. 

\textbf{Future work:} We conducted our experiments for particular physical and virtualized resources in the edge-cloud framework. However, repeating these experiments for all existing and new flavors of physical and virtual resources is daunting work and to a large extent is impossible. To tackle this challenge,  we can leverage AI and ML to discover patterns of resource utilization in the edge-cloud landscape based on the data collected in our experimental scenarios \cite{GILL20221}. This can aid in predicting whether full offloading of database workloads from edge to cloud should be conducted. 
While we empirically measured resource utilization of databases under non-/full-offloading, partial offloading and optimal resource management might save energy consumption of distributed databases in edge-cloud framework \cite{GILL2019}\cite{LIU2022}. Furthermore, we can also exploit ML models to find a correlation between resource consumption in a wide range of computing devices and custom database parameter settings, such as replication number, consistency model, and data size in order to analyze offloading possibility more precisely. Ultimately, we can create ML models to predict resource utilization given a combination of hardware, database parameters, and distance between database client and servers. This can enable determining when and where to offload database workloads in a given configuration. Lastly, we can evaluate big data frameworks (e.g., Spark \footnote{Spark: \url{https://spark.apache.org/}} and Flink \footnote{flink: \url{https://flink.apache.org/}}) \cite{GILL20221} using our experimental framework to find potential correlations between parameter settings and resource utilization. This may provide further insights into the feasibility of data processing on edge devices compared to sending and processing data in a centralized cloud.



%


\ifCLASSOPTIONcompsoc
\else
\fi


\ifCLASSOPTIONcaptionsoff
  \newpage
\fi

\bibliographystyle{IEEEtran}
\bibliography{references.bib}


\vspace{-11 mm}
\begin{IEEEbiography}[{\includegraphics[width=1in,height=1.25in,clip, keepaspectratio]{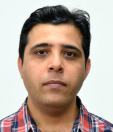}}]{Yaser Mansouri} is a researcher with the Centre for Research on Engineering Software Technologies (CREST) at the University of Adelaide.  Yaser obtained his Ph.D. from Cloud Computing and Distributed Systems (CLOUDS) Laboratory, at the University of Melbourne, Australia. Yaser was awarded a first-class scholarship supporting his Ph.D. studies. His research interests cover the broad area of data management in cloud storage services.
\end{IEEEbiography}

\vspace{-11 mm}
\begin{IEEEbiography}[{\includegraphics[width=1in,height=1.25in,clip, keepaspectratio]{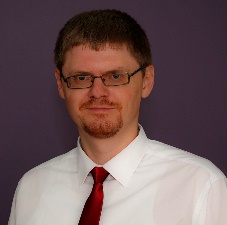}}]{Victor Prokorenko} is a researcher with the Centre for Research on Engineering Software Technologies (CREST) at the University of Adelaide. Victor has more than 14 years of experience in software engineering with main areas of expertise including the investigation of technologies related to software resilience, trust management, and big data solutions hosted within OpenStack and Microsoft Azure cloud platforms. Victor has obtained a Ph.D. in Computer Science from the University of South Australia.
\end{IEEEbiography}

\vspace{-11 mm}
\begin{IEEEbiography}[{\includegraphics[width=1in,height=1.25in,clip, keepaspectratio]{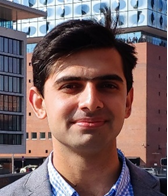}}]{Faheem Ullah} is a lecturer and cyber security program coordinator at the University of Adelaide, Australia. Faheem Ullah is also a member of CREST - Centre for Research on Engineering Software Technologies. He  completed his Ph.D. and Postdoc at the University of Adelaide. Faheem's research and teaching interests include big data analytics, cyber security, software engineering, and cloud computing. Faheem is a two times gold medalist, one-time silver medalist, and receiver of 6 academic distinctions. 
\end{IEEEbiography}

\vspace{-11 mm}
\begin{IEEEbiography}[{\includegraphics[width=1in,height=1.25in,clip, keepaspectratio]{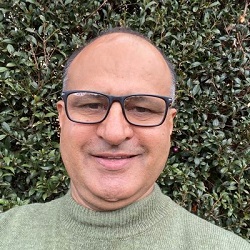}}]{M. Ali Babar}
is a Professor in the School of Computer Science, University of Adelaide, Australia. He leads a theme on architecture and platform for security as service in Cyber Security Cooperative Research Centre (CSCRC), a large initiative funded by the Australian government, industry, and research institutes. Professor Babar is the technical project lead of one of the largest projects on 
“Software Security” in ANZEC region funded by the CSCRC. SOCRATES brings more than 75 researchers and practitioners from 6 research providers and 4 industry partners for developing and evaluating novel knowledge and AI-based platforms, methods, and tools for software security. After joining the University of Adelaide, Prof Babar established an interdisciplinary research centre called CREST, Centre for Research on Engineering Software Technologies, where he directs the research, development and education activities of more than 25 researchers and engineers in the areas of Software Systems Engineering, Security and Privacy, and Social Computing. Professor Babar’s research team draws a significant amount of funding and in-kind
resources from governmental and industrial organisations. Professor Babar has authored/co-authored more than 275 peer-reviewed research papers at premier Software journals and conferences. Professor Babar obtained a Ph.D. in Computer Science and Engineering from the school of computer science and engineering of University of New South Wales, Australia. He also holds a M.Sc. degree in Computing Sciences from University of Technology, Sydney, Australia. More information on Professor Babar can be found at \url{http://malibabar.wordpress.com}.
\end{IEEEbiography}

\appendix
\appendices


\begin{figure*}[h]
  \centering
  \subfloat[Cassandra]{\label{figur:server-cass-af}\includegraphics[height=5cm,width=0.5\textwidth]{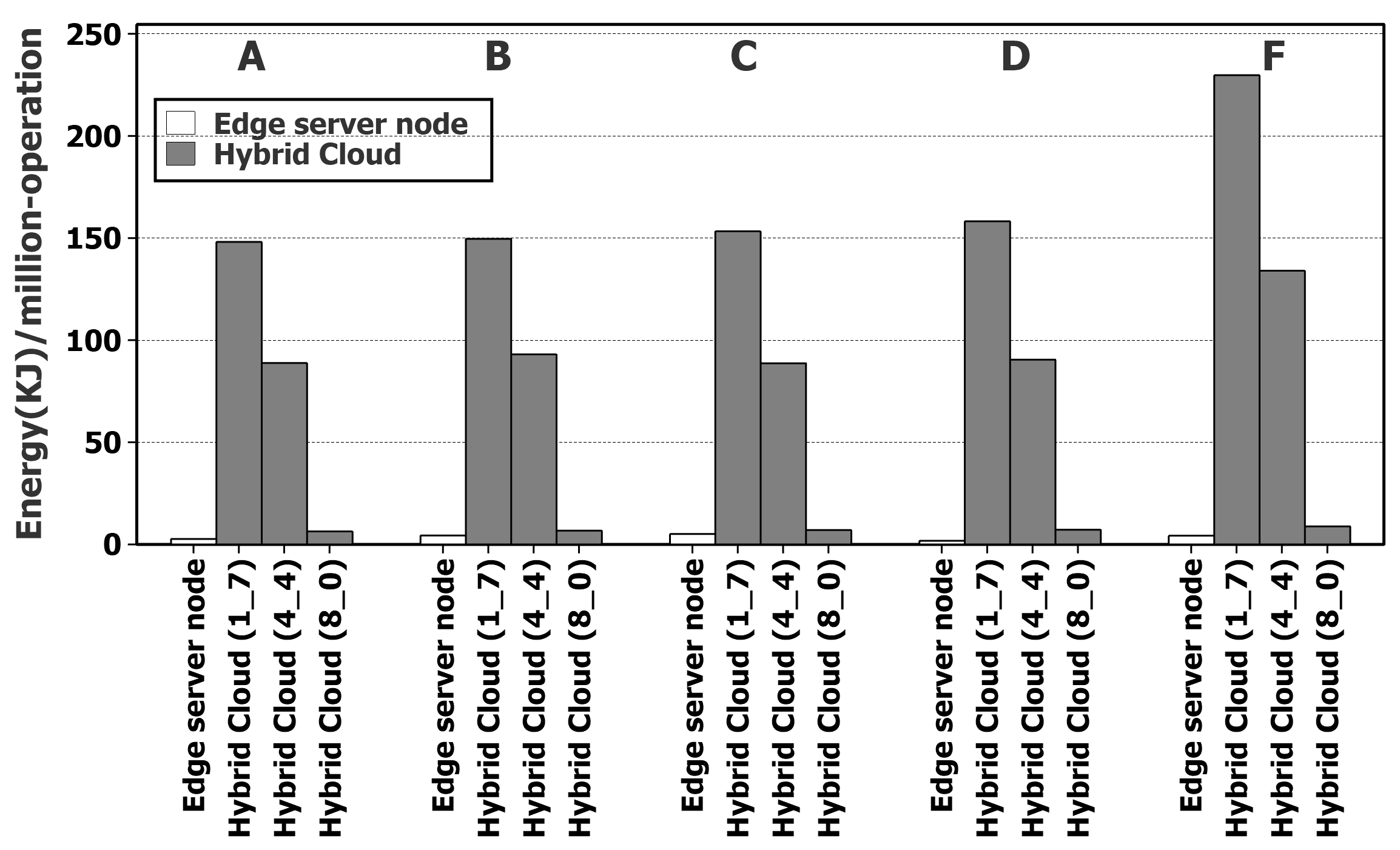}}
  \subfloat[Mongo]{\label{figur:server-mongo-af}\includegraphics[height=5cm,width=0.5\textwidth]{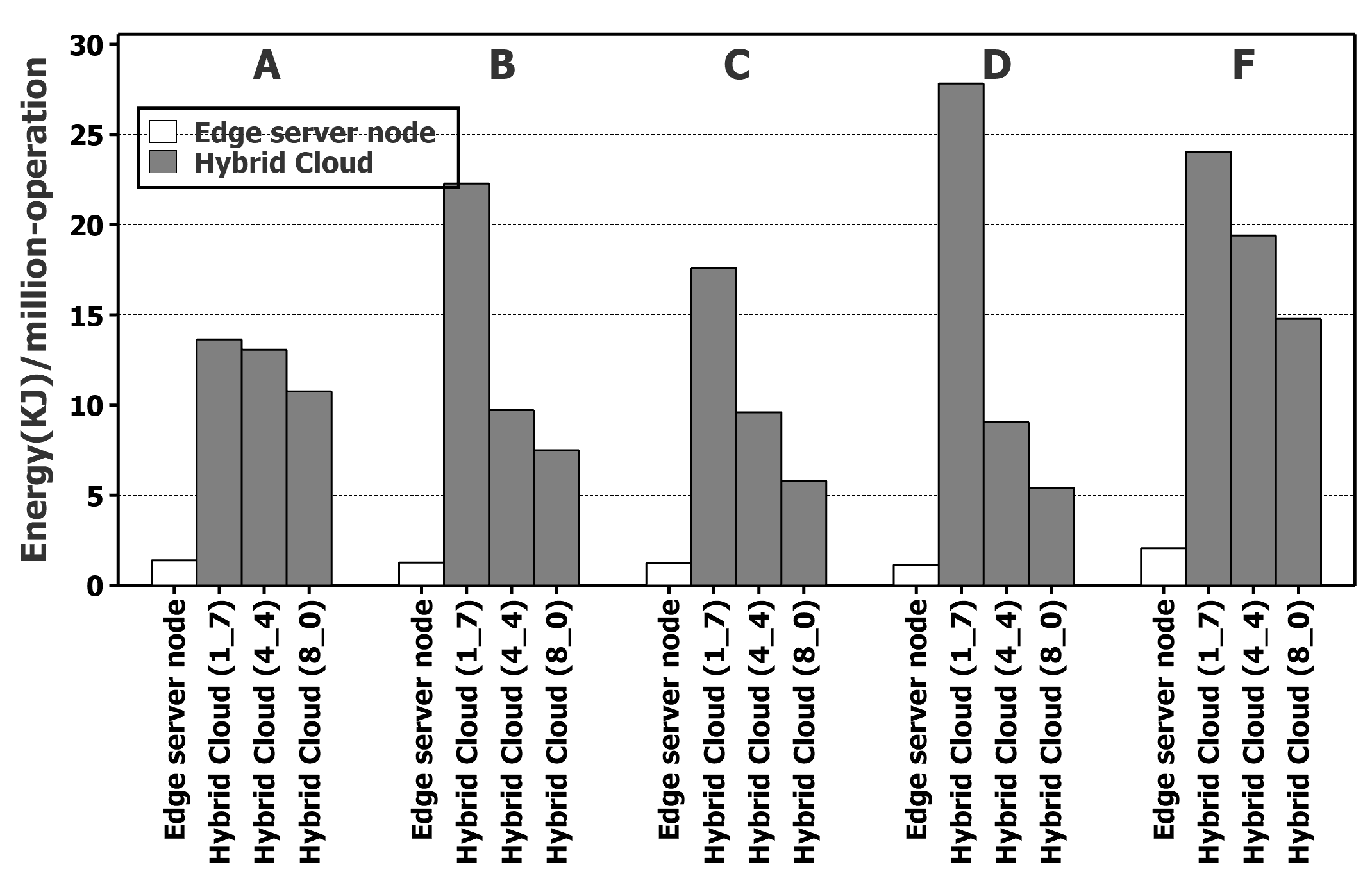}}\\
  \subfloat[Redis]{\label{figur:server-redis-af}\includegraphics[height=5cm,width=0.5\textwidth]{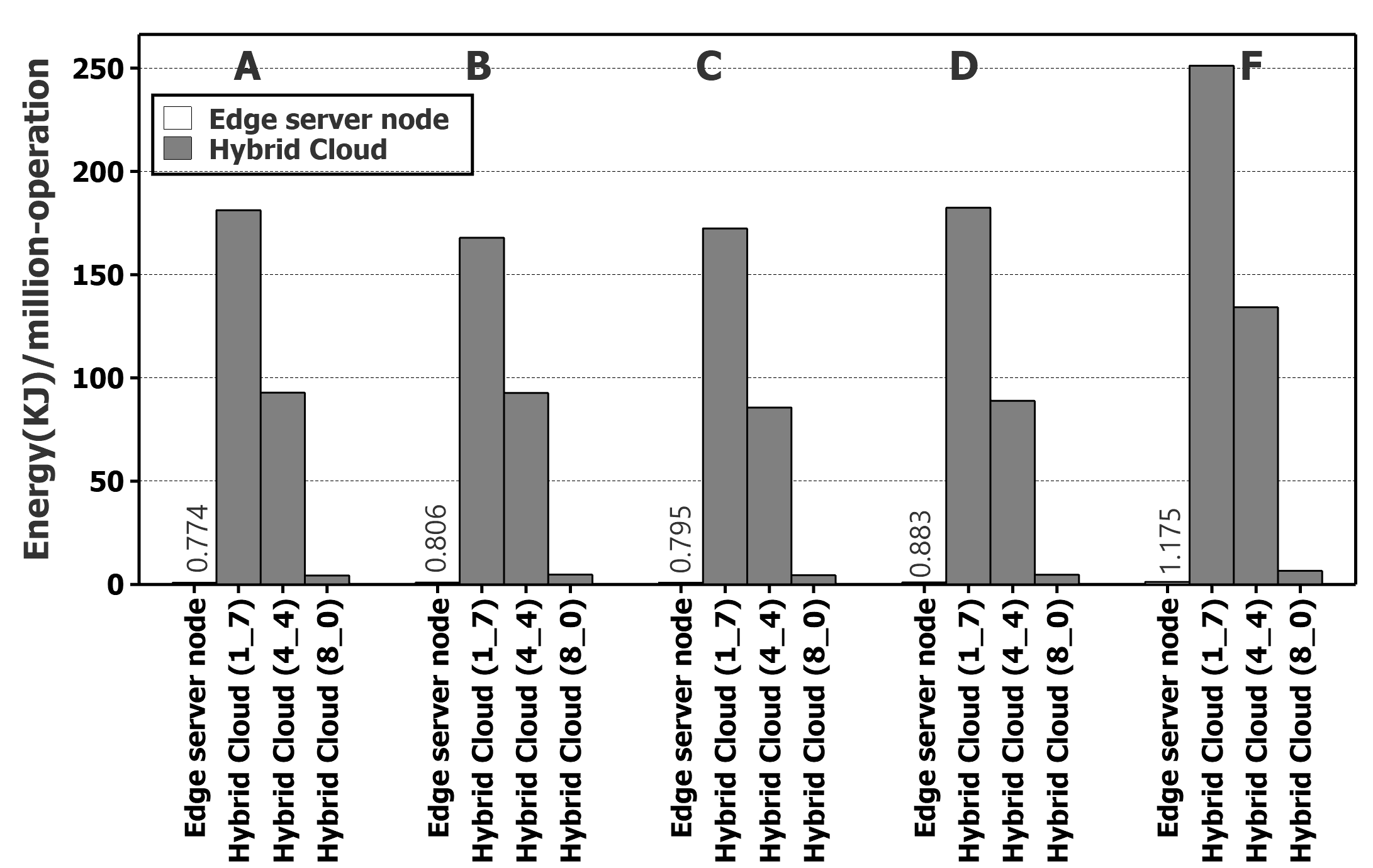}}
  \subfloat[MySQL]{\label{figur:server-mysql-af}\includegraphics[height=5cm,width=0.5\textwidth]{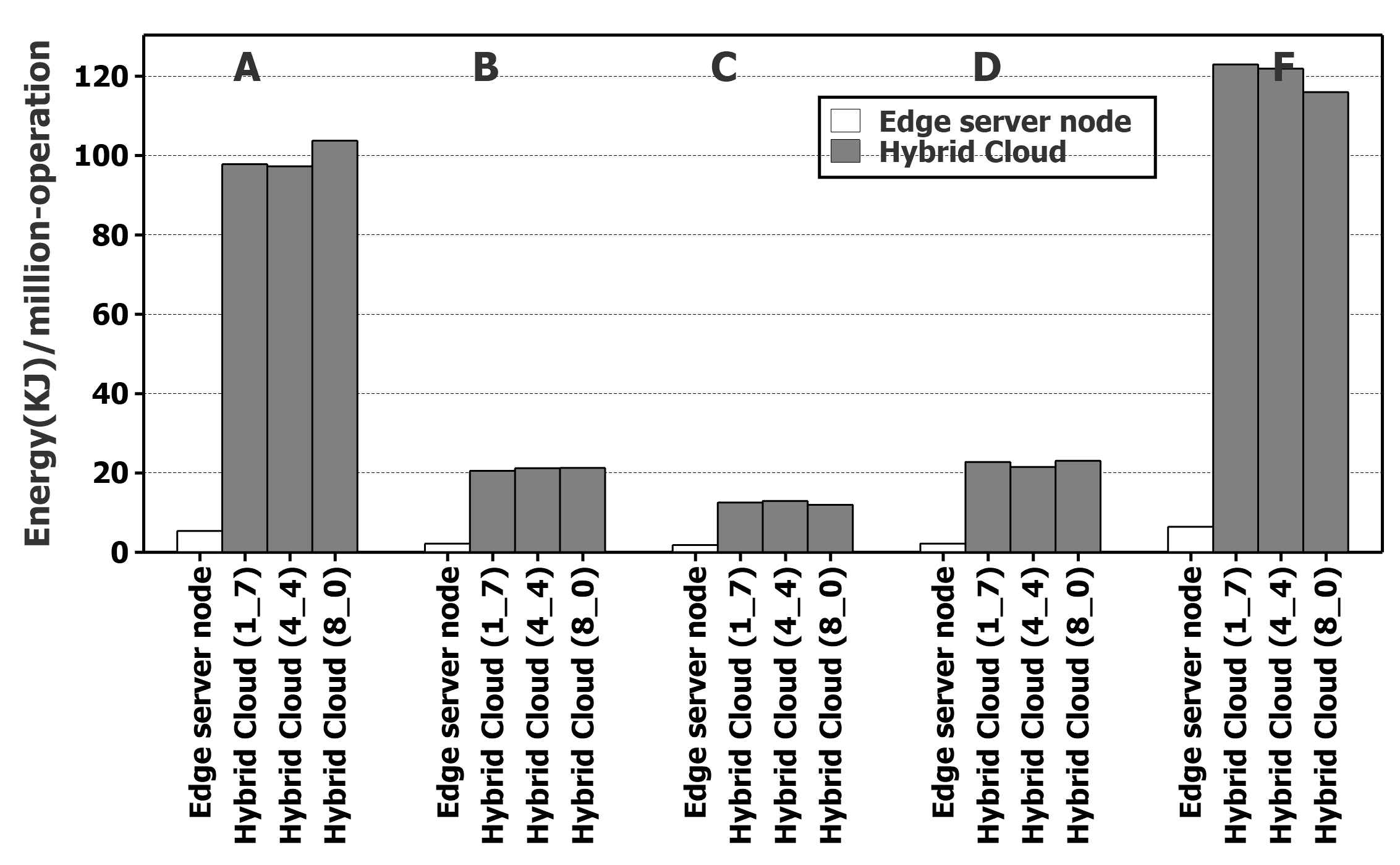}}
  \caption{Energy consumption of data offloading from \textbf{edge server node} to the edge node, edge server node, and different hybrid cloud configurations for Workloads \textbf{A}, \textbf{B}, \textbf{C}, \textbf{D}, and \textbf{F}}
\label{fig:server-af}
\vspace{-1mm}
\end{figure*}
\section{Energy Consumption of Running YCSB workloads on the edge server node}

\begin{figure*}[h]
  \centering
  \subfloat[Cassandra]{\label{figur:server-cass-e}\includegraphics[height=5cm,width=0.25\textwidth]{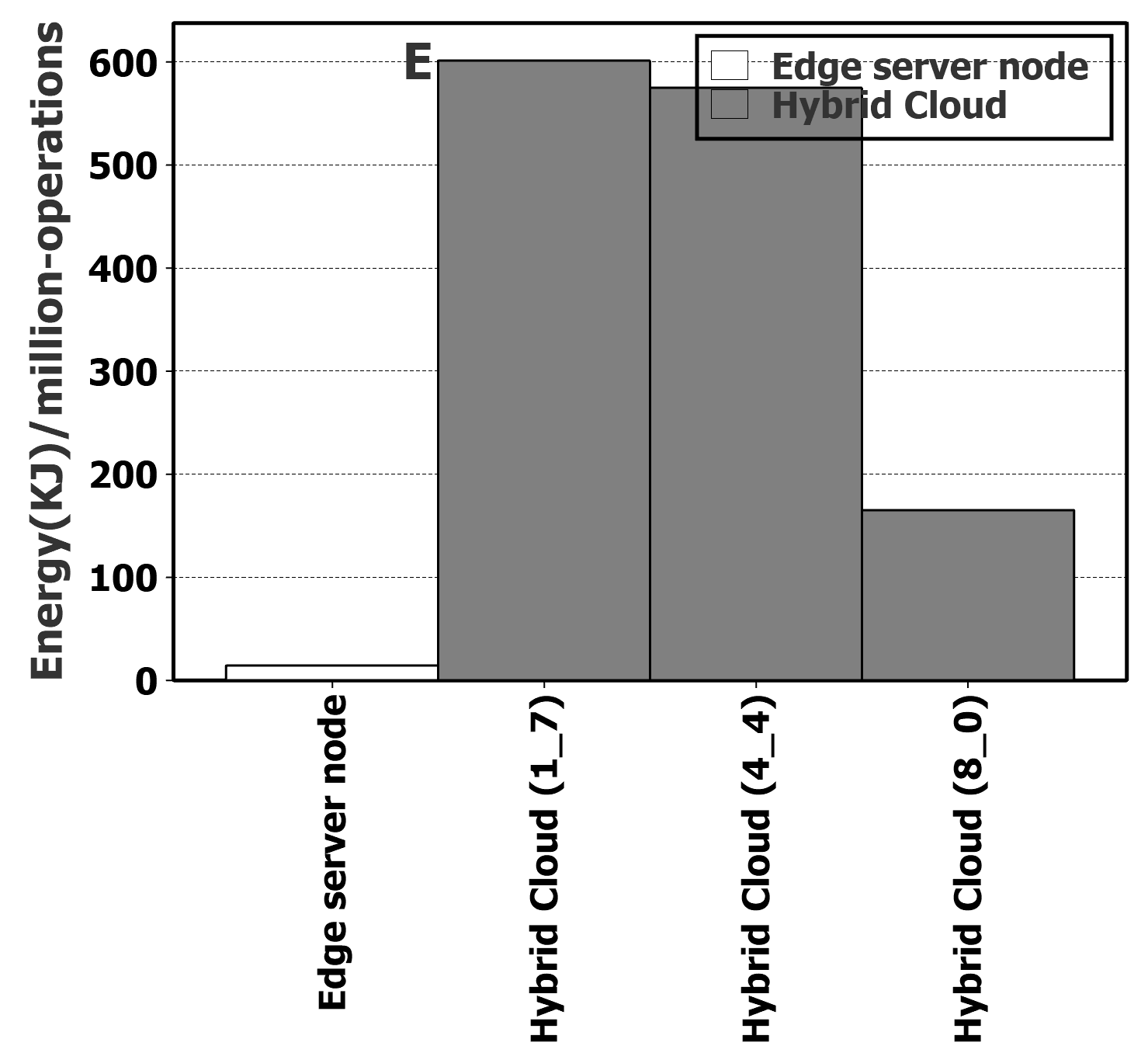}}
  \subfloat[Mongo]{\label{figur:server-mongo-e}\includegraphics[height=5cm,width=0.25\textwidth]{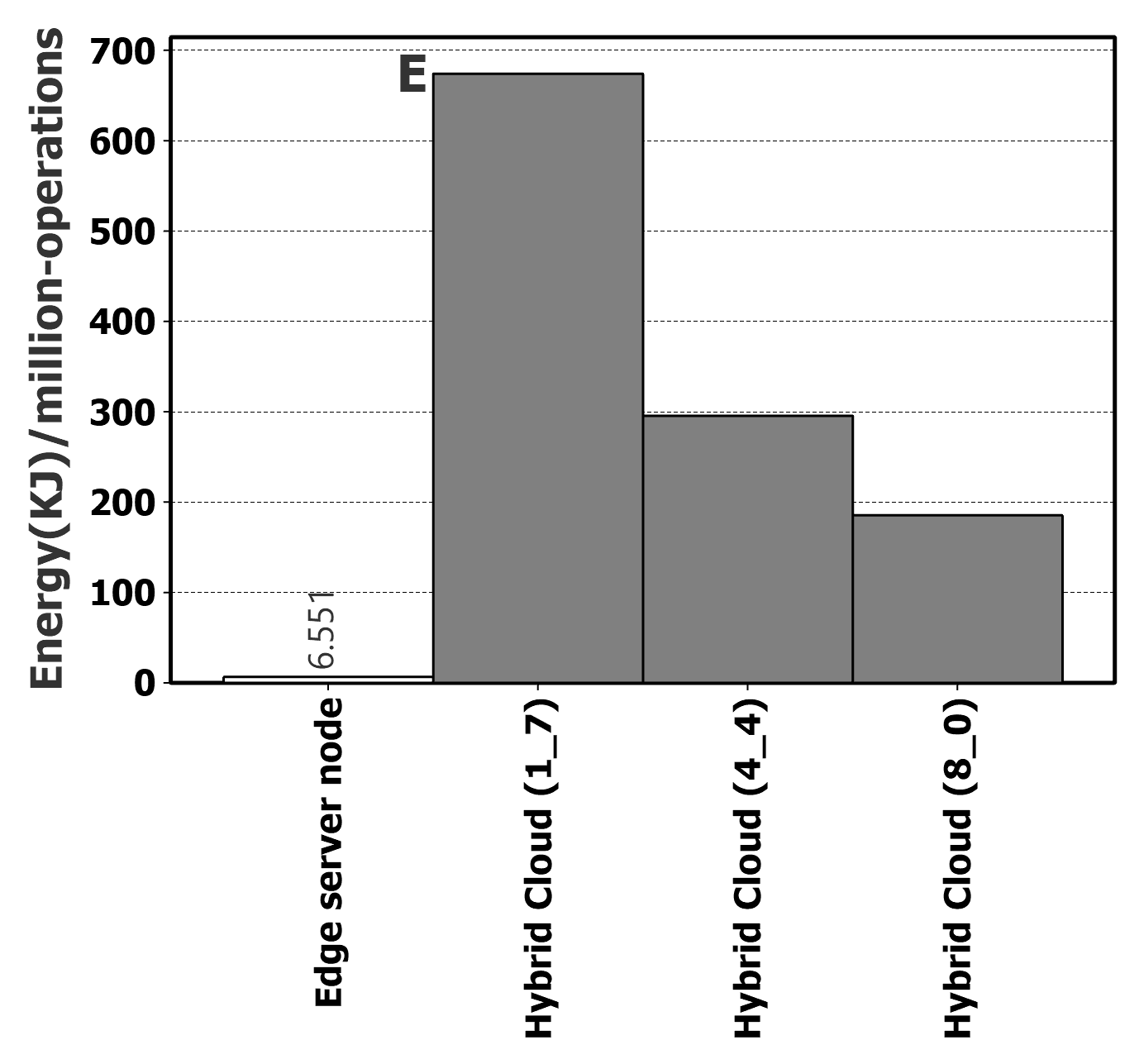}}
  \subfloat[Redis]{\label{figur:server-redis-e}\includegraphics[height=5cm,width=0.25\textwidth]{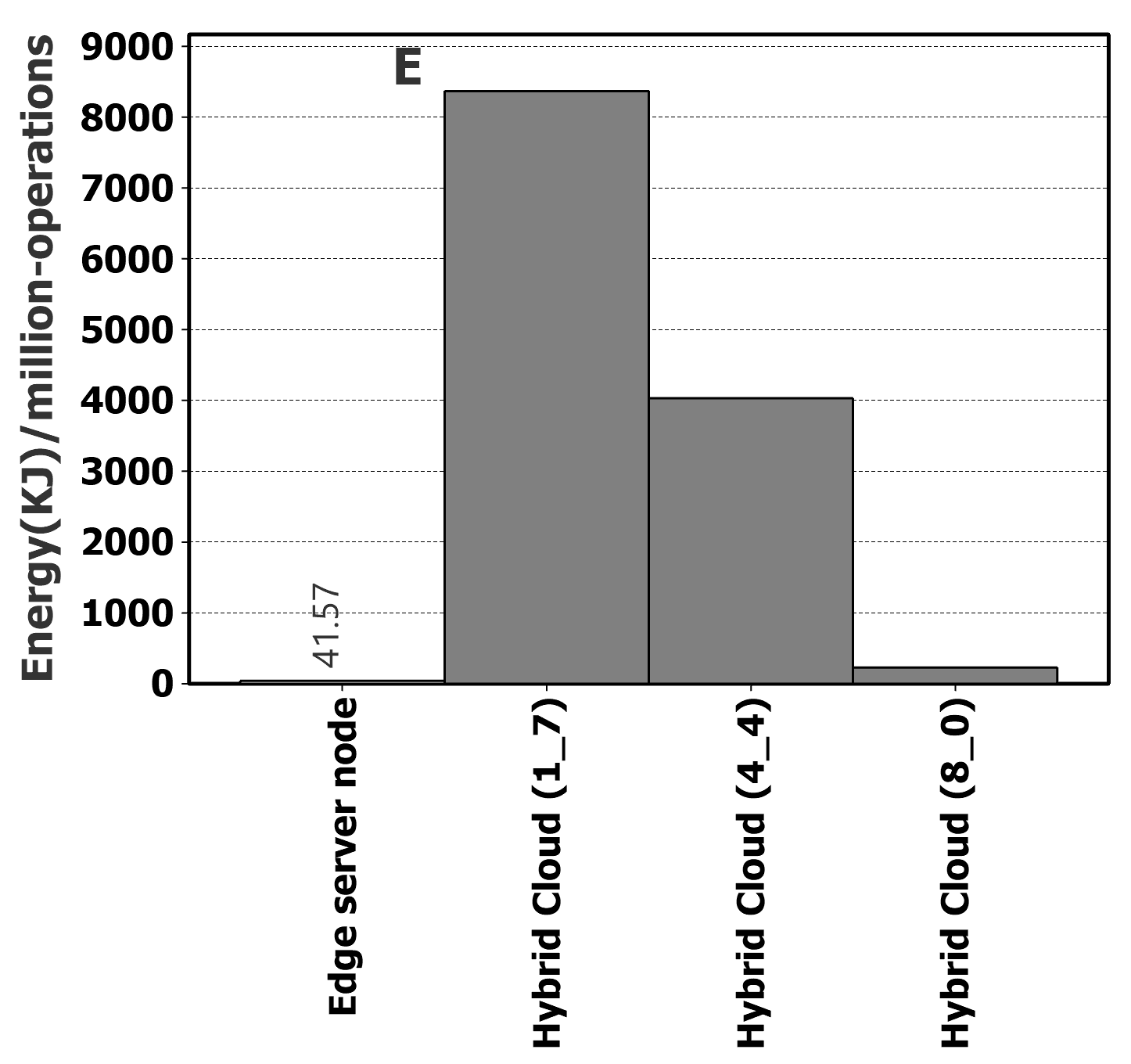}}
  \subfloat[MYSQL]{\label{figur:server-mysql-e}\includegraphics[height=5cm,width=0.25\textwidth]{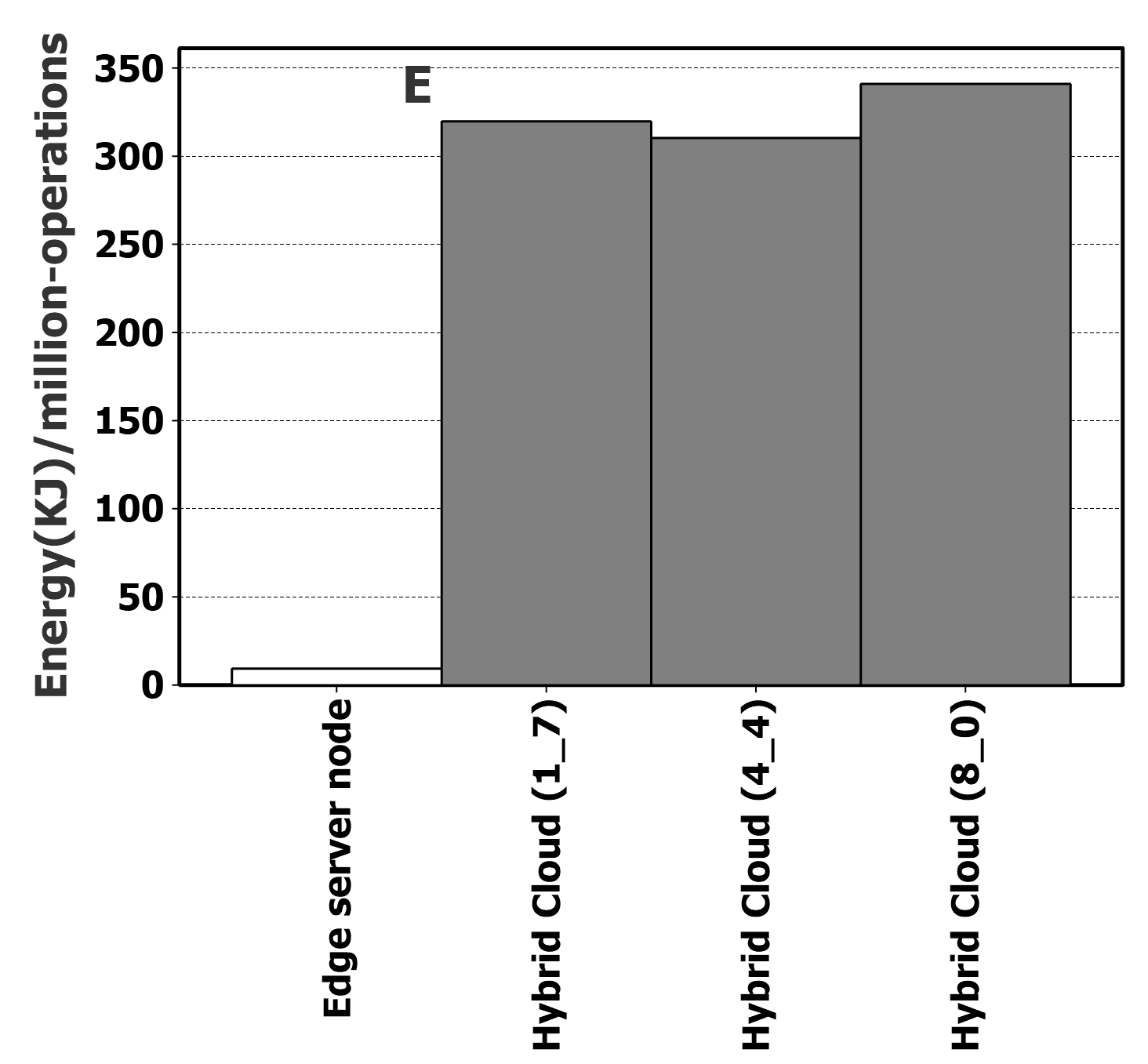}}
  \caption{Energy consumption of data offloading from \textbf{edge server node} to the edge node, edge server node, and different hybrid cloud configurations for Workloads \textbf{E}}
\label{fig:server-e}
\vspace{-6mm}
\end{figure*}

Fig. \ref{fig:server-af} shows the energy consumption of databases,  where the YSCB workload (A, B, C, D, and F) is run on the edge server node, and the hybrid cloud hosts databases.  From the results, we observe that Redis outperforms other databases in energy consumption (743-1175 J/MOPs) since it achieves the highest throughput due to being RAM-based. By contrast, disk-based databases, MySQL and Cassandra exhibit the highest energy consumption for workloads (A, D, F) and (B, C) respectively. As the databases are moved into the hybrid cluster configuration of (8\_0), Redis still outperforms all databases and consumes energy at the level of (4200-6500 J/MOPs), while MySQL has the largest values of energy consumption (11 KJ/MOPs for workload C - 116 KJ/MOPs for workload F). Cassandra and Mongo respectively are positioned between Redis and MySQL, where Monogo outperforms Cassandra for read-related workloads (B, C, D) due to using eventual consistency rather than quorum-based consistency. With the hybrid cluster configuration of (1\_7), Redis, however, has the highest value in energy consumption between 167 KJ/MOPs for workload B and 251 KJ/MOPs for workload F. This is mainly because more data are transferred across the WAN network. With the same cluster configuration, Mongo  requires the lowest amount of energy for write-related workloads (A and F); Likewise MySQL for read-related workloads (B, C, and D). This implies that Mongo is faster than Cassandra to complete write operations due to eventual consistency support. 
As expected from Fig. \ref{fig:server-e},  workload E is again the most expensive workload for all databases particularly for Redis (41.5 JK/MOPs), while Mongo achieves the best for this workload (6.5 KJ/MOPs) as databases are locally run on the edge server node. Redis still keeps the same performance, where its energy consumption is 8366 KJ/MOPs for the hybrid cluster configuration of (1\_7), and this value drops by 98\% for the hybrid cluster configuration of (8\_0). This indicates Redis must be used on the VMs located across the LAN network, not WAN. With the same conditions, Cassandra and Mongo exhibit a 27\% reduction in energy consumption as the hybrid cluster configuration changes from (1\_7) to (8\_0), while MySQL exhibits energy consumption almost at the same level for all cluster configurations.

\section*{Appendix B} \label{Appen:B}
\section*{Transmit (TX) and Receive (RX) data across hybrid cloud VMs and worker }
Tables \ref{tab:txrx-1-7}-\ref{tab:txrx-8-0} summarize the total data Transmitted (TX) and Received (RX) in MB between the edge node and the hybrid cloud. For the hybrid cluster configuration of (n\_m), the first $n$ nodes from the left in tables represent the nodes in the private cloud and the rest (m nodes) refer to the nodes in the public cloud. It should be noted that our hybrid cloud consists of 8 nodes for Cassandra, Mongo, and Redis and 9 nodes for MYSQL. We installed MySQL with NDB (stands for network database) cluster, in which one node is ndb management server and 8 nodes are data nodes. In tables, the first node is the ndb management server that connects the worker and data nodes.  

Table \ref{tab:txrx-1-7} shows the results for the hybrid cluster configuration of (1\_7). As can be seen, workload E generates the highest amount of data transferred and received by nodes in the hybrid cloud. Redis data servers send and receive data in the scale of about 5000 MB  by all public and private nodes with the same order of magnitude, while for Cassandra three nodes are highly involved to serve workload E  (node-1, node-6, and node-8 more than  13000 MB, 6000 MB, and  15000 MB RX data receptively). For Mongo, only one node in the public cloud is significantly involved in handling workload E, while  for MySQL two nodes: one in the private cloud as the ndb node, and the other in the public cloud. For other workloads (A, B, C, D, and F), a small number of nodes mainly serve operations for Mongo and MySQL, while for Cassandra and Redis almost all nodes do. From a size perspective, Mongo server nodes send and receive  more data than  other databases, particularly Redis and MySQL. Therefore, we see that Mongo transmits more data across clouds, while Redis and MySQL send and receive the least amount of data for these workloads.

As the hybrid cluster configuration was changed to (4\_4) -- see Table \ref{tab:txrx-4-4} -- it is still the same scale of data amount that is transmitted across nodes. For example, workload E requires more bandwidth between nodes in comparison to other workloads. The main difference in the transmitted data trend is that more local nodes in the private cloud are engaged to serve workloads. This is the main reason to have fewer data sent and received  across the clouds for (4\_4) compared to (1\_7). This fact correlates with other experimental results discussed in \S 5.5.2. 

Table \ref{tab:txrx-8-0} summarises TX and RX values for (8\_0). Clearly, in this case, all nodes are local and there is no transmitted data across clouds. In this set of experiments, the summation of TX and RX  for each database and each workload are calculated (last column of the table). As can be seen, the total TX/RX for workload E is more than  all workloads, where  this value for Cassandra is more than two times (62937 MB) in comparison to Mongo and Redis. For workloads (B, C, D, and F), TX and RX values are in the same magnitude (i.e., several 100 MBs) for Redis and MySQL, while for Cassandra and Mongo,  these values change in the range of 1000-2000 MB. For write-intensive workload A, Mongo and Cassandra are very close at  TX (1728 MB) and RX (1821 MB). As shown in Table \ref{tab:txrx-8-0}, Redis transfers data  slightly more than MySQL.

\begin{table*}[h!]
\caption{Transmitted (TX) and Received (RX) data in MB between the edge node as the worker and database servers for the hybrid cluster configuration of \textbf{(1\_7)}} \label{tab:txrx-1-7}

\begin{subtable}{1\textwidth}
\centering
\rowcolors{1}{}{lightgray}
    \begin{tabular}{rrcccccccccc}
		\hline
		\multicolumn{1}{p{1.5cm}}{Database} & \multicolumn{1}{p{2cm}}{Data Direction}   &Client  &node-1 &node-2 &node-3 &node-4 &node-5 &node-6 &node-7 &node-8 &node-9\\\hline\hline
		\multicolumn{1}{c}{\multirow{2}[0]{*}{\textbf{Cassandra}}} & \multicolumn{1}{c}{\textbf{TX}} 
        &1222& 49& 54& 57& 53& 58& 50& 55& 71 &-\\
		\multicolumn{1}{c}{} & \multicolumn{1}{c}{\textbf{RX}} 
        &383& 125& 154& 167& 153& 170& 145& 159& 215 &-\\
		
		\multicolumn{1}{c}{\multirow{2}[0]{*}{\textbf{Mongo}}} & \multicolumn{1}{c}{\textbf{TX}} 
        &1811& 467& 536& 538& 1448& 535& 1186& 2601& 532 &-\\
		\multicolumn{1}{c}{} & \multicolumn{1}{c}{\textbf{RX}} 
        &896& 363& 448& 448& 1601& 443& 1168& 3844& 443 &-\\
		
		\multicolumn{1}{c}{\multirow{2}[0]{*}{\textbf{Redis}}} & \multicolumn{1}{c}{\textbf{TX}} 
        &417& 224& 270& 262& 256& 253& 264& 259& 278 &- \\
		\multicolumn{1}{c}{} & \multicolumn{1}{c}{\textbf{RX}} 
        &412& 203& 228& 237& 344& 220& 229& 358& 253 &-\\
        
		\multicolumn{1}{c}{\multirow{2}[0]{*}{\textbf{MySQL}}} & \multicolumn{1}{c}{\textbf{TX}} 
        &3612& 153& 50& 53& 667& 55& 42& 51& 50& 42 \\
		\multicolumn{1}{c}{} & \multicolumn{1}{c}{\textbf{RX}} 
        &485& 2108& 51& 1827& 54& 46& 53& 50& 50& 52\\
   \end{tabular}
   \caption{Workload A}\label{tab:detailnet-1-7-a}
\end{subtable}

\begin{subtable}{1\textwidth}
\centering
\rowcolors{1}{}{lightgray}
    \begin{tabular}{rrcccccccccc}
		\hline
		\multicolumn{1}{p{1.5cm}}{Database} & \multicolumn{1}{p{2cm}}{Data Direction}   &Client  &node-1 &node-2 &node-3 &node-4 &node-5 &node-6 &node-7 &node-8 &node-9\\\hline\hline
		\multicolumn{1}{c}{\multirow{2}[0]{*}{\textbf{Cassandra}}} & \multicolumn{1}{c}{\textbf{TX}} 
        &1154 &21   &22 &23 &22 &24 &21 &22 &29 &- \\
		\multicolumn{1}{c}{} & \multicolumn{1}{c}{\textbf{RX}} 
        & 151& 112 &143& 153& 142& 158 & 131 & 146& 203&-\\
		
		\multicolumn{1}{c}{\multirow{2}[0]{*}{\textbf{Mongo}}} & \multicolumn{1}{c}{\textbf{TX}} 
        &1565&100 &107& 107& 316& 107& 276 & 900& 107 &-\\
		\multicolumn{1}{c}{} & \multicolumn{1}{c}{\textbf{RX}} 
        &470 &104 &112& 112& 329& 109& 278 &1960& 110 &-\\
		
		\multicolumn{1}{c}{\multirow{2}[0]{*}{\textbf{Redis}}} & \multicolumn{1}{c}{\textbf{TX}} 
        &686 &110 &120& 115  &114& 119& 116 & 118& 116 &- \\
		\multicolumn{1}{c}{} & \multicolumn{1}{c}{\textbf{RX}} 
        &401 &253 &274 &388 &267& 323& 432 & 300& 361 &-\\
        
		\multicolumn{1}{c}{\multirow{2}[0]{*}{\textbf{MySQL}}} & \multicolumn{1}{c}{\textbf{TX}} 
        &2935 &44  &306 &14  &13 &8 &11 &14 &14 &14 \\
		\multicolumn{1}{c}{} & \multicolumn{1}{c}{\textbf{RX}} 
        &255 &1117 &1901 &15  &15  &15 &14 &15  &14& 14 \\
   \end{tabular}
   \caption{Workload B}\label{tab:detailnet-1-7-b}
\end{subtable}

\begin{subtable}{1\textwidth}
\centering
\rowcolors{1}{}{lightgray}
    \begin{tabular}{rrcccccccccc}
		\hline
		\multicolumn{1}{p{1.5cm}}{Database} & \multicolumn{1}{p{2cm}}{Data Direction}   &Client  &node-1 &node-2 &node-3 &node-4 &node-5 &node-6 &node-7 &node-8 &node-9\\\hline\hline
		\multicolumn{1}{c}{\multirow{2}[0]{*}{\textbf{Cassandra}}} & \multicolumn{1}{c}{\textbf{TX}} 
         &1149 &19 &21 &21 &20 &22 &19  &21& 27 &-\\
		\multicolumn{1}{c}{} & \multicolumn{1}{c}{\textbf{RX}} 
        &139 &112 &144 &150 &140 &155 &130  &148 &201 &-\\
		
		\multicolumn{1}{c}{\multirow{2}[0]{*}{\textbf{Mongo}}} & \multicolumn{1}{c}{\textbf{TX}} 
        & 1439& 22 &22& 22& 27& 22& 26& 404 &22 &-\\
		\multicolumn{1}{c}{} & \multicolumn{1}{c}{\textbf{RX}} 
        & 375& 22 &22& 23& 28& 24& 26& 1463& 24 &-\\
		
		\multicolumn{1}{c}{\multirow{2}[0]{*}{\textbf{Redis}}} & \multicolumn{1}{c}{\textbf{TX}} 
        &571 &202 &214& 204& 209& 211& 567& 213& 1408 &-\\
		\multicolumn{1}{c}{} & \multicolumn{1}{c}{\textbf{RX}} 
        &150 &113 &265& 222 &105 &141 &98 &105 &105 &-\\
        
		\multicolumn{1}{c}{\multirow{2}[0]{*}{\textbf{MySQL}}} & \multicolumn{1}{c}{\textbf{TX}} 
        & 675 &39 &288 &13  &12  &12 &13 &13 &13 &13 \\
		\multicolumn{1}{c}{} & \multicolumn{1}{c}{\textbf{RX}} 
        &241 &164 &594 &14  &13  &12& 13& 12& 13& 13\\
   \end{tabular}
   \caption{Workload C}\label{tab:detailnet-1-7-c}
\end{subtable}

\begin{subtable}{1\textwidth}
\centering
\rowcolors{1}{}{lightgray}
    \begin{tabular}{rrcccccccccc}
		\hline
		\multicolumn{1}{p{1.5cm}}{Database} & \multicolumn{1}{p{2cm}}{Data Direction}   &Client  &node-1 &node-2 &node-3 &node-4 &node-5 &node-6 &node-7 &node-8  &node-9\\\hline\hline
		\multicolumn{1}{c}{\multirow{2}[0]{*}{\textbf{Cassandra}}} & \multicolumn{1}{c}{\textbf{TX}} 
        &1151 &27 &27& 31& 29& 28& 26  & 30& 29  &-\\
		\multicolumn{1}{c}{} & \multicolumn{1}{c}{\textbf{RX}} 
         &196 &124 &143& 164& 154& 149& 134  &160& 155 &-\\
		
		\multicolumn{1}{c}{\multirow{2}[0]{*}{\textbf{Mongo}}} & \multicolumn{1}{c}{\textbf{TX}} 
        &1929 &202 &201& 201& 223& 201& 223  &1190& 202 &-  \\
		\multicolumn{1}{c}{} & \multicolumn{1}{c}{\textbf{RX}} 
        &1096 &63 &58& 61& 368& 60& 227  &2582& 58 &-\\
		
		\multicolumn{1}{c}{\multirow{2}[0]{*}{\textbf{Redis}}} & \multicolumn{1}{c}{\textbf{TX}} 
        &911 &129 &134& 136  &137& 134& 135  & 134& 145 &-\\
		\multicolumn{1}{c}{} & \multicolumn{1}{c}{\textbf{RX}} 
        &230 &150  &262& 190 &263& 190& 183   &252& 274 &-\\
         
		\multicolumn{1}{c}{\multirow{2}[0]{*}{\textbf{MySQL}}} & \multicolumn{1}{c}{\textbf{TX}} 
        &293   &56 &380  &10  &12    &17 &17  &18 &11\\
		\multicolumn{1}{c}{} & \multicolumn{1}{c}{\textbf{RX}} 
        &315 &158 &227 &18  &17  &18 &19& 18  &18 &18 \\
   \end{tabular}
   \caption{Workload D}\label{tab:detailnet-1-7-d}
\end{subtable}

\begin{subtable}{1\textwidth}
\centering
\rowcolors{1}{}{lightgray}
    \begin{tabular}{rrcccccccccc}
		\hline
		\multicolumn{1}{p{1.5cm}}{Database} & \multicolumn{1}{p{2cm}}{Data Direction}   &Client  &node-1 &node-2 &node-3 &node-4 &node-5 &node-6 &node-7 &node-8 &node-9\\\hline\hline
		\multicolumn{1}{c}{\multirow{2}[0]{*}{\textbf{Cassandra}}} & \multicolumn{1}{c}{\textbf{TX}} 
        &54805 &2600 &3235& 3244& 3245& 3224& 2233   & 3236& 2000 &-  \\
		\multicolumn{1}{c}{} & \multicolumn{1}{c}{\textbf{RX}} 
        &574 &13306 &6872& 6873& 6844& 6927& 14055  &6871& 15499 &-\\
		
		\multicolumn{1}{c}{\multirow{2}[0]{*}{\textbf{Mongo}}} & \multicolumn{1}{c}{\textbf{TX}} 
        &49337 &145 &145& 146& 169& 146& 167  & 1111& 146 &-\\
		\multicolumn{1}{c}{} & \multicolumn{1}{c}{\textbf{RX}} 
        &984 &98 &97 & 93& 228& 92& 170  & 49657& 92 &-\\
		
		\multicolumn{1}{c}{\multirow{2}[0]{*}{\textbf{Redis}}} & \multicolumn{1}{c}{\textbf{TX}} 
        &8086 &4320 &4443& 4468& 4454& 4467& 4508  &4404& 4384 &- \\
		\multicolumn{1}{c}{} & \multicolumn{1}{c}{\textbf{RX}} 
        &2308 &4885 &4970& 5686& 4980& 5143& 5786  &4978& 4799 &-\\
        
		\multicolumn{1}{c}{\multirow{2}[0]{*}{\textbf{MySQL}}} & \multicolumn{1}{c}{\textbf{TX}} 
        &98345 &284 &1263  &89  &87  &72 &89 &89  &77& 89\\
		\multicolumn{1}{c}{} & \multicolumn{1}{c}{\textbf{RX}} 
        &925 &44067 &54844 &93  &95  &85 &91 &94   &98 &92\\
   \end{tabular}
   \caption{Workload E}\label{tab:detailnet-1-7-e}
\end{subtable}

\begin{subtable}{1\textwidth}
\centering
\rowcolors{1}{}{lightgray}
    \begin{tabular}{rrcccccccccc}
		\hline
		\multicolumn{1}{p{1.5cm}}{Database} & \multicolumn{1}{p{2cm}}{Data Direction}   &Client  &node-1 &node-2 &node-3 &node-4 &node-5 &node-6 &node-7 &node-8 &node-8\\\hline\hline
		\multicolumn{1}{c}{\multirow{2}[0]{*}{\textbf{Cassandra}}} & \multicolumn{1}{c}{\textbf{TX}} 
         &790 &23 &25& 26& 24& 26& 23  & 25& 32 &- \\
		\multicolumn{1}{c}{} & \multicolumn{1}{c}{\textbf{RX}} 
        &172 &78 &99& 106& 98& 108& 91  & 102& 139 &- \\
		
		\multicolumn{1}{c}{\multirow{2}[0]{*}{\textbf{Mongo}}} & \multicolumn{1}{c}{\textbf{TX}} 
        &1017 &212 &248& 247& 710& 245& 578  & 1283& 244 &-\\
		\multicolumn{1}{c}{} & \multicolumn{1}{c}{\textbf{RX}} 
        &401 &189 &229& 226& 762& 228& 575  & 1942& 230 &-\\
		 
		\multicolumn{1}{c}{\multirow{2}[0]{*}{\textbf{Redis}}} & \multicolumn{1}{c}{\textbf{TX}} 
        &107 &271 &116& 112  &112 &117 &112  & 116& 114 &- \\
		\multicolumn{1}{c}{} & \multicolumn{1}{c}{\textbf{RX}} 
        &103 &187 &104& 102  &95 &194 &193  &102 & 97 &-\\
        
		\multicolumn{1}{c}{\multirow{2}[0]{*}{\textbf{MySQL}}} & \multicolumn{1}{c}{\textbf{TX}} 
        &1681 &66 &322 &15  & 20  & 18& 20& 20  &16 &14  \\
		\multicolumn{1}{c}{} & \multicolumn{1}{c}{\textbf{RX}} 
        &245 &460 &1326 &21  &21  &20 &21 &20  &19 &23 \\
   \end{tabular}
   \caption{Workload F}\label{tab:detailnet-1-7-f}
\end{subtable}
\end{table*}

\begin{table*}[h!]
\caption{Transmitted (TX) and Received (RX) data in MB between the edge node as the worker and database server for the hybrid cluster configuration of \textbf{(4\_4)}} \label{tab:txrx-4-4}

\begin{subtable}{1\textwidth}
\centering
\rowcolors{1}{}{lightgray}
    \begin{tabular}{rrcccccccccc}
		\hline
		\multicolumn{1}{p{1.5cm}}{Database} & \multicolumn{1}{p{2cm}}{Data Direction}   &Client  &node-1 &node-2 &node-3 &node-4 &node-5 &node-6 &node-7 &node-8 &node-9\\\hline\hline
		
		\multicolumn{1}{c}{\multirow{2}[0]{*}{\textbf{Cassandra}}} & \multicolumn{1}{c}{\textbf{TX}} 
        &1231 &49 &57 &54 &74 &52 &59 &48 &44 &-\\
		\multicolumn{1}{c}{} & \multicolumn{1}{c}{\textbf{RX}} 
        &390 &139 &168 &156 &216 &154 &178 &140 &126 &-\\
		
		\multicolumn{1}{c}{\multirow{2}[0]{*}{\textbf{Mongo}}} & \multicolumn{1}{c}{\textbf{TX}} 
        &1801 &962 &2497 &584 &2222 &288 &288 &289 &286 &-\\
		\multicolumn{1}{c}{} & \multicolumn{1}{c}{\textbf{RX}} 
        &923 &1224 &3089 &506 &2872 &151 &151 &154 &149 &-\\
		
		\multicolumn{1}{c}{\multirow{2}[0]{*}{\textbf{Redis}}} & \multicolumn{1}{c}{\textbf{TX}} 
        &293 &173 &177 &177 &185 &185 &176 &173 &190 &- \\
		\multicolumn{1}{c}{} & \multicolumn{1}{c}{\textbf{RX}} 
        &418 &156 &161 &154 &166 &163 &159 &188 &162 &-\\
        
		\multicolumn{1}{c}{\multirow{2}[0]{*}{\textbf{MySQL}}} & \multicolumn{1}{c}{\textbf{TX}} 
        &196 &166 &57 &53 &618 &58 &71 &53 &72 &52 \\
		\multicolumn{1}{c}{} & \multicolumn{1}{c}{\textbf{RX}} 
        &427 &197 &61 &132 &315 &55 &49 &52 &55 &53\\
   \end{tabular}
   \caption{Workload A}\label{tab:detailnet-4-4-a}
\end{subtable}

\begin{subtable}{1\textwidth}
\centering
\rowcolors{1}{}{lightgray}
    \begin{tabular}{rrcccccccccc}
		\hline
		\multicolumn{1}{p{1.5cm}}{Database} & \multicolumn{1}{p{2cm}}{Data Direction}   &Client  &node-1 &node-2 &node-3 &node-4 &node-5 &node-6 &node-7 &node-8 &node-9\\\hline\hline
		\multicolumn{1}{c}{\multirow{2}[0]{*}{\textbf{Cassandra}}} & \multicolumn{1}{c}{\textbf{TX}} 
        &1165 &20 &23 &22 &31 &21 &24 &20 &18 &- \\
		\multicolumn{1}{c}{} & \multicolumn{1}{c}{\textbf{RX}} 
        &155 &130 &156 &146 &202 &140 &167 &131 &118 &-\\
		
		\multicolumn{1}{c}{\multirow{2}[0]{*}{\textbf{Mongo}}} & \multicolumn{1}{c}{\textbf{TX}} 
        &1515 &271 &647 &102 &827 &72 &72 &72 &72 &-\\
		\multicolumn{1}{c}{} & \multicolumn{1}{c}{\textbf{RX}} 
        &480 &307 &700 &107 &1764 &73 &73 &73 &73 &-\\
		
		\multicolumn{1}{c}{\multirow{2}[0]{*}{\textbf{Redis}}} & \multicolumn{1}{c}{\textbf{TX}} 
        &480 &84 &86 &91 &83 &84 &90 &93 &87 &- \\
		\multicolumn{1}{c}{} & \multicolumn{1}{c}{\textbf{RX}} 
        &152 &99 &85 &107 &85 &182 &105 &212 &151 &-\\
        
		\multicolumn{1}{c}{\multirow{2}[0]{*}{\textbf{MySQL}}} & \multicolumn{1}{c}{\textbf{TX}} 
        &375 &58 &20 &272 &21 &19 &22 &22 &22 &21 \\
		\multicolumn{1}{c}{} & \multicolumn{1}{c}{\textbf{RX}} 
        &205 &75 &24 &425 &18 &34 &18 &17 &19 &17 \\
   \end{tabular}
   \caption{Workload B}\label{tab:detailnet-4-4-b}
\end{subtable}

\begin{subtable}{1\textwidth}
\centering
\rowcolors{1}{}{lightgray}
    \begin{tabular}{rrcccccccccc}
		\hline
		\multicolumn{1}{p{1.5cm}}{Database} & \multicolumn{1}{p{2cm}}{Data Direction}   &Client  &node-1 &node-2 &node-3 &node-4 &node-5 &node-6 &node-7 &node-8 &node-9\\\hline\hline
		\multicolumn{1}{c}{\multirow{2}[0]{*}{\textbf{Cassandra}}} & \multicolumn{1}{c}{\textbf{TX}} 
         &1156 &18 &21 &20 &27 &19 &22 &18 &16 &-\\
		\multicolumn{1}{c}{} & \multicolumn{1}{c}{\textbf{RX}} 
        &141 &129 &155 &145 &198 &138 &165 &130 &116 &-\\
		
		\multicolumn{1}{c}{\multirow{2}[0]{*}{\textbf{Mongo}}} & \multicolumn{1}{c}{\textbf{TX}} 
        &1492 &14 &19 &11 &469 &11 &11 &11 &11 &-\\
		\multicolumn{1}{c}{} & \multicolumn{1}{c}{\textbf{RX}} 
        &455 &15 &19 &11 &3550 &12 &12 &10 &12 &-\\
		
		\multicolumn{1}{c}{\multirow{2}[0]{*}{\textbf{Redis}}} & \multicolumn{1}{c}{\textbf{TX}} 
        &3448 &83 &85 &88 &83 &84 &87 &89 &84 &-\\
		\multicolumn{1}{c}{} & \multicolumn{1}{c}{\textbf{RX}} 
        &150 &202 &115 &242 &114 &108 &102 &105 &110 &-\\
        
		\multicolumn{1}{c}{\multirow{2}[0]{*}{\textbf{MySQL}}} & \multicolumn{1}{c}{\textbf{TX}} 
        &187 &56 &18 &256 &18 &17 &26 &26 &26 &24 \\
		\multicolumn{1}{c}{} & \multicolumn{1}{c}{\textbf{RX}} 
        &192 &65 &25 &234 &17 &52 &18 &17 &19 &16\\
   \end{tabular}
   \caption{Workload C}\label{tab:detailnet-4-4-c}
\end{subtable}

\begin{subtable}{1\textwidth}
\centering
\rowcolors{1}{}{lightgray}
    \begin{tabular}{rrcccccccccc}
		\hline
		\multicolumn{1}{p{1.5cm}}{Database} & \multicolumn{1}{p{2cm}}{Data Direction}   &Client  &node-1 &node-2 &node-3 &node-4 &node-5 &node-6 &node-7 &node-8  &node-9\\\hline\hline
		\multicolumn{1}{c}{\multirow{2}[0]{*}{\textbf{Cassandra}}} & \multicolumn{1}{c}{\textbf{TX}} 
        &1163 &28 &32 &29 &31 &28 &27 &30 &25 &-\\
		\multicolumn{1}{c}{} & \multicolumn{1}{c}{\textbf{RX}} 
         &205 &143 &171 &152 &155 &145 &142 &154 &125 &-\\
		
		\multicolumn{1}{c}{\multirow{2}[0]{*}{\textbf{Mongo}}} & \multicolumn{1}{c}{\textbf{TX}} 
        &1872 &165 &188 &153 &1064 &155 &155 &155 &155 &-  \\
		\multicolumn{1}{c}{} & \multicolumn{1}{c}{\textbf{RX}} 
        &1028 &306 &593 &22 &2031 &20 &21 &21 &20 &-\\
		
		\multicolumn{1}{c}{\multirow{2}[0]{*}{\textbf{Redis}}} & \multicolumn{1}{c}{\textbf{TX}} 
        &432 &104 &113 &102 &102 &104 &102 &101 &104 &-\\
		\multicolumn{1}{c}{} & \multicolumn{1}{c}{\textbf{RX}} 
        &229 &94 &111 &104 &92 &225 &123 &113 &173 &-\\
         
		\multicolumn{1}{c}{\multirow{2}[0]{*}{\textbf{MySQL}}} & \multicolumn{1}{c}{\textbf{TX}} 
        &188 &60 &21 &326 &21 &19 &21 &21 &21 &20\\
		\multicolumn{1}{c}{} & \multicolumn{1}{c}{\textbf{RX}} 
        &258 &71 &20 &242 &18 &25 &20 &17 &21 &19 \\
   \end{tabular}
   \caption{Workload D}\label{tab:detailnet-4-4-d}
\end{subtable}

\begin{subtable}{1\textwidth}
\centering
\rowcolors{1}{}{lightgray}
    \begin{tabular}{rrcccccccccc}
		\hline
		\multicolumn{1}{p{1.5cm}}{Database} & \multicolumn{1}{p{2cm}}{Data Direction}   &Client  &node-1 &node-2 &node-3 &node-4 &node-5 &node-6 &node-7 &node-8 &node-9\\\hline\hline
		\multicolumn{1}{c}{\multirow{2}[0]{*}{\textbf{Cassandra}}} & \multicolumn{1}{c}{\textbf{TX}} 
        &54332 &1323 &1316 &1309 &1389 &1180 &92 &1119 &1163 &-  \\
		\multicolumn{1}{c}{} & \multicolumn{1}{c}{\textbf{RX}} 
        &627 &6824 &6988 &6990 &6997 &6766 &14712 &6604 &6716 &-\\
		
		\multicolumn{1}{c}{\multirow{2}[0]{*}{\textbf{Mongo}}} & \multicolumn{1}{c}{\textbf{TX}} 
        &50911 &291 &383 &236 &1960 &234 &237 &236 &237 &-\\
		\multicolumn{1}{c}{} & \multicolumn{1}{c}{\textbf{RX}} 
        &1688 &335 &487 &217 &51131 &234 &206 &219 &209 &-\\
		
		\multicolumn{1}{c}{\multirow{2}[0]{*}{\textbf{Redis}}} & \multicolumn{1}{c}{\textbf{TX}} 
        &10732 &3703 &3917 &3711 &3694 &3725 &3755 &3720 &3754 &- \\
		\multicolumn{1}{c}{} & \multicolumn{1}{c}{\textbf{RX}} 
        &5536 &3912 &4071 &7800 &3148 &3866 &4533 &3914 &3930 &-\\
        
		\multicolumn{1}{c}{\multirow{2}[0]{*}{\textbf{MySQL}}} & \multicolumn{1}{c}{\textbf{TX}} 
        &53923 &1391 &125 &132 &133 &369 &115 &109 &113 &112\\
		\multicolumn{1}{c}{} & \multicolumn{1}{c}{\textbf{RX}} 
        &962 &54251 &126 &121 &115 &485 &110 &107 &116 &129\\
   \end{tabular}
   \caption{Workload E}\label{tab:detailnet-4-4-e}
\end{subtable}

\begin{subtable}{1\textwidth}
\centering
\rowcolors{1}{}{lightgray}
    \begin{tabular}{rrcccccccccc}
		\hline
		\multicolumn{1}{p{1.5cm}}{Database} & \multicolumn{1}{p{2cm}}{Data Direction}   &Client  &node-1 &node-2 &node-3 &node-4 &node-5 &node-6 &node-7 &node-8 &node-8\\\hline\hline
		\multicolumn{1}{c}{\multirow{2}[0]{*}{\textbf{Cassandra}}} & \multicolumn{1}{c}{\textbf{TX}} 
         &795 &22 &26 &25 &34 &23 &27 &22 &20 &- \\
		\multicolumn{1}{c}{} & \multicolumn{1}{c}{\textbf{RX}} 
        &174 &91 &106 &100 &138 &97 &114 &90 &82 &- \\
		
		\multicolumn{1}{c}{\multirow{2}[0]{*}{\textbf{Mongo}}} & \multicolumn{1}{c}{\textbf{TX}} 
        &1102 &536 &1459 &310 &1245 &132 &131 &131 &131 &-\\
		\multicolumn{1}{c}{} & \multicolumn{1}{c}{\textbf{RX}} 
        &462 &649 &1694 &301 &1706 &92 &91 &91 &91 &-\\
		 
		\multicolumn{1}{c}{\multirow{2}[0]{*}{\textbf{Redis}}} & \multicolumn{1}{c}{\textbf{TX}} 
        &266 &79 &82 &85 &80 &81 &85 &86 &81 &- \\
		\multicolumn{1}{c}{} & \multicolumn{1}{c}{\textbf{RX}} 
        &186 &71 &159 &74 &73 &152 &71 &71 &68 &-\\
        
		\multicolumn{1}{c}{\multirow{2}[0]{*}{\textbf{MySQL}}} & \multicolumn{1}{c}{\textbf{TX}} 
        &2985 &73 &17 &291 &26 &25 &33 &33 &33 &32  \\
		\multicolumn{1}{c}{} & \multicolumn{1}{c}{\textbf{RX}} 
         &207 &93 &28 &449 &23 &57 &22 &23 &25 &22 \\
   \end{tabular}
   \caption{Workload F}\label{tab:detailnet-4-4-f}
\end{subtable}
\end{table*}

\begin{table*}[h!]
\caption{Transmitted (TX) and Received (RX) data in MB between the edge node as the worker and database server for the hybrid cluster configuration of \textbf{(8\_0)}} \label{tab:txrx-8-0}

\begin{subtable}{1\textwidth}
\centering
\rowcolors{1}{}{lightgray}
    \begin{tabular}{rrccccccccccc}
		\hline
		\multicolumn{1}{p{1.5cm}}{Database} & \multicolumn{1}{p{2cm}}{Data Direction}   &Client  &node-1 &node-2 &node-3 &node-4 &node-5 &node-6 &node-7 &node-8 &node-9 &Total\\\hline\hline
		
		\multicolumn{1}{c}{\multirow{2}[0]{*}{\textbf{Cassandra}}} & \multicolumn{1}{c}{\textbf{TX}} 
        &1286 &55 &61 &55 &49 &53 &67 &49 &54 &- &1728\\
		\multicolumn{1}{c}{} & \multicolumn{1}{c}{\textbf{RX}} 
        &419 &164 &181 &159 &144 &158 &201 &144 &160 &- &1728\\
		
		\multicolumn{1}{c}{\multirow{2}[0]{*}{\textbf{Mongo}}} & \multicolumn{1}{c}{\textbf{TX}} 
        &1789 &1555 &393 &392 &394 &392 &2085 &731 &393 &- &1821\\
		\multicolumn{1}{c}{} & \multicolumn{1}{c}{\textbf{RX}} 
        &902 &1889 &269 &275 &270 &273 &3183 &791 &270 &- &1821\\
		
		\multicolumn{1}{c}{\multirow{2}[0]{*}{\textbf{Redis}}} & \multicolumn{1}{c}{\textbf{TX}} 
        &407 &108 &106 &109 &109 &120 &118 &116 &105 &- &1297 \\
		\multicolumn{1}{c}{} & \multicolumn{1}{c}{\textbf{RX}} 
        &411 &90 &145 &94 &209 &97 &79 &81 &92 &- &1297\\
        
		\multicolumn{1}{c}{\multirow{2}[0]{*}{\textbf{MySQL}}} & \multicolumn{1}{c}{\textbf{TX}} 
         &392 &176 &30 &29 &30 &29 &29 &19 &29 &654 &1024\\
		\multicolumn{1}{c}{} & \multicolumn{1}{c}{\textbf{RX}} 
        &484 &208 &55 &55 &52 &57 &56 &60 &57 &541 &1140\\
   \end{tabular}
   \caption{Workload A}\label{tab:detailnet-8-0-a}
\end{subtable}

\begin{subtable}{1\textwidth}
\centering
\rowcolors{1}{}{lightgray}
    \begin{tabular}{rrccccccccccc}
		\hline
		\multicolumn{1}{p{1.5cm}}{Database} & \multicolumn{1}{p{2cm}}{Data Direction}   &Client  &node-1 &node-2 &node-3 &node-4 &node-5 &node-6 &node-7 &node-8 &node-9 &Total\\\hline\hline
		\multicolumn{1}{c}{\multirow{2}[0]{*}{\textbf{Cassandra}}} & \multicolumn{1}{c}{\textbf{TX}} 
        &1223 &23 &26 &24 &21 &22 &28 &21 &23 &- &1411 \\
		\multicolumn{1}{c}{} & \multicolumn{1}{c}{\textbf{RX}} 
        &173 &155 &171 &151 &136 &147 &190 &138 &151&- &1411\\
		
		\multicolumn{1}{c}{\multirow{2}[0]{*}{\textbf{Mongo}}} & \multicolumn{1}{c}{\textbf{TX}} 
        &1525 &94 &94 &644 &94 &94 &480 &805 &94 &- &3922\\
		\multicolumn{1}{c}{} & \multicolumn{1}{c}{\textbf{RX}} 
        &502 &97 &96 &700 &98 &94 &493 &1746 &96 &- &3922\\
		
		\multicolumn{1}{c}{\multirow{2}[0]{*}{\textbf{Redis}}} & \multicolumn{1}{c}{\textbf{TX}} 
        &506 &57 &56 &58 &57 &61 &61 &61 &56 &- &973 \\
		\multicolumn{1}{c}{} & \multicolumn{1}{c}{\textbf{RX}} 
        &161 &52 &82 &184 &54 &57 &93 &208 &82 &- &973\\
        
		\multicolumn{1}{c}{\multirow{2}[0]{*}{\textbf{MySQL}}} & \multicolumn{1}{c}{\textbf{TX}} 
        &186 &54 &9 &9 &10 &9 &9 &6&9 &301 &602\\
		\multicolumn{1}{c}{} & \multicolumn{1}{c}{\textbf{RX}} 
        &249 &80 &17 &17 &18 &19 &16 &17 &17 &233 &681 \\
   \end{tabular}
   \caption{Workload B}\label{tab:detailnet-8-0-b}
\end{subtable}

\begin{subtable}{1\textwidth}
\centering
\rowcolors{1}{}{lightgray}
    \begin{tabular}{rrccccccccccc}
		\hline
		\multicolumn{1}{p{1.5cm}}{Database} & \multicolumn{1}{p{2cm}}{Data Direction}   &Client  &node-1 &node-2 &node-3 &node-4 &node-5 &node-6 &node-7 &node-8 &node-9 &Total\\\hline\hline
		\multicolumn{1}{c}{\multirow{2}[0]{*}{\textbf{Cassandra}}} & \multicolumn{1}{c}{\textbf{TX}} 
         &1221 &22 &24 &22 &20 &21 &26 &19 &21 &- &1396\\
		\multicolumn{1}{c}{} & \multicolumn{1}{c}{\textbf{RX}} 
        &159 &154 &171 &151 &136 &148 &189 &136 &151 &- &1396\\
		
		\multicolumn{1}{c}{\multirow{2}[0]{*}{\textbf{Mongo}}} & \multicolumn{1}{c}{\textbf{TX}} 
        &1503 &6 &6 &10 &6 &6 &9 &483 &6 &- &2033\\
		\multicolumn{1}{c}{} & \multicolumn{1}{c}{\textbf{RX}} 
        &475 &7 &5 &10 &6 &7 &8 &1509 &6 &- &2033\\
		
		\multicolumn{1}{c}{\multirow{2}[0]{*}{\textbf{Redis}}} & \multicolumn{1}{c}{\textbf{TX}} 
        &430 &53 &52 &53 &53 &57 &57 &56 &52 &- &862\\
		\multicolumn{1}{c}{} & \multicolumn{1}{c}{\textbf{RX}} 
        &147 &178 &79 &82 &80 &55 &106 &88 &48 &- &862\\
        
		\multicolumn{1}{c}{\multirow{2}[0]{*}{\textbf{MySQL}}} & \multicolumn{1}{c}{\textbf{TX}} 
        &375 &52 &9 &9 &9 &9 &9 &6 &9 &284 &394\\
		\multicolumn{1}{c}{} & \multicolumn{1}{c}{\textbf{RX}} 
        &234 &64 &16 &15 &18 &16 &17 &17 &16 &421 &599\\
   \end{tabular}
   \caption{Workload C}\label{tab:detailnet-8-0-c}
\end{subtable}

\begin{subtable}{1\textwidth}
\centering
\rowcolors{1}{}{lightgray}
    \begin{tabular}{rrccccccccccc}
		\hline
		\multicolumn{1}{p{1.5cm}}{Database} & \multicolumn{1}{p{2cm}}{Data Direction}   &Client  &node-1 &node-2 &node-3 &node-4 &node-5 &node-6 &node-7 &node-8  &node-9 &Total\\\hline\hline
		\multicolumn{1}{c}{\multirow{2}[0]{*}{\textbf{Cassandra}}} & \multicolumn{1}{c}{\textbf{TX}} 
        &1188 &29 &34 &31 &28 &29 &30 &28 &32 &- &1428\\
		\multicolumn{1}{c}{} & \multicolumn{1}{c}{\textbf{RX}} 
         &216 &143 &173 &152 &140 &148 &150 &142 &164 &- &1428\\
		
		\multicolumn{1}{c}{\multirow{2}[0]{*}{\textbf{Mongo}}} & \multicolumn{1}{c}{\textbf{TX}} 
        &1817 &137 &137 &169 &137 &137 &161 &978 &137 &-  &3811\\
		\multicolumn{1}{c}{} & \multicolumn{1}{c}{\textbf{RX}} 
        &951 &14 &16 &664 &16 &15 &161 &1959 &16 &- &3811\\
		
		\multicolumn{1}{c}{\multirow{2}[0]{*}{\textbf{Redis}}} & \multicolumn{1}{c}{\textbf{TX}} 
        &335 &67 &67 &67 &76 &68 &67 &67 &68 &- &831\\
		\multicolumn{1}{c}{} & \multicolumn{1}{c}{\textbf{RX}} 
        &226 &56 &58 &57 &88 &71 &57 &71 &198 &- &831\\
         
		\multicolumn{1}{c}{\multirow{2}[0]{*}{\textbf{MySQL}}} & \multicolumn{1}{c}{\textbf{TX}} 
        &189 &55 &20 &20 &20 &19 &19 &18 &19 &364 &744\\
		\multicolumn{1}{c}{} & \multicolumn{1}{c}{\textbf{RX}} 
        &303 &82 &16 &17 &17 &18 &18 &19 &18 &237 &744\\
   \end{tabular}
   \caption{Workload D}\label{tab:detailnet-8-0-d}
\end{subtable}

\begin{subtable}{1\textwidth}
\centering
\rowcolors{1}{}{lightgray}
    \begin{tabular}{rrccccccccccc}
		\hline
		\multicolumn{1}{p{1.5cm}}{Database} & \multicolumn{1}{p{2cm}}{Data Direction}   &Client  &node-1 &node-2 &node-3 &node-4 &node-5 &node-6 &node-7 &node-8 &node-9 &Total\\\hline\hline
		\multicolumn{1}{c}{\multirow{2}[0]{*}{\textbf{Cassandra}}} & \multicolumn{1}{c}{\textbf{TX}} 
        &54183 &109 &1194 &1230 &1245 &1238 &1261 &1248 &1229 &-  &62937\\
		\multicolumn{1}{c}{} & \multicolumn{1}{c}{\textbf{RX}} 
        &707 &14682 &6526 &6854 &6747 &6800 &6967 &6856 &6799 &- &62937\\
		
		\multicolumn{1}{c}{\multirow{2}[0]{*}{\textbf{Mongo}}} & \multicolumn{1}{c}{\textbf{TX}} 
        &17059 &1375 &1372 &978 &1378 &1385 &318 &1368 &1399 &- &26633\\
		\multicolumn{1}{c}{} & \multicolumn{1}{c}{\textbf{RX}} 
        &516 &1363 &1406 &17626 &1329 &1395 &312 &1373 &1314 &- &26633\\
		
		\multicolumn{1}{c}{\multirow{2}[0]{*}{\textbf{Redis}}} & \multicolumn{1}{c}{\textbf{TX}} 
        &8933 &2589 &2557 &2602 &2794 &2587 &2581 &2611 &2597 &-  &29851\\
		\multicolumn{1}{c}{} & \multicolumn{1}{c}{\textbf{RX}} 
        &6966 &2577 &2528 &1781 &2692 &2619 &5585 &2604 &2499 &- &29851\\
        
		\multicolumn{1}{c}{\multirow{2}[0]{*}{\textbf{MySQL}}} & \multicolumn{1}{c}{\textbf{TX}} 
        &5436 &390 &63 &62 &62 &62 &62 &39 &62 &1319 &7557\\
		\multicolumn{1}{c}{} & \multicolumn{1}{c}{\textbf{RX}} 
        &940 &529 &124 &123 &124 &121 &125 &133 &126 &305 &2649\\
   \end{tabular}
   \caption{Workload E}\label{tab:detailnet-8-0-e}
\end{subtable}

\begin{subtable}{1\textwidth}
\centering
\rowcolors{1}{}{lightgray}
    \begin{tabular}{rrccccccccccc}
		\hline
		\multicolumn{1}{p{1.5cm}}{Database} & \multicolumn{1}{p{2cm}}{Data Direction}   &Client  &node-1 &node-2 &node-3 &node-4 &node-5 &node-6 &node-7 &node-8 &node-9 &Total\\\hline\hline
		\multicolumn{1}{c}{\multirow{2}[0]{*}{\textbf{Cassandra}}} & \multicolumn{1}{c}{\textbf{TX}} 
         &822 &26 &29 &27 &23 &25 &31 &24 &26 &- &1032 \\
		\multicolumn{1}{c}{} & \multicolumn{1}{c}{\textbf{RX}} 
        &188 &106 &115 &104 &92 &101 &129 &95 &103 &- &1032\\
		
		\multicolumn{1}{c}{\multirow{2}[0]{*}{\textbf{Mongo}}} & \multicolumn{1}{c}{\textbf{TX}} 
        &1085 &196 &196 &1174 &195 &195 &860 &988 &196 &- &5084\\
		\multicolumn{1}{c}{} & \multicolumn{1}{c}{\textbf{RX}} 
        &471 &167 &167 &1401 &167 &164 &881 &1498 &168 &- &5084\\
		 
		\multicolumn{1}{c}{\multirow{2}[0]{*}{\textbf{Redis}}} & \multicolumn{1}{c}{\textbf{TX}} 
        &351 &53 &52 &53 &53 &58 &57 &56 &51 &- &783\\
		\multicolumn{1}{c}{} & \multicolumn{1}{c}{\textbf{RX}} 
        &183 &129 &125 &132 &42 &44 &44 &44 &42 &- &783\\
        
		\multicolumn{1}{c}{\multirow{2}[0]{*}{\textbf{MySQL}}} & \multicolumn{1}{c}{\textbf{TX}} 
        &128 &69 &12 &12 &12 &12 &12 &8 &12 &310 &585\\
		\multicolumn{1}{c}{} & \multicolumn{1}{c}{\textbf{RX}} 
         &243 &103 &24 &21 &21 &21 &22 &22 &23 &189 &688 \\
   \end{tabular}
   \caption{Workload F}\label{tab:detailnet-8-0-f}
\end{subtable}
\end{table*}






\end{document}